\documentclass[12pt,preprint]{aastex}

\usepackage{graphics,epsfig}
\usepackage{natbib}
\usepackage{lscape}

\def\spose#1{\hbox to 0pt{#1\hss}}
\def\lta{\mathrel{\spose{\lower 3pt\hbox{$\mathchar"218$}}
     \raise 2.0pt\hbox{$\mathchar"13C$}}}
\def\gta{\mathrel{\spose{\lower 3pt\hbox{$\mathchar"218$}}
     \raise 2.0pt\hbox{$\mathchar"13E$}}}

\def\figure#1#2 {\par{\narrower\noindent {\bf Fig. #1}
   \hskip 2mm #2\par}\bigskip\noindent}
\def\table#1#2 {\par{\narrower\noindent {\bf Tab. #1}
   \hskip 2mm #2\par}\bigskip\noindent}

\def\registered{{\ooalign{\hfil\raise .00ex\hbox{\scriptsize R}\hfil\crcr\mathhexbox20D}}}

\shorttitle{Habitability in Binary Systems I}

\shortauthors{Cuntz}

\begin{document}


\title{S-Type and P-Type Habitability in Stellar Binary Systems: \\
A Comprehensive Approach \\
I.~Method and Applications}

\author{M. Cuntz}

\affil{Department of Physics}
\affil{University of Texas at Arlington, Arlington, TX 76019-0059;}
\email{cuntz@uta.edu}

\begin{abstract}
A comprehensive approach is provided to the study of both
S-type and P-type habitability in stellar binary systems, which in
principle can also be expanded to systems of higher order.  P-type
orbits occur when the planet orbits both binary components, whereas
in case of S-type orbits the planet orbits only one of the binary
components with the second component considered a perturbator.  The
selected approach encapsulates a variety of different aspects,
which include:  (1) The consideration of a joint constraint including
orbital stability and a habitable region for a putative system
planet through the stellar radiative energy fluxes (``radiative
habitable zone"; RHZ) needs to be met.  (2) The treatment of
conservative, general and extended zones of habitability for the
various systems as defined for the Solar System and beyond.
(3) The providing of a combined formalism for the assessment of both
S-type and P-type habitability; in particular, mathematical criteria
are presented for which kind of system S-type and P-type habitability
is realized.  (4) Applications of the attained theoretical approach to
standard (theoretical) main-sequence stars.  In principle, five different
cases of habitability are identified, which are: S-type and P-type
habitability provided by the full extent of the RHZs; habitability, where
the RHZs are truncated by the additional constraint of planetary orbital
stability (referred to as ST and PT-type, respectively); and cases of no
habitability at all.  Regarding the treatment of planetary orbital stability,
we utilize the formulae of Holman \& Wiegert (1999) [AJ 117, 621] as
also used in previous studies.
In this work we focus on binary systems in circular orbits.  Future
applications will also consider binary systems in elliptical orbits
and provide thorough comparisons to other methods and results
given in the literature.
\end{abstract}



\section{Introduction}

Starting more than a decade ago considerable observational evidence has
been obtained indicating that planets are able to exist in stellar binary
(and higher order) systems; see results and discussions by, e.g.,
\cite{pat02}, \cite{egg04}, and \cite{egg07}.  These observations are
in line with the empirical finding that binary (and higher order)
systems occur in high frequency in the local Galactic neighborhood
\citep{duq91,lad06,rag06,bon07,rag10}.  For example, \cite{rag10}
presented results of a detailed analysis of companions to solar-type
stars, based on a sample size of 454, and concluded that the overall
fractions of double and triple systems are about 33\% and 8\%, respectively,
if all confirmed stellar and brown dwarf companions are accounted for.
Updated results were meanwhile given by \cite{roe12}.  This study
shows that 57 exoplanet host stars are identified having a stellar
companion.

The fairly frequent occurrence of
planets in binary systems is furthermore consistent with the presence
of debris disks in a considerably large number of main-sequence star
binary systems \citep[e.g.,][]{tri07}.  In principle, as discussed by
\cite{per11}, planets in binary systems can be identified through
two different venues:  First, binaries or multiple star systems can
be surveyed for the presence of planets by utilizing the
established detection methods.  Second, stars with detected
planets can be scrutinized afterward to check if they possess
one or more widely separated stellar companion(s); in this case,
the planet(s) will also be categorized as belonging to a binary
(or higher order) system.

From the view point of orbital mechanics, there are two different kinds
of possible orbits (notwithstanding positions near the
Lagrangian points L$_4$ and L$_5$) for planets in binary systems:
S-type and P-type orbits \citep{dvo82}.
A P-type orbit is given when the planet orbits both binary components, whereas
in case of an S-type orbit the planet orbits only one of the binary components
with the second component behaving as a perturbator.  \cite{egg04} presented
a list of 15 planet-bearing binary systems with all planets in S-type orbits.
They constitute mostly wide binaries with separation distances of up to
$\sim$6400~AU; however, smaller separation distances on the order of
20~AU or less have also been identified.  In the meantime, systems
with planets in P-type orbits have also been identified.  Arguably,
the most prominent case is Kepler-16, as reported by \cite{doy11} and
previously suggested by \cite{sla11}, containing a Saturnian mass 
circumbinary planet.  \cite{qua12} have subsequently studied this system
regarding the possibility of habitable exoplanets and habitable exomoons.
Recently, a transiting circumbinary multiplanet system, i.e., Kepler-47,
has also been identified \citep{oro12}.

There is a significant body of literature devoted to the study of
habitability\footnote{The notion of habitability adopted in this study
follows the conventional concept of \cite{kas93} and related work,
where habitability is defined based on the principal possibility that
liquid water is able to exist on the surface of an Earth-type planet
possessing a CO$_2$/H$_2$O/N$_2$ atmosphere (see Sect.~2 for details).
More sophisticated approaches to habitability have been given in the
meantime taking into account additional aspects, such as the planet's
size and mass, atmospheric structure and composition, magnetic field,
geodynamic properties, ionizing stellar UV and X-ray fluxes, and
tidal locking (if existing)
\citep[e.g.,][]{kas03,sca07,tar07,zah07,sel08,lam09,lam10,kal10,hor10,cun12,for13,luc13}.
Additionally, as pointed out by \cite{wil02}, planets with sufficiently
thick atmospheres may remain habitable even when temporarily absent from
their HZs due to orbits of considerable ellipticity.  This possibility is
disregarded in the following as well, as planets will be required to stay
permanently in the CHZ, GHZ, or EHZ (see Sect.~2 for definitions),
as applicable, to be considered habitable.}
in binary systems as well as in multiplanetary systems,
which often also encompass stellar evolutionary considerations.  Examples
include the work by \cite{jon01}, \cite{nob02}, \cite{men03}, \cite{asg04},
\cite{san07}, \cite{tak08}, \cite{dvo10}, \cite{jon10}, and
\cite{kop10}.  An important aspect that has received increased recognition
in the literature is that in order for habitability to exist and to be maintained,
a joint constraint that includes both orbital stability and a habitable environment
for a system planet through the stellar radiative energy fluxes needs to be met.
In the framework of this paper, the zone related to this latter requirement will
subsequently be referred to as {\it radiative habitable zone} (RHZ), which
constitutes a necessary, though often insufficient, condition for the existence
of circumstellar habitability.

Previous work, mostly concentrated on the existence of habitability in single star
multi-planetary systems, rendered the publication of detailed ``stability catalogs"
for the habitability zones of extrasolar planetary systems
\citep[e.g.,][]{men03,san07,tak08,dvo10,kop10}.  For example, \cite{men03}
quantified the dynamical habitability of 85 planetary systems by considering the
perturbing influence of giant planets beyond the traditional Hill sphere for
close encounters with the theoretical terrestrial planets.  They concluded that
a significant fraction of the identified extrasolar planetary systems are unable
to harbor habitable terrestrial planets.  A statistical study on the stability of
Earth-mass planets orbiting solar-mass stars in presence of stellar companions
focusing on both the statistical properties of ejection times and the general
prospects of planetary habitability was given by \cite{fat06}.  Additional work
providing stability assessments for various observed extrasolar planetary systems
based on detailed stability maps was given by \cite{san07}.  \cite{tak08} explored
the orbital stability or planets in double-planet systems for binaries by supplying
an analytic framework based on secular perturbation theory; they also provided
dynamical classification categories.  Additional stability analyses to
assess the habitability of planetary systems based on detailed numerical
simulations were given by \cite{dvo10} and \cite{kop10}; note that the
study of \cite{dvo10} also dealt with a limited cases of planets in double
star systems in orbit either around one stellar component (S-type) or around
both components (P-type).

The study of planetary dynamics and habitable planet formation has meanwhile
been described by, e.g., \cite{qui10} and \cite{hag10}.  They show that
Earth-mass planets are, in principle, able to form in stellar binary systems,
although many details of the relevant processes are not fully understood.
The overarching
conclusion of those investigations is that habitable planets in stellar binary
(and, as anticipated, in higher-order systems) are, in general, possible, 
which is a stark motivation for providing a comprehensive study of S-type
and P-type habitability in binary systems.  The approach adopted in this
study will be entirely analytic.  Specifically, it will consider both
S-type and P-type habitable orbits in the view of the joint constraint
including orbital stability and a habitable region for a system planet
through the appropriate amount of the stellar radiative energy fluxes.
In an earlier study, \cite{egg12} focused on S-type habitability in binary
systems taking into account both circular and elliptical orbits for the
stellar binary components; this latter aspect is however beyond the scope
of the present work as we solely focus on systems in circular orbits.
Numerical studies for P-type habitable environments with applications to
Kepler-16, Kepler-34, Kepler-35 and Kepler-47 have been given by \cite{kan13}.

Our paper is structured as follows: In Sect.~2, we comment on the
adopted main-sequence star parameters and single star habitability.
In Sect.~3, we introduce our theoretical approach suitable for
stellar systems of the order of $N$, although our focus will be on
binary systems.  In this regard both S-type and P-type orbits
will be examined, and detailed mathematical criteria for the existence
of S-type and P-type RHZs will be derived.  In Sect.~4, we consider
the additional constraint of planetary orbital stability for the
establishment of circumstellar habitability.  Applications regarding
S-type and P-type systems are given in Sect.~5, whereas the habitability
classifications S, P, ST, and PT are introduced in Sect.~6.  Section~7
conveys our summary and conclusions.


\section{Stellar Parameters and Single Star Habitability}

In this study, S-type and P-type habitability is investigated mostly pertaining
to standard (i.e., theoretical) main-sequence stars.  The adopted stellar
parameters, which are the stellar effective temperatures $T_{\rm eff}$, the
stellar radii $R_\ast$ (which together allow to define the stellar luminosities
$L_\ast$), and the stellar masses $M_\ast$ are mainly based on the work by
\cite{gra05} (see his Table B.1) that assumes detailed photospheric spectral
analyses.  For stellar spectral types with no data available, the missing data
were computed by employing biparabolic interpolation.

The exception, however, are data for stars of spectral type K5~V and below.
In this regard we relied on the results from the spectral models of
R.~L. Kurucz and collaborators.  They took into account hundreds of millions
of spectral lines for a large set of atoms and molecules; see \cite{cas04}
and \cite{kur05} for details.  The effective temperatures implied by these
models are in close similarity to those given by \cite{gra05} for most
types of stars; however, \cite{gra05} reports consistently higher effective
temperatures for stars of spectral type late-K and M; for the latter, the
difference amounts to nearly 300~K.  Table~1 depicts the stellar
parameters adopted for the present work.

An alternative approach expected to provide very similar results
for either the stellar luminosity or the stellar mass (with the other
parameter taken as fixed) is the employment of a mass--luminosity
relationships applicable to main-sequence stars.  The work by
\cite{rei87}, as well as data from subsequent studies, yield
\begin{equation}
\frac{L_\ast}{L_\odot} \ = \ \eta \Big(\frac{M_\ast}{M_\odot}\Big)^\alpha
\end{equation}
with $\eta = 0.23$ and $\alpha = 2.3$ for $M_\ast < 0.43~M_\odot$ and 
     $\eta = 1$    and $\alpha = 4.0$ for $M_\ast \ge 0.43~M_\odot$.
At the high-mass end, this relationship holds until about
$M_\ast = 2~M_\odot$.  It also becomes increasingly inaccurate for
low-mass M dwarfs.  Fortunately, the domains of applicability for
Eq.~(1) is consistent with most studies of binary habitability;
see, e.g., \cite{egg12} as example.

Next we focus on single star habitability, i.e., the evaluation
of various limits of habitable zones (HZs), which in the solar
case shall be referred to as $s_\ell$.  Previous work by,
e.g., \cite{kas93} distinguished between conservative (CHZ) and
the generalized habitable zone (GHZ), which can also be evaluated
for general main-sequence stars, and other types of stars as well.
For the Sun, the limits of the CHZ are given as 0.95 and 1.37~AU
($\ell$ = 2 and 4, respectively), whereas for the GHZ, they
are given as 0.84 and 1.67~AU ($\ell$ = 1 and 5, respectively);
see Table~2.

The physical significance of the various kinds of HZs obtained
by \cite{kas93} can be summarized as follows:  The GHZ is defined
as bordered by the runaway greenhouse effect (inner limit) and
the maximum greenhouse effect (outer limit).  Concerning the latter
it is assumed that a cloud-free CO$_2$ atmosphere is still able to
provide a surface temperature of 273~K.  The inner limit of the
CHZ is defined by the onset of water loss.  In this case, a
wet stratosphere is assumed to exist where water is lost by
photodissociation and subsequent hydrogen escape to space.
Furthermore, the outer limit of the CHZ is defined by the first
CO$_2$ condensation attained by the onset of formation of CO$_2$
clouds at a temperature of 273~K; see, e.g., \cite{und03}
and \cite{sel07} for further details.  Table~3 conveys the
results for the HZs for the different types of main-sequence
stars of the present study with the different limits referred
to as HZ$(s_\ell)$.
 
For the outer edge of circumstellar habitability, even less
stringent limits have been introduced in the meantime
\citep[e.g.,][]{for97,mis00}.  They are based on the assumption of
relatively thick planetary CO$_2$ atmospheres as well as strong
backwarming that may further be enhanced by CO$_2$ crystals and
clouds.  These limits, which in case of the Sun correspond to
2.4~AU $(s_\ell = s_6)$, conform to the extended habitable zone
(EHZ), have also been taken into account in our study, although
the significance of the EHZ has meanwhile been criticized as a
result of detailed planetary radiative transfer models
\citep{hal09}.  Moreover, in the framework of the present
study, we also consider planetary Earth-equivalent positions
defined as $R_{\oplus,{\rm eqv}} \simeq \sqrt{{L_\ast}/{L_\odot}}$
and labelled as $s_\ell = s_3$; see Table~3.  It is meant as
an intriguing reference distance of habitability both
regarding single stars and stellar binary systems.


\section{Theoretical Approach}

\subsection{Basic Equations}

Next we introduce the governing equations for investigating the RHZs of
binary systems pertaining to both S-type and P-type orbits.  This approach
targets the requirement of providing a habitable region for a
system planet based on the radiative energy fluxes of the stellar components.
The requirement of planetary orbital stability will be disregarded for now;
it will be revisited in Sect.~4.  The importance of orbital stability for
allowing circumstellar habitability in stellar binaries will, however, be
considered in an appropriate and consistent manner in the main body of the study.

For a star of luminosity $L_i$, given in units of solar
luminosity $L_\odot$, the distance $d_i$ of the habitability limit
$s_\ell$ as identified for the Sun, which may constitute either an inner
or outer limit of habitability (except $\ell = 3$), is given as
\begin{equation}
d_i \ = \ s_\ell \sqrt{\frac{L_i}{S_{{\rm rel}, i\ell} L_\odot}}
\end{equation}
In case of a multiple star system of order $N$ with distances $d_i$, the
limit of habitability related to $s_\ell$ is given as\footnote{
This equation is analogous to an equivalent equation of electrostatics
relating a general distribution of charges to the resulting electrostatic
potential in free space \citep[][see p.~40, Eq.~(1.48)]{jac99}; a modified
version of Eq.~(3) has previously been considered by, e.g., \cite{egg12}.}
\begin{equation}
\sum_{i=1}^N \frac{L_i}{S_{{\rm rel}, i\ell} d_i^2} \ = \ \frac{L_\odot}{s_\ell^2} \ .
\end{equation}

In Eq. (2) and (3), $S_{{\rm rel}, i\ell} = S_{{\rm rel}, i\ell}(T_{\rm eff})$
(see Table~1) describes
the stellar flux in units of the solar constant that is a function of
the stellar effective temperature $T_{\rm eff}$ \citep[e.g.,][]{kas93,und03}.
Specifically, using the formalism\footnote{$S_{{\rm rel}, i\ell}$ represents the
normalized stellar flux in units of the solar constant, 1368 W~m$^{-2}$, given by
the stellar spectral energy distribution.  Therefore, ordinarily, no $s_\ell$ dependence
for $S_{{\rm rel}, i\ell}$ should exist.  However, the formulae by \cite{sel07} utilize
previous results by \cite{kas93} who provided numerical values for limits of
habitability for different types of stars considering various limit definitions
(i.e., $s_\ell$ values identified for the Sun).  But \cite{kas93} used for the
solar effective temperature an unusually low value of 5700~K instead of 5777~K
as currently accepted \citep[e.g.,][]{sti04}.  Hence, transforming the
polynomial fit based on the work by \citeauthor{kas93} aimed at considering
the correct solar effective temperature renders a weak dependence on $s_\ell$ for
the $S_{{\rm rel}, i\ell}$ values.  In contrast, the method by \cite{und03} provides a
polynomial fit for $S_{{\rm rel}, i\ell}$ without considering the solar temperature
revision.  An alternative method has been used by \cite{cun09} and subsequent work.
In this approach the polynomial fit by \cite{und03} is corrected via a triangular
function based on data for stars of spectral type F0~V, G0~V, and K0~V.  As a result
the corresponding $S_{{\rm rel}, i\ell}$ values do also not depend on
${s_\ell}$.} by \cite{sel07}, we find that
\begin{equation}
S_{{\rm rel}, i\ell} \ = \ \Big({\frac{s_\ell}{s_\ell - a_z T_\ast - b_z T_\ast^2}}\Big)^2
\end{equation}
with $s_\ell$, $a_z$, and $b_z$ in AU, and $T_\ast = T_{\rm eff} - 5700$ in K.
\cite{sel07} also found that for $s_\ell < 1$, corresponding to
inner limits of habitability, the fitting parameters are given as
$a_z = 2.7619 \times 10^{-5}$ and $b_z = 3.8095 \times 10^{-9}$, whereas for
$s_\ell > 1$, corresponding to outer limits of habitability, they are given as
$a_z = 1.3786 \times 10^{-4}$ and $b_z = 1.4286 \times 10^{-9}$; note that
$s_\ell \equiv 1$ corresponds to the customary notion of Earth-equivalent
positions.  Appropriate values for $s_\ell$ are given in Table~2. 

In the following we will focus on the case of binary systems, i.e., $N=2$.  In
this case Eq. (3) reads
\begin{equation}
\frac{L_1}{S_{{\rm rel}, 1\ell} d_1^2} + \frac{L_2}{S_{{\rm rel}, 2\ell} d_2^2} \ = \ \frac{L_\odot}{s_\ell^2}
\end{equation}
with
\begin{mathletters}
\begin{eqnarray}
d_1^2 \ & = & \ a^2 + z^2 + 2 a z \cos{\varphi} \\ 
d_2^2 \ & = & \ a^2 + z^2 - 2 a z \cos{\varphi} \ .
\end{eqnarray}
\end{mathletters}
Here $a$ denotes the semidistance of binary separation, $z$ the distance of
a position at the habitability limit contour (which later on will be referred
to as ``radiative habitable limit", see below), and $\varphi$ the associated angle;
see Fig.~1 for information on the coordinate set-up for both S-type and P-type orbits.
We will also assume $L_1 \ge L_2$ without loss of generality.

With $L_{i\ell}'$ defined as
\begin{equation}
L_{i\ell}' = \frac{L_i}{L_\odot S_{{\rm rel}, i\ell}} ,
\end{equation}
henceforth referred to as {\it recast stellar luminosity} (see Table~4), $z(\varphi)$
is given as
\begin{equation}
z^4 + A_2 z^2 + A_1 z + A_0 \ = \ 0
\end{equation}
with
\begin{mathletters}
\begin{eqnarray}
A_2 & = & 2 a^2 (1  - 2 \cos^2 \varphi) - s_\ell^2 (L_{1\ell}' + L_{2\ell}') \\
A_1 & = & 2 a s_\ell^2 \cos \varphi (L_{1\ell}'-L_{2\ell}') \\ 
A_0 & = & a^4 - a^2 s_\ell^2 (L_{1\ell}' + L_{2\ell}') \ .
\end{eqnarray}
\end{mathletters}

Equation (8) constitutes a fourth-order algebraic equation that is known to
possess four possible solutions \citep{bro97}, although some (or all) of them may
constitute unphysical solutions, i.e., $z(\varphi)$ having a complex or
imaginary value.  The adopted coordinate system constitutes, in essence,
a polar coordinate system except that negative values for $z$ are permitted;
in this case the position of $z$ is found on the opposite side of angle $\varphi$.

In principle, it is possible to consider for Eq.~(8) to only have solutions
for $z(\varphi)$ given as $z \ge 0$; in this case the entire interval for
$z(\varphi)$, which is  $0 \le \varphi < 2\pi$, needs to be examined.
The following types of solutions are identified:  For S-type orbits, two
solutions exist in the intervals centered at $\varphi = 0$ and at
$\varphi = \pi$ (or one coinciding solution at each tangential point);
see Fig.~1.  However, there will be no solution in the typically relatively
large intervals containing $\varphi = \pi/2$ and $\varphi = 3\pi/2$.
Clearly, the size of any of those intervals critically depends on the
system parameters $a$, $s_\ell$, $L_{1\ell}'$, and $L_{2\ell}'$, as expected.
For P-type orbits, on the other hand, there will be one solution for
each value of $\varphi$ in the range of $0 \le \varphi < 2\pi$.

However, in general, negative values for the solutions of $z(\varphi)$ also
exist.  If taken into account, it will be sufficient to restrict the evaluation
of Eq. (8) to the range $0 \le \varphi \le \pi$.  In this case, for S-type
orbits, there will be four solutions in the intervals with endpoints
$\varphi = 0$ and $\varphi = \pi$, as well as two solutions
(if $L_{1\ell}' \ne L_{2\ell}'$)
in a more extended interval containing these points.  Also, a pair of
solutions will become one coinciding solution at each tangential point.
However, again, there will be no solution in the interval containing
$\varphi = \pi/2$.  In case of P-type orbits, there will be two solutions
for any value of $\varphi$ in the range of $0 \le \varphi \le \pi$.  We will
revisit this assessment in conjunction with the algebraic method for attaining
the solution $z_i$; additionally, detailed mathematical criteria will be
given for the existence of RHZ for S-type and P-type orbits.

Next we will focus on equal-star binary systems.  Detailed solutions for general
binary systems (i.e., systems of stellar components with by default unequal masses,
luminosities, and effective temperatures) pertaining to both S-type and P-type
orbits will be given in Sect.~3.3.  Both subsections will be aimed at deriving
RHZs; see, e.g., \cite{wil02} for general discussions on the role of RHZs
for the attainment of habitability in star--planet systems.
However, strictly speaking, they will deal with identifying {\it radiative
habitable limits} (RHLs) connected to a distinct value of $s_\ell$ noting that
manifesting a RHZ requires that the RHL for $s_{\ell,{\rm out}}$ to be located
{\it completely  outside} of the RHL for $s_{\ell,{\rm in}}$ with $s_{\ell,{\rm in}}$
and $s_{\ell,{\rm out}}$ appropriately paired.  A summary about the existence
and structure of the RHZs, encompassing the radiative CHZs, GHZs, and EHZs,
will be given in Sect.~3.4; this subsection will also convey cases where
no RHZs exist due to the behavior of the RHLs owing to the choices of
$s_{\ell,{\rm in}}$ and $s_{\ell,{\rm out}}$.


\subsection{Equal-Star Binary Systems}

Now we focus on the special case of equal-star binary systems, i.e., stars of
identical recast luminosities, i.e., $L_{1\ell}' = L_{2\ell}' = L_\ell'$.
For theoretical main-sequence stars
this assumption also implies $S_{{\rm rel}, 1\ell} = S_{{\rm rel}, 2\ell}$ and
$M_1 = M_2$; this latter assumption about the stellar masses is relevant for
the orbital stability constraint of system planets.  With $A_1 = 0$, Eq. (8) now
constitutes a biquadratic equation that can be solved in a straightforward manner.
The other coefficients are given as
\begin{mathletters}
\begin{eqnarray}
A_2 & = & 2 a^2 ( 1 - 2 \cos^2 \varphi ) - 2 s_\ell^2 L_\ell' \\
A_0 & = & a^4 - 2 a^2 s_\ell^2 L_\ell' \ .
\end{eqnarray}
\end{mathletters}
Thus, the solution of Eq. (8) is given as
\begin{equation}
z \ = \ \pm \sqrt{ - a^2 ( 1 - 2 \cos^2 \varphi ) + s_\ell^2 L_\ell' \pm \sqrt{ D_2 }}
\end{equation}
with
\begin{equation}
D_2 \ = \  s_\ell^4 {L_\ell'}^2 + 4 a^4 \cos^4 \varphi - 4 a^2 \cos^2 \varphi (a^2 - s_\ell^2 L_\ell') .
\end{equation}
With known systems parameters, which are $a$, $s_\ell$, and $L_\ell'$, the function
$z(\varphi)$, describing the habitability limits for the binary system associated
with inner limit and outer limit values $s_\ell$ derived for the Sun (see Sect.~2
for details) can be obtained in a straightforward manner.

Owing to the system symmetry, the existence of S-type and P-type RHLs can be identified
by attaining the solutions of Eq.~(11) for $\varphi=0$ and $\varphi=\pi/2$.
First we examine the solutions of Eq. (11) for $\varphi=0$, i.e., $\cos \varphi = 1$,
which are given as
\begin{equation}
z \ = \ \pm \sqrt{a^2 + s_\ell^2 L_\ell' \pm s_\ell \sqrt{s_\ell^2{L_\ell'}^2 + 4 a^2 L_\ell'}} \ .
\end{equation}
This allows us to explore the existence of S-type RHLs.
The total number of solutions for $z_i$ (if existing) is four as expected, which can be
ordered as $z_1 < z_2 < z_3 < z_4$. Due to symmetry it is found that $z_3 \ge 0$, which
implies that
\begin{equation}
a^2 + s_\ell^2 L_\ell' - s_\ell \sqrt{s_\ell^2{L_\ell'}^2 + 4 a^2 L_\ell'} \ \ge \ 0 \ .
\end{equation}
Thus, the condition for the existence of S-type RHLs is given as 
\begin{equation}
a \ \ge \ s_\ell \sqrt{2 L_\ell'} \ .
\end{equation}

Next we examine the solutions of Eq. (11) for $\varphi=\pi/2$, i.e., $\cos \varphi = 0$.
This allows us to explore the existence of P-type RHLs; the latter implies two solutions
of Eq. (11) regardless of the value for $\varphi$.  If the positive root of $D_2$
(see Eq.~12) is considered, the solution is given as
\begin{equation}
z \ = \ \pm \sqrt{- a^2 + 2 s_\ell^2 L_\ell'} \ .
\end{equation}
Thus, the condition for the existence of P-type RHLs is given as 
\begin{equation}
a \ \le \ s_\ell \sqrt{2 L_\ell'} \ .
\end{equation}

Therefore, Eqs. (15) and (17) allow to identify the conditions for S-type and P-type RHLs,
respectively, for equal-star binary systems, which depend on the systems parameters
$a$, $s_\ell$, and $L_\ell'$; note that the equal signs in these equations carry little
relevance.  Comparing Eqs. (15) and (17) also implies that the joint existence of
S-type and P-type habitability in equal-star binary systems in circular orbits is
not possible, irrespectively of the system parameters and the planetary orbital
stability requirement (see Sect.~4), noting
that the latter imposes an additional constraint on habitability even when the
RHZ-related conditions are met.  Figure~2 depicts the borders of the S-type and P-type
radiative habitable limits, i.e., RHLs, for different values of $s_\ell$ in regard
to $a$ and $L_\ell'$.


\subsection{General Binary Systems}

\subsubsection{Method of Solution}

We now focus on obtaining solutions for our key equation, Eq. (8), pertaining
to S-type and P-type RHLs for general binary systems, i.e., $L_{1\ell}' \ne L_{2\ell}'$.
Following, e.g., \cite{abr72} and \cite{bey87}, the set of solutions for a
fourth-order polynomial reads
\begin{mathletters}
\begin{eqnarray}
z_1 & = & - \frac{1}{2} {\cal C} - \frac{1}{2} {\cal D}  \\
z_2 & = & - \frac{1}{2} {\cal C} + \frac{1}{2} {\cal D}  \\
z_3 & = & + \frac{1}{2} {\cal C} - \frac{1}{2} {\cal E}  \\ 
z_4 & = & + \frac{1}{2} {\cal C} + \frac{1}{2} {\cal E}
\end{eqnarray}
\end{mathletters}
with
\begin{mathletters}
\begin{eqnarray}
  {\cal C} & = & \sqrt{ - 2 a^2 (1  - 2 \cos^2 \varphi) + s_\ell^2 (L_{1\ell}' + L_{2\ell}') + y_1 }  \\
  {\cal D} & = & \sqrt{ s_\ell^2 (L_{1\ell}' + L_{2\ell}') + 4 a s_\ell^2 (L_{1\ell}' - L_{2\ell}') {\cal C}^{-1} \cos \varphi - 2 a^2 (1  - 2 \cos^2 \varphi) - y_1} \\
  {\cal E} & = & \sqrt{ s_\ell^2 (L_{1\ell}' + L_{2\ell}') - 4 a s_\ell^2 (L_{1\ell}' - L_{2\ell}') {\cal C}^{-1} \cos \varphi - 2 a^2 (1  - 2 \cos^2 \varphi) - y_1}
\end{eqnarray}
\end{mathletters}
with $y_1$ as a solution of the resolvent cubic equation
\begin{equation}
y^3 - A_2 y^2 - A_0 y + ( 4 A_2 A_0 - A_1^2 ) \ = \ 0
\end{equation}
with $A_0$, $A_1$, and $A_2$ given by Eqs. (9a) to (9c); here the term
of $(L_{1\ell}' - L_{2\ell}') {\cal C}^{-1}$ in Eqs.~(19b) and (19c) corresponds to
the case of non-equal star binaries assumed in the following.  Note
that for equal-star binaries a more straightforward method of solution
is available (see Sect. 3.2).  The solutions
$z_i$ given through Eqs. (18a) to (18d), if existing, are ordered as
$z_1 < z_2 < z_3 < z_4$ for $\varphi = 0$; this order is also maintained
for any other value of $\varphi$ as identified in all model simulations
pursued.  Although Eq. (20)
has three possible solutions, there is only one appropriate choice for $y$, named $y_1$,
because it is necessary to avoid that all $z_i$ obtained through Eq. (8) are
of imaginary or conjugate complex value in cases where S-type or P-type RHLs
exist.

The acceptable solution for $y$ is given as
\begin{equation}
y_1 \ = \ - \frac{1}{3} \hat{a}_2 + (S + T)
\end{equation}
with the substitutions
\begin{mathletters}
\begin{eqnarray}
S   & = & \sqrt[3]{R + \sqrt{D_3}} \\
T   & = & \sqrt[3]{R - \sqrt{D_3}} \\
D_3 & = & Q^3 + R^2
\end{eqnarray}
\end{mathletters}
and with $Q$ and $R$ given as
\begin{mathletters}
\begin{eqnarray}
Q & = &   \frac{1}{3} \hat{a}_1 - \frac{1}{9} {\hat{a}_2}^2                                    \\
R & = & - \frac{1}{2} \hat{a}_0 + \frac{1}{6} \hat{a}_1 \hat{a}_2 - \frac{1}{27} {\hat{a}_2}^3
\end{eqnarray}
\end{mathletters}
while noting that
\begin{mathletters}
\begin{eqnarray}
\hat{a}_0 & = &   4 A_0 A_2 - A_1^2 \\
\hat{a}_1 & = & - 4 A_0             \\
\hat{a}_2 & = & - A_2   \ .
\end{eqnarray}
\end{mathletters}
These sets of equation can be solved and appropriate values for $z(\varphi)$ can be
obtained.  The results will depend on the system parameters $a$, $s_\ell$, $L_{1\ell}'$,
and $L_{2\ell}'$, as expected.

Next we describe the solutions for S-type and P-type RHLs in more detail.  It is important
to recognize that a priori choices about the existence of S-type and P-type RHLs are
neither necessary nor possible as the existence of any of those RHLs is determined by
the fulfillment of well-defined mathematical conditions; they will also be given in the
following.


\subsubsection{S-type Orbits}

An analysis of the possible solutions for Eqs. (18a) to (18d) shows that for S-type RHLs
valid solutions are obtained based on
\begin{mathletters}
\begin{eqnarray}
S & = & \sqrt[6]{R^2 + K^2} \ \big( \cos \xi + i \sin \xi \big) \\
T & = & \sqrt[6]{R^2 + K^2} \ \big( \cos \xi - i \sin \xi \big)
\end{eqnarray}
\end{mathletters}
with
\begin{mathletters}
\begin{eqnarray}
K   & = & \sqrt{\vert D_3 \vert}                             \\
\xi & = & \frac{1}{3} {\rm arctan} \Big( \frac{K}{R} \Big)
\end{eqnarray}
\end{mathletters}
with $R$ given by Eq.~(23b). Therefore, the solution of the resolvent cubic equation,
Eq. (20), is given as 
\begin{equation}
y_1 \ = \ - \frac{1}{3} \hat{a}_2 + 2 \sqrt[6]{R^2 + K^2} \cos \xi
\end{equation}

For the values of $z_i$ it is found that for the interval centered at
$\varphi = 0$, $z_1$ and $z_2$ exhibit negative values, whereas $z_3$ and $z_4$
exhibit positive values with $z_1 < z_2 < z_3 < z_4$.  Thus, $z_1$ and $z_2$
describe the RHL regarding star S1, whereas $z_3$ and $z_4$ describe the RHL
regarding star S2 (see Fig.~1).  Conversely, for the interval centered at
$\varphi = \pi$, $z_1$ and $z_2$ again exhibit negative values and $z_3$ and
$z_4$ exhibit positive values.  In this case, $z_1$ and $z_2$
describe the RHL for star S2, whereas $z_3$ and $z_4$ describe the RHL
for star S1.  No solutions are obtained in the vicinity of
$\varphi = \pi/2$ and $\varphi = 3\pi/2$, as expected.  Thus, for each
angle $\varphi$ in the range of $0 \le \varphi < 2\pi$ the appropriate
number of solutions is attained to describe S-type RHLs.  However,
due to symmetry, solutions are only needed for $0 \le \varphi \le \pi$.

The existence of S-type RHLs requires that $z_2 \le z_3$ because otherwise
the two distinct S-type RHL contours about the two binary components would
not be separated, which corresponds to the condition
\begin{equation}
{\cal C} \ \ge \ \frac{1}{2} \big( {\cal D} + {\cal E} \big) \ .
\end{equation}
Equation (28) can be rewritten to provide an expression based on the
system parameters $A_0$, $A_1$, and $A_2$ defined through Eqs. (9a) to (9c).
It is found that
\begin{equation}
2 y_1 (A_2 - y_1)^2 - A_1^2 \ge \ 0 \
\end{equation}
with $y_1 = y_1 (A_0,A_1,A_2)$.  The expression for $y_1$ is highly complicated;
however, it can be obtained based on Eqs. (21) to (24c) by using, e.g.,
MATHEMATICA$^\registered$ in a straightforward manner.

In conclusion, for S-type RHLs to exist for the system parameters $a$, $s_\ell$,
$L_{1\ell}'$, and $L_{2\ell}'$, it is necessary that the relations (28) and (29),
which are equivalent, must be fulfilled for any angle of $\varphi$, though the
evaluation can be limited to $\varphi=0$.  Furthermore, through analytical
transformations it can be shown that the condition depicted as Eqs.~(28) and
(29) requires 
\begin{equation}
6912 A_0^3 - 3456 A_0^2 A_2^2 + 432 A_0 A_2^4 - 729 A_1^4  \ \le \ 0 \ .
\end{equation}
In the limiting case of equal-star binary systems, attained as $A_1 \rightarrow 0$,
Eq.~(30) can be simplified as
\begin{equation}
\big( 4 A_0 - A_2^2 )^2 \ge 0 \ ;
\end{equation}
this relationship is fulfilled in a trivial manner.


\subsubsection{P-type Orbits}

An analysis of the possible solutions for Eqs. (18a) to (18d) also shows that for
P-type RHLs valid solutions require
\begin{equation}
D_3 \ \ge \ 0 \ ;
\end{equation}
see Eq.~(22c).  The detailed evaluation of this condition requires the evaluation
of various sets of equations denoted as Eqs. (9a) to (9c), (22a) to (22c), and
(23a) to (23b); see  Sect.~3.1 and 3.3.1.

In terms of the solutions for P-type RHLs it is found that for
$0 \le \varphi < \pi/2$ and $3\pi/2 < \varphi \le 2\pi$, $z_1$ exhibit negative
values and $z_2$ exhibit positive values, whereas $z_3$ and $z_4$ are undefined;
they are also not needed for outlining P-type RHLs.  Moreover, for the range
of $\pi/2 < \varphi < 3\pi/2$, $z_3$ exhibit negative values and $z_4$ exhibit
positive values, noting that $z_1$ and $z_2$ remain undefined.
For $\pi/2$ and $3\pi/2$, removable singularities are identified, which can
easily be fixed through interpolation taking values of $z$ for neighboring
angles of $\varphi$.  In summary, for each angle $\varphi$ in the range of
$0 \le \varphi < 2\pi$ two values of $z_i$ (i.e., one positive and one
negative value) are identified allowing to determine P-type RHLs.  However,
due to symmetry, solutions are only needed for $0 \le \varphi \le \pi$.

Moreover, through analytical transformations it can be shown that the condition
depicted as Eq.~(32) can be rewritten as
\begin{equation}
16 A_0 A_2^4 + 144 A_0 A_1^2 A_2 - 128 A_0^2A_2^2 + 256 A_0^3 - 4 A_1^2 A_2^3 - 27 A_1^4 \ \le \ 0 \ .
\end{equation}
with $A_0$, $A_1$, and $A_2$ defined through Eqs. (9a) to (9c) with the
left hand side of
Eq. (33) representing $-108 \cdot D_3$.  In conclusion, for P-type RHLs to exist
for the system parameters $a$, $s_\ell$, $L_1'$, and $L_2'$, it is necessary that
the relations (32) and (33), which are equivalent, must be fulfilled for any angle
of $\varphi$,  though the evaluation can be limited to $\varphi=\pi/2$.
In the limiting case of equal-star binary systems, attained as $A_1 \rightarrow 0$,
Eq.~(33) can be simplified as
\begin{equation}
\big( 4 A_0 - A_2^2 )^2 \ge 0 \ ;
\end{equation}
this relationship, already given as Eq.~(31), is fulfilled in a trivial manner.


\subsection{Calculation of RHZs for Binary Systems}

The identification of the RHZs in binary systems requires the calculation of
limits of habitable zones, i.e., RHLs, as pointed out in Sect.~3.1.  The RHZs need to be
established for values of $s_\ell$ with $\ell$ = 1, 2, 4, 5, and 6 (see Sect.~2 and
Table~2), which are informed by model-dependent physical limits of habitability
for the solar environment \citep[e.g.,][]{kas93}. As part of the process, the
parameters of $s_\ell$ need to be appropriately paired in terms of the inner
and outer limits of habitability.  For the CHZ the parameters
$(s_{\ell,{\rm in}},s_{\ell,{\rm out}})$ need to be paired as $(s_2,s_4)$,
whereas for the GHZ, they need to paired as $(s_1,s_5)$.  For the EHZ, the
parameters of $s_\ell$ need to paired as $(s_1,s_6)$, considering that both
the CHZ and the GHZ shall be viewed as subdomains of the EHZ.

For S-type and P-type orbits, the radiative zones of habitability ${\rm RHZ}(z)$,
which constitutes a circular region (annulus) around each star S1 and S2 (S-type) 
or both stars (P-type) can be determined as
\begin{equation}
{\rm RHZ}(z) \ = \ {\rm Min}\Big({\cal R}\big(z,{\alpha}\big)\Big)\Big|_{s_{\ell,{\rm out}}} -
                   {\rm Max}\Big({\cal R}\big(z,{\alpha}\big)\Big)\Big|_{s_{\ell,{\rm in}}}
\end{equation}
and
\begin{equation}
{\rm RHZ}(z) \ = \ {\rm Min}\Big({\cal R}\big(z,{\varphi}\big)\Big)\Big|_{s_{\ell,{\rm out}}} -
                   {\rm Max}\Big({\cal R}\big(z,{\varphi}\big)\Big)\Big|_{s_{\ell,{\rm in}}} ,
\end{equation}
respectively; see Fig.~1 for coordinate information.  Here ${\cal R}(z,\alpha)$
and ${\cal R}(z,\varphi)$ describe the areas bordered by the RHLs defined by
${s_{\ell,{\rm in}}}$ and ${s_{\ell,{\rm out}}}$.  The calculation of the
extrema is applied to the angles $\alpha$ and $\varphi$ for the intervals
$0 \le \alpha \le \pi$ and $0 \le \varphi \le \pi/2$, respectively; note
that we assumed $L_1 \ge L_2$ without loss of generality.  In the S-type case 
the calculation of the extrema pertaining to the RHZ values is based on the
angular coordinate $\alpha$ instead of $\varphi$; however, the angular
coordinate $\varphi$ is still needed for the calculation of $z_i$ as part
of the overall approach toward identifying S-type and P-type habitability.

Figure~3 depicts examples of RHLs and RHZs for different types of systems.
In the S-type case the RHLs are bended toward the center of the system,
whereas in the P-type case they are of notable elliptical shape.
The RHZs always constitute circular annuli obtained through inspecting
the appropriate minima and maxima of the RHLs.  The examples as depicted
include S-type and P-type systems with separation distances $2a$ of
0.5~AU and 5.0~AU, respectively.  Cases of both equal-star and non-equal
star binaries are selected.  The focus of this figure is the identification
of the appropriate circular region (i.e., annulus) for each case.  The figure
also indicates the portions within the ${\cal R}(z,{\alpha})$ and 
${\cal R}(z,{\varphi})$ domains that are not part of the RHZ$(z)$ annuli.

Next we determine the values for the extrema pertaining to ${\rm RHZ}(z)$
following Eqs.~(35) and (36) based on the solutions of Eq.~(8) given as
Eqs. (18a) to (18d).  In cases where four solutions $z_i$ exist, it is
found that they are ordered as $z_1 < z_2 < z_3 < z_4$ with $z_1$ and
$z_2$ constituting negative values, and $z_3$ and $z_4$
constituting positive values.  If the negative solutions for $z_i$ are
permitted, it is sufficient for S-type orbits, both for star S1 and S2,
to only consider solutions for $\varphi = 0$.  For P-type orbits, a more
detailed assessment is required (see below).  For S-type orbits, regarding
star S1, the extrema are obtained as
\begin{mathletters}
\begin{eqnarray}
{\rm RHZ}_{\rm in}  \ = \ {\rm Max}\big({\cal R}\big(z,{\alpha}{\big)}\big)\Big|_{s_{\ell,{\rm in}}}  & = & \big\vert a + z_2(0) \big\vert \Big|_{s_{\ell,{\rm in}}}  \\
{\rm RHZ}_{\rm out} \ = \ {\rm Min}\big({\cal R}\big(z,{\alpha}{\big)}\big)\Big|_{s_{\ell,{\rm out}}} & = & \big\vert a + z_1(0) \big\vert \Big|_{s_{\ell,{\rm out}}} ,
\end{eqnarray}
\end{mathletters}
and for star S2, they are obtained as
\begin{mathletters}
\begin{eqnarray}
{\rm RHZ}_{\rm in}  \ = \ {\rm Max}\big({\cal R}\big(z,{\alpha}{\big)}\big)\Big|_{s_{\ell,{\rm in}}}  & = & \big\vert a - z_3(0) \big\vert \Big|_{s_{\ell,{\rm in}}}  \\
{\rm RHZ}_{\rm out} \ = \ {\rm Min}\big({\cal R}\big(z,{\alpha}{\big)}\big)\Big|_{s_{\ell,{\rm out}}} & = & \big\vert a - z_4(0) \big\vert \Big|_{s_{\ell,{\rm out}}} .
\end{eqnarray}
\end{mathletters}

The size of each annulus $\Delta_{\rm RHZ}$ for pairs $(s_{\ell,{\rm in}},s_{\ell,{\rm out}})$
is given as $\Delta_{\rm RHZ} = {\rm RHZ}_{\rm out} - {\rm RHZ}_{\rm in}$.  ${\rm RHZ}_{\rm in}$
also constitutes a generalization of HZ$(s_\ell)$ with $\ell = 1,2$ previously defined
for single stars (see Sect.~2 and Table~3).  Likewise, ${\rm RHZ}_{\rm out}$ constitutes
a generalization of HZ$(s_\ell)$ with $\ell =$ 4, 5, and 6. 

For P-type orbits, the extrema are given as follows:
\begin{mathletters}
\begin{eqnarray}
{\rm RHZ}_{\rm in}  \ = \ {\rm Max}\big({\cal R}\big(z,{\varphi}{\big)}\big)\Big|_{s_{\ell,{\rm in}}}  & = & \big\vert z_1(0)                 \big\vert \Big|_{s_{\ell,{\rm in}}}  \\
{\rm RHZ}_{\rm out} \ = \ {\rm Min}\big({\cal R}\big(z,{\varphi}{\big)}\big)\Big|_{s_{\ell,{\rm out}}} & = & \big\vert z_1(\varphi_{\rm out}) \big\vert \Big|_{s_{\ell,{\rm out}}} ,
\end{eqnarray}
\end{mathletters}
We note that for the angle $\varphi_{\rm out}$ for ${\rm RHZ}_{\rm out}$
no straightforward expression\footnote{Due to the nature of the underlying
equations for $z_i$, an analytic expression for $\varphi_{\rm out}$ is deemed possible.
However, it will be highly complicated and thus a numerical solution may be
preferred.} exists; it is located in the interval $0 < \varphi_{\rm out} \le \pi/2$.
It can be found numerically as it is given by the angle where the minimum of
$\vert z(\varphi) \vert$ occurs.  In the special case of ${L_{2\ell}'}/{L_{1\ell}'} \ll 1$
with $L_{1\ell}'$ and $L_{2\ell}'$ denoting the stellar primary and secondary, respectively,
it is found that $\varphi_{\rm out} \rightarrow 0$, whereas for $L_{1\ell}'=L_{2\ell}'$
it is found that $\varphi_{\rm out} = \pi/2$ (see Sect.~3.2).

There is also another complication in the identification of RHZ P-type orbits.
Generally, it is required for the RHL of $s_{\ell,{\rm out}}$ to be located
completely outside of the RHL of $s_{\ell,{\rm in}}$, i.e.,
\begin{equation}
{\rm Min}\Big({\cal R}\big(z,\varphi\big)\Big)\Big|_{s_{\ell,{\rm out}}} \ \ge \
{\rm Max}\Big({\cal R}\big(z,\varphi\big)\Big)\Big|_{s_{\ell,{\rm in}}} \ .
\end{equation}
This condition is however violated in some models, especially for relatively
large values of $a$ as well as relatively small ratios of ${L_{2\ell}'}/{L_{1\ell}'}$.
In this case the RHZ for $(s_{\ell,{\rm in}},s_{\ell,{\rm out}})$
is nullified, a behavior that may occur for the pairings $(s_2,s_4)$,
$(s_1,s_5)$, and $(s_1,s_6)$, corresponding to the CHZ, GHZ, and EHZ,
respectively.  In this regard the existence of the CHZ is in most jeopardy
as $(s_4-s_2)$ constitutes the smallest bracket among the various kinds
of HZs (see Table~2).  Also note that if the GHZ is nullified, the CHZ
will be nullified as well considering that the CHZ (if existing) is
entirely located within the GHZ.  Likewise, if the EHZ is nullified
the existence of both the CHZ and the GHZ will be nullified.
Detailed examples will be given in the application segment of this paper;
see Sect.~5.2 for details.  However, this type of phenomenon does not
occur for RHZs pertaining to S-type orbits.

For equal-star binary systems, with the property of
$L_{1\ell}' = L_{2\ell}' = L_\ell'$,
expressions for RHZ$_{\rm in}$ and RHZ$_{\rm out}$ for S-type and P-type orbits
can be obtained based on Eqs.~(13) and (16).  For S-type orbits we find
\begin{mathletters}
\begin{eqnarray}
{\rm RHZ}_{\rm in}  & = & {\Big\vert} a - \sqrt{a^2 + s_{\ell,{\rm in}}^2  L_{\ell,{\rm in}}'  + s_{\ell,{\rm in}}  \sqrt{s_{\ell,{\rm in}}^2  {L_{\ell,{\rm in}}'^2}  + 4 a^2 L_{\ell,{\rm in}}'}}  {\Big\vert} \\
{\rm RHZ}_{\rm out} & = & {\Big\vert} a - \sqrt{a^2 + s_{\ell,{\rm out}}^2 L_{\ell,{\rm out}}' - s_{\ell,{\rm out}} \sqrt{s_{\ell,{\rm out}}^2 {L_{\ell,{\rm out}}'^2} + 4 a^2 L_{\ell,{\rm out}}'}} {\Big\vert} \ .
\end{eqnarray}
\end{mathletters}
It is also intriguing to explore the limits $a \gg {s_{\ell,{\rm in}} \sqrt{L_{\ell,{\rm in}}'}}$
and $a \gg {s_{\ell,{\rm out}} \sqrt{L_{\ell,{\rm out}}'}}$.  If these limits are met, it is found that
\begin{mathletters}
\begin{eqnarray}
{\rm RHZ}_{\rm in}  & = &  {\Big\vert} \frac{1}{2a} \Big( s_{\ell,{\rm in}}^2  L_{\ell,{\rm in}}'  + s_{\ell,{\rm in}}  \sqrt{s_{\ell,{\rm in}}^2  {L_{\ell,{\rm in}}'^2}  + 4 a^2 L_{\ell,{\rm in}}'}   \Big) {\Big\vert} \\
{\rm RHZ}_{\rm out} & = &  {\Big\vert} \frac{1}{2a} \Big( s_{\ell,{\rm out}}^2 L_{\ell,{\rm out}}' - s_{\ell,{\rm out}} \sqrt{s_{\ell,{\rm out}}^2 {L_{\ell,{\rm out}}'^2} + 4 a^2 L_{\ell,{\rm out}}'} \Big) {\Big\vert} \ .
\end{eqnarray}
\end{mathletters}
Moreover, in the limit of $a \rightarrow \infty$, the expressions for single star HZs
regarding ${\rm RHZ}_{\rm in}$ and ${\rm RHZ}_{\rm out}$ are recovered, as expected,
which are given as ${s_{\ell,{\rm in}} \sqrt{L_{\ell,{\rm in}}'}}$ and
${s_{\ell,{\rm out}} \sqrt{L_{\ell,{\rm out}}'}}$, respectively.
They are in agreement with the expressions previously obtained by
\cite{kas93}, \cite{und03}, \cite{sel07}, and others.

Results for P-type orbits can be obtained considering
\begin{mathletters}
\begin{eqnarray}
{\rm RHZ}_{\rm in}  & = & {\rm Max}\Big({\cal R}(z,0),{\cal R}\big(z,\frac{\pi}{2}\big)\Big)\Big|_{s_{\ell,{\rm in}}}  \ = \ {\cal R}\big(z,0\big)\Big|_{s_{\ell,{\rm in}}}              \\
{\rm RHZ}_{\rm out} & = & {\rm Min}\Big({\cal R}(z,0),{\cal R}\big(z,\frac{\pi}{2}\big)\Big)\Big|_{s_{\ell,{\rm out}}} \ = \ {\cal R}\big(z,\frac{\pi}{2}\big)\Big|_{s_{\ell,{\rm out}}} \ .
\end{eqnarray}
\end{mathletters}
In this case we find
\begin{mathletters}
\begin{eqnarray}
{\rm RHZ}_{\rm in}  & = &  \sqrt{a^2 + s_{\ell,{\rm in}}^2 L_{\ell,{\rm in}}' + s_{\ell,{\rm in}} \sqrt{s_{\ell,{\rm in}}^2{L_{\ell,{\rm in}}'^2} + 4 a^2 L_{\ell,{\rm in}}'}} \\
{\rm RHZ}_{\rm out} & = &  \sqrt{- a^2 + 2 s_{\ell,{\rm out}}^2 L_{\ell,{\rm out}}'}
\end{eqnarray}
\end{mathletters}
based on the system parameters $a$, $s_{\ell,{\rm in}}$, $s_{\ell,{\rm out}}$,
$L_{\ell,{\rm in}}'$, and $L_{\ell,{\rm out}}'$.  Additionally, the requirement to
avoid that the RHL for $s_{\ell,{\rm out}}$ to be partially or completely located
inside of the RHL for $s_{\ell,{\rm in}}$, see Eq.~(40), entails
\begin{equation}
a^2 + \frac{1}{2} s_{\ell,{\rm in}}^2 L_{\ell,{\rm in}}'
 + s_{\ell,{\rm in}} \sqrt{a^2{L_{\ell,{\rm in}}'} + \frac{1}{4} s_{\ell,{\rm in}}^2{L_{\ell,{\rm in}}'^2}}
 - s_{\ell,{\rm out}}^2 L_{\ell,{\rm out}}' \ \le \ 0 ,
\end{equation}
which allows to set constraints on the separation distance $2a$ of the binary system
noting that the values of $s_{\ell,{\rm in}}$, $s_{\ell,{\rm out}}$,
$L_{\ell,{\rm in}}'$, and $L_{\ell,{\rm out}}'$ are subject to
distinct restrictions, particularly in case of main-sequence stars
(see Tables~1 and 2).  Depictions of the condition (45) for equal-star binaries
for the pairings $(s_2,s_4)$, $(s_1,s_5)$, and $(s_1,s_6)$, corresponding
to the CHZ, GHZ, and EHZ, respectively, are given in Fig.~4.  A similar
expression is expected to hold for nonequal-star binaries, albeit it will
be highly complicated.  Hence, for those systems a numerical assessment of
${\rm RHZ}_{\rm in}$ and ${\rm RHZ}_{\rm out}$ (see Eqs. 39a and 39b) may be preferred
to accommodate condition (40); see Sect. 5.2.2 for additional information and data.


\section{Constraints on Habitability due to Planetary Orbital Stability}

A primary constraint on planetary habitability is that planets are required
to exist in the HZ for a sufficient amount of time allowing basic forms of
life to emerge and develop.  In order to adhere to this criterion, planetary
orbital stability is required.  There is a significant body of literature
devoted to this topic, including studies of binary and multi-planetary systems,
which often also consider aspects of stellar evolution
\citep[e.g.,][]{jon01,nob02,men03,san07,tak08,dvo10,hag10,kop10}.

Early studies of planetary orbital stability pertaining to planets in
both S-type and P-type orbits demonstrated that planets can exist in systems
of binary stars for 3000 binary periods
\cite{dvo84,dvo86}.  Although these investigations considered relatively
short integration times, \citeauthor{dvo86} determined upper and lower bounds
of planetary orbital stability considering the orbital elements,
semimajor axis and eccentricity, of the proposed binary stars.  Since this
pioneering work, many additional studies have been performed.  The foremost
investigation extended the original study by a factor of 10 in integration
times and an extended range of orbital elements \citep{hol99}.  In addition,
the nature of the bounding formula was derived and discussed using a more
statistical framework.  \citeauthor{hol99} developed fitting formulae
for both S-type and P-type planets in binary systems given as
\begin{equation}
\frac{a_{\rm cr}}{a} \ = \ 0.464 - 0.38 \mu + {\cal F}_{\rm S}(\mu, e_{\rm b})
\end{equation}
and
\begin{equation}
\frac{a_{\rm cr}}{a} \ = \ 1.60 + 4.12 \mu + {\cal F}_{\rm P}(\mu, e_{\rm b}) \ ,
\end{equation}
respectively.

These equations give the critical semimajor axis $a_{\rm cr}$ in units of the
semimajor axis $a$ in case of S-type and P-type orbits.  For an S-type orbit,
the ratio ${a_{\rm cr}}/a$, see Eq.~(46), conveys the {\it upper limit} of
planetary orbital stability, whereas for a P-type orbit, the ratio
${a_{\rm cr}}/a$, see Eq.~(47), conveys the {\it lower limit} of planetary
orbital stability.  Moreover, $\mu$ denotes the stellar mass ratio given as
$\mu = M_2 / (M_1 + M_2)$, where $M_1$ and $M_2$ constitute the two masses
of the binary components with $M_2 \le M_1$.
Equations (46) and (47) also contain the parameter functions
${\cal F}_{\rm S}(\mu, e_{\rm b})$ and ${\cal F}_{\rm P}(\mu, e_{\rm b})$,
which depend on the aforementioned mass ratio $\mu$ and the eccentricity
of the stellar binary, $e_{\rm b}$.  Considering that this paper is solely
aimed at stellar binaries in circular orbits (i.e., $e_{\rm b}=0$),
it is found that ${\cal F}_{\rm S} = {\cal F}_{\rm P} = 0$.

Planetary orbital stability has been investigated by many authors using
chaos indicators, such as the maximal Lyapunov exponents (MLE),
fast Lyapunov indicator (FLI), and the mean exponential growth factor
of nearby orbits (MEGNO), to name those commonly used; see, e.g., \cite{saty13}
for details, recent applications, and references.  These methods have also
been used to characterize the transition from stable to unstable orbits
within the framework of the circular and elliptical 3-body problems; see,
e.g., \cite{cun07}, \cite{ebe08}, and \cite{sze08} for details.

Previously, \cite{mus05} studied the stability of both
S-type and P-type orbits in stellar binary systems, and deduced orbital stability
limits for planets.  These limits were found to depend on the mass ratio between
the stellar components and the distance ratio between planetary and
binary semimajor axes.  This topic was revisited by \cite{ebe08},
who used  the concept of Jacobi's integral and Jacobi's constant
to deduce stringent criteria for the stability of planetary orbits in binary systems
for the special case of the coplanar circular restricted three-body problem.  Recently
the planetary orbital stability was studied through the perspective of a chaos indicator,
the MLE by, e.g., \cite{qua11}.  From the use of a chaos indicator a cutoff value for the
maximum Lyapunov exponent was determined as an additional stability criterion for
S-type planets in the circular restricted 3-body problem.


\section{Case Studies}

\subsection{S-Type Habitability in Binary Systems}

Next we investigate S-type habitability for selected
binary systems, including systems of equal and non-equal masses (see Table~5).
Our main intent is to demonstrate the functionality of the method-as-proposed\footnote{The
method has also successfully been used for determining the radiative habitable zone
of Kepler-16, a binary system with $M_1 = 0.69~M_\odot$ and $M_2 = 0.20~M_\odot$,
and a Saturnian planet in a P-type orbit; see \cite{qua12} for a detailed study of
the system's habitability.}; an extensive parameter study will be given in Sect.~6. 
Figure 5 allows comparative insight into S-type habitability for selected binary
systems, i.e., systems with masses of $M_1 = M_2 = 1.0~M_\odot$ and
$M_1 = 1.5~M_\odot$, $M_2 = 1.0~M_\odot$; the binary separation distances are chosen
as 10~AU and 20~AU.  For single stars of $1.0~M_\odot$, the radiative CHZ extends
from 1.049 to 1.498~AU, and the radiative GHZ extends from 0.927 to 1.831~AU; these
values are slightly higher than those for G2~V stars given in Table~3 owing to
a minuscule difference in mass (i.e., 0.99 versus 1.0~$M_\odot$).

For an equal-mass binary system of 1.0~$M_\odot$ with a separation distance $2a$
of 10~AU, the radiative CHZ and GHZ extend from 1.056 to 1.511~AU, and from
0.932 to 1.853~AU, respectively, for each component.  Furthermore, the outer
limit of the radiative EHZ is altered from 2.64 to 2.70~AU.  However, there is now
an upper orbital stability limit of 1.37~AU imposed on each star.  Consequently,
significant portions of the radiative CHZ and GHZ are unavailable as circumstellar
habitable regions.  If the second star is placed at a distance of 20 AU, the
alteration of the radiative CHZ and GHZ relative to single stars is
very minor.  Specifically, for binary separations of 10~AU and
20~AU, the sizes of the radiative GHZ increase by 1.9\% and 0.6\% relative
to the case of single stars.  Moreover, for the system with a separation
distance of 20~AU, the imposed orbital stability limit is found at 2.74~AU;
consequently, the full extents of the radiative CHZ, GHZ, and EHZ are now
available for planetary habitability.

Figure 5 also shows results for the pairs $M_1 = 1.5$~$M_\odot$ and $M_2 =
1.0$~$M_\odot$.  In case of a single 1.5~$M_\odot$ mass star, the radiative CHZ
and GHZ extend from 1.88 to 2.49~AU, and from 1.65 to 3.11~AU, respectively,
whereas the radiative EHZ extends up to 4.61~AU.  In this type of system, a
secondary star of 1.0~$M_\odot$ placed at a separation distance of 10~AU again
modifies the extents of the radiative CHZ and GHZ, which now extend from
1.89 to 2.49~AU and from 1.66 to 3.14~AU, respectively; however, the planetary
orbital stability limit now occrs at 1.56~AU.  Therefore, the entire domains
of the radiative CHZ, GHZ, and EHZ of the primary are unavailable as
circumstellar habitable regions.  If the secondary star is placed at
a separation distance of 20~AU, the radiative CHZ, GHZ, and EHZ of the
primary star are again similar to those of a 1.5~$M_\odot$ mass star.
However, the orbital stability limit is now found at a distance of 3.12~AU
from the primary; therefore, the entire supplement of the radiative EHZ,
given by the bracket $(s_6-s_5)$ is now considered habitable.

In summary, for potentially habitable S-type binaries, owing to the
implied requirement of  the relatively large separations of the stellar
components, the effect of the stellar secondary on the extents of the RHZs
is often minor, i.e., about a few percent or less, with the biggest impact
occurring in F-type systems.  For most systems, the secondary's main influence
on circumstellar habitability thus consists in limiting planetary orbital
stability rather than offering significant augmentations of the RHZs, a
feature most pronounced in close binaries.


\subsection{P-Type Habitability in Binary Systems}

\subsubsection{Case Studies}

Various sets of models have been pursued to examine P-type habitability
(see Fig.~6).  As examples we considered systems with masses of
$M_1 = M_2 = 1.0~M_\odot$ and $M_1 = 1.5~M_\odot$ and $M_2 = 0.5~M_\odot$;
additionally, we also focused on models of $M_1 = 1.25~M_\odot$ and
$M_2 = 0.75~M_\odot$ (see Tables 5 to 8 for details).  The separation
distances $2a$ were chosen as 0.5, 1.0, and 2.0~AU, respectively.  Our
approach consists again of two steps.  First, we explore the existence
and extent of the radiative CHZs, GHZs, and EHZs.  Subsequently, we
consider the additional constraint of planetary orbital stability,
which in the case of P-type orbits constitutes a lower limit (see Sect.~4).
Our results can be summarized as follows.

For systems with masses of $M_1 = M_2 = 1.0~M_\odot$, the following behavior
is found.  For separation distances $2a$ of 0.5~AU, the inner limit (i.e.,
RHL; see Sect.~3.4) of the radiative CHZ varies between 1.46 and 1.54~AU
as function of polar angle $\varphi$ with 1.54~AU to be considered as acceptable
inner limit; see Eq.~(39a).  Furthermore, the outer limit of the radiative
CHZ varies between 2.10 and 2.16~AU with 2.10~AU as acceptable outer
limit; see Eq.~(39b).  In consideration of the orbital stability limit
at 0.92~AU (see Eq.~45), constituting an inner limit of orbital stability,
the entire extent of the radiative CHZ is available as a circumbinary
habitable region.  The acceptable inner limit of the radiative GHZ is given
as 1.38~AU, whereas the acceptable outer limit occurs at 2.58~AU; hence,
the entire radiative GHZ is again identified as habitable.

For separation distances of 1.0~AU, the orbital stability limit is given at
1.83~AU, which falls inside the domain of the radiative CHZ ranging from
1.69 to 2.06~AU; therefore, only about half of the radiative CHZ is available
for circumbinary habitability, whereas the other half is not.  Since the
radiative CHZ is fully embedded into the radiative GHZ, only a fraction of
the radiative GHZ offers circumbinary habitability.  However, the full extent
of the supplementary radiative EHZ, given by the bracket $(s_6-s_5)$, with
an acceptable outer limit of 3.83~AU, offers habitability.  We also considered
models with binary separations of 2.0~AU.  In this case, the orbital stability
limit is found at 3.66~AU.  Therefore, both the radiative CHZ and GHZ are
unavailable for providing habitability; the latter has an outer limit that
varies between 2.39 and 3.05~AU with 2.39~AU as acceptable limit.  The outer
limit of the radiative EHZ varies between 3.60 and 4.09~AU with 3.60~AU to
be ruled acceptable as the conservatively selected (i.e., inner) limit of
the radiative EHZ.  Hence, the entire radiative EHZ is also not considered
available for providing circumbinary habitability.

Most significantly, we also pursued case studies for systems of unequal
distributions of mass, and by implication of unequal distributions of
luminosity as, for example, the system  $M_1 = 1.5~M_\odot$ and
$M_2 = 0.5~M_\odot$.  According to the mass--luminosity relationship
for main-sequence stars, it is found that a 1.5~$M_\odot$ star possesses
a luminosity $L_\ast$ about 3.5 times higher than a 1.0~$M_\odot$ star;
a similar factor of difference exists for the recast stellar luminosity
$L'_{i\ell}$ (see Tables 4 and 5).  Thus, the combined luminosity
of the $(1.5~M_{\odot},0.5~M_{\odot})$ system is considerably higher
than the combined luminosity of the $(1.0~M_{\odot},1.0~M_{\odot})$ system,
as expected.  On the other hand, following the work by \cite{dvo86} and
\cite{hol99}, an unequal distribution of stellar mass, i.e., a smaller
value of $\mu$ (see Sect.~4), entails a smaller orbital stability limit.
Since it constitutes a lower limit, i.e., positioned more closely to the
stellar system,  it offers larger ``windows of opportunity" for planets
in the RHZs (if existing) to be orbitally stable.

Results for separation distances $2a$ of 0.5, 1.0, and 2.0~AU are given
in Fig.~6.  For a binary separation of 0.5~AU, it is found that both the
radiative CHZ and GHZ exist, and habitability in these domains is fully
permitted according to the planetary orbital stability constraint, although
the width of the CHZ is relatively small.  The CHZ extents from 2.14 to
2.28~AU, whereas the GHZ extents from 1.91 to 2.90~AU; the orbital
stability limit is given at 0.66~AU.  In this type of system, there are
extreme variations for the inner and outer limits of both the radiative
CHZ and GHZ.  For example, the inner RHL for the CHZ varies between
1.65 and 2.14~AU, whereas its outer RHL varies between 2.28 and 2.76~AU
as function of polar angle $\varphi$.  There is also a considerably
large domain of the supplementary portion of the radiative EHZ, which
has an outer limit that varies between 4.40 and 4.89~AU.  Detailed
depictions of the variations of the inner and outer limits of the RHZs
for the various systems are given in Fig.~7.  This figure indicates
relatively small bars of variations for equal-mass systems such as
$M_1 = M_2 = 1.0~M_\odot$ with small separation distances as, e.g., 
$2a = 0.5$~AU.  However, large bars of variations are obtained for
non-equal mass binaries or for equal-mass binaries with large separation
distances as, e.g., $2a = 2.0$~AU.  

In systems with a binary separation of 1.0~AU, the radiative
CHZ is nullified; note that the orbital stability limit in this system
is given at 1.32~AU.  The reason for the disallowance of the CHZ is that
the RHL for $s_4$, which is at 2.38~AU, is located inside of the RHL for
$s_2$, given as 2.05~AU.  The same criterion (see Eq.~40) also leads to
a relatively small width of the radiative GHZ, which extends between
2.16 and 2.66~AU.  At a binary separation of 2.0~AU, the situation is
even more drastic as both the radiative CHZ and GHZ are disallowed.  The
only type of circumbinary habitable region remaining is that provided by
the relatively large supplementary portion of the radiative EHZ
given by the bracket $(s_6-s_5)$.  In this zone, habitable planets are
expected to be possible as their existence would be consistent with the
planetary orbital stability constraint.


\subsubsection{Additional Analyses}

Next we explore the existence of P-type RHZs, both for equal-mass and
non-equal mass binaries, in a more systematic manner through the means of
{\it numerical experiments}.  Specifically, we pursue sets of model calculations
with the binary separation distance $2a$ considered as an independent
variable; see Table~9 for results.  The stellar masses are altered
between $0.5~M_\odot$ and $1.5~M_\odot$ in increments of $0.25~M_\odot$
(see Table 5).  Results are given for the pairings $(s_2,s_4)$ (CHZ),
$(s_1,s_5)$ (GHZ), and $(s_1,s_6)$ (EHZ).  Note that for equal-mass
binaries, it is sufficient to solve Eq.~(45), whereas for general binary
systems a more thorough assessment is needed to satisfy relation (40).
This approach allows us to explore the maximum binary separation
distances, which are upper limits for permitting RHZs for each case
(i.e., combination of binary masses and choice of CHZ, GHZ, or EHZ).

Generally, it is found that for any binary system, the greatest
permissible binary separation distance is attained for the EHZ, and
furthermore that value-as-attained is greater for the GHZ than for the
CHZ; these findings are as expected.  For example, for the system
$M_1 = M_2 = 1.0~M_\odot$, the expiration distance for the radiative
EHZ is given as 4.25~AU, whereas for the radiative GHZ and CHZ, the
distances are given as 2.57 and 1.64~AU, respectively (see Table~9).
As another example, the system $M_1 = 1.25~M_\odot$ and
$M_2 = 0.75~M_\odot$, the expiration distances for the radiative
EHZ, GHZ, and CHZ are given as 3.80, 1.93, and 0.96~AU, respectively.

It is also intriguing to compare results for stellar pairs
as, e.g., $M_1 = M_2 = 1.0~M_\odot$ to stellar pairs such as
$M_1 = 1.5~M_\odot$ and $M_2 = 0.5~M_\odot$.  For the system of
$M_1 = M_2 = 1.0~M_\odot$, it is found that the radiative CHZ and GHZ
are nullified --- as defined by the limit of validity of Eq.~(45) ---
at binary separation distances of 1.64 and 2.57~AU, respectively (see
Table~9).  At those binary separations, the distances of the vanishing
RHZ-CHZ and RHZ-GHZ (as measured from the geometrical center of the system;
see Fig.~1) are given as 1.95 and 2.25~AU, respectively.  In comparison,
the limits of planetary orbital stability (to be interpreted as lower
limits) are identified as 3.00 and 4.71~AU, respectively.  Thus, we
conclude that for equal-mass binary systems such as $M_1 = M_2 = 1.0~M_\odot$,
habitability for widely spaced binaries is lost due to the lack of orbital
stability already at binary separations where the circumbinary CHZ-RHZs
and GHZ-RHZs are still in place.

The same type of study has been pursued for systems with highly unequal
intrabinary distributions of masses and, by implication, stellar luminosities
as, e.g., $M_1 = 1.5~M_\odot$ and $M_2 = 0.5~M_\odot$.   In this case it is
found that the radiative CHZ and GHZ vanish at binary separation distances
$2a$ of 0.65 and 1.55~AU, respectively.  Furthermore, the distances of
the vanishing radiative CHZ and GHZ (as measured from the system center,
see Fig.~1) are given as 2.21 and 2.43~AU, respectively (see Table 9).
The respective limits of planetary orbital stability are identified
as 0.85 and 2.04~AU.  Thus, for this type of system it is found that
habitability is lost due to the vanishing RHZs, even though circumstellar
habitability would still be permitted according to the planetary orbital
stability criterion.  The fact that circumbinary habitability is lost
already for systems of relatively small binary separations is a consequence
of the extreme radiative imbalance caused by the highly unequal distribution
of stellar luminosities, which determine the circumbinary RHLs.

Radiative imbalance within binary systems
may cause the RHL for $s_{\ell,{\rm out}}$ to be
partially or completely located inside of the RHL for $s_{\ell,{\rm in}}$;
see Sect.~3.4.  In fact, when the pairs $M_1 = M_2 = 1.0~M_\odot$ and
$M_1 = 1.5~M_\odot$ and $M_2 = 0.5~M_\odot$ are compared to one another,
it is found that although the unequal-mass binary system has almost
twice the combined stellar luminosity of the equal-mass binary system
(i.e., 3.85 versus 2.0 $L_\odot$), it still possesses much narrower
CHZ, GHZ, and EHZ RHZs.  In fact, it is found that the condition
expressed as Eq.~(40) is most readily met in cases of equal-mass binary
systems of relatively small separation distances and mostly violated
in systems of relatively large separation distances and/or unequal
distributions of masses and, by implication, luminosities.  Various
examples have been depicted in Fig.~7; see discussion in Sect.~5.2.1.

In summary, although an unequal distribution of stellar masses within
binary systems is identified as advantageous for facilitating planetary
orbital stability, in consideration of that lower stability limits
for P-type orbits occur for smaller mass ratios $\mu$ (see Sect.~4),
the situation for the existence of the RHZs is much less ideal,
even for systems where the stellar primary is highly luminous
owing to the behavior of the RHLs.
In this regard, the radiative CHZ is in most jeopardy as $(s_4-s_2)$
constitutes the smallest bracket among the various kinds of HZs
(see Table~2).  More fortunate scenarios are expected to occur for
the radiative GHZ and EHZ, with the brackets given as $(s_5-s_1)$ and
$(s_6-s_1)$, respectively; they are characterized by considerably
larger widths, especially in case of equal-mass systems of stars with
relatively high luminosities.


\section{Proposed Habitability Classification: Habitability Types S, P, ST, and PT}

Another aspect of this study is to provide an appropriate classification of
habitability applicable to general binary systems.  Previously, \cite{dvo82}
introduced the terminology of S-type and P-type orbits for system planets,
which is now widely used by the orbital stability, planetary, and the
astrobiology science communities.  Evidently, besides the assessment of
orbital stability behaviors, these terms are also appropriate for classifying
binary system RHZs, if existing.  However, following previous investigations
\citep[e.g.,][]{tak08,dvo10,hag10,kop10,egg12}, as well as the results of
the present work, the spatial domain of S-type and P-type habitability
depicted by the RHZs is often adversely affected, and in some cases even
nullified, by the requirement that system planets must be orbitally stable.
Thus, if the available extent of the S-type and P-type RHZs for the
manifestation of habitability is truncated owing to the additional constraint
of planetary orbital stability, these zones shall be referred to as ST-type
and PT-type, respectively, in the following.

Detailed results are given in Table~10, which provides an extensive summary
of P, PT, ST, and S-type habitability for both equal-mass and non-equal
mass binary systems.  The stellar masses are varied between
$0.5~M_\odot$ and $1.5~M_\odot$ in increments of $0.25~M_\odot$ amounting
to a total of 15 combinations.  Table~10 features the results
for the pairings $(s_2,s_4)$ (CHZ), $(s_1,s_5)$ (GHZ), and $(s_1,s_6)$ (EHZ).
In principle, it is found that --- with the secondary taken as fixed ---
the higher the mass and, by implication, the luminosity of the stellar
primary, the larger values are obtained for P, PT, ST, and S-type
habitability.  Additionally, larger values for the limits of P, PT, ST,
and S-type habitability are obtained regarding the GHZ relative to the
CHZ, as expected.  The largest values are obtained for PT and S-type
habitability for the EHZ; in this regard, there is no change for P
and ST-type habitability relative to the GHZ since both types of HZs
are based on the same inner bracket value of $s_1$ (see above).

The results of Table~10 are in line with the previously discussed findings
about highly unequal intrabinary distributions of masses and, by
implication, stellar luminosities as, e.g., $M_1 = 1.25~M_\odot$ and
$M_2 = 0.75~M_\odot$ or $M_1 = 1.5~M_\odot$ and $M_2 = 0.5~M_\odot$
compared to the case of $M_1 = M_2 = 1.0~M_\odot$.   For systems of
highly unequal mass distributions, the domains of P-type and PT-type
habitability are typically relatively small as the RHL for $s_{\ell,{\rm out}}$
crosses the RHL for $s_{\ell,{\rm in}}$ in relatively close proximity
to the primary, such allowing only small distance ranges to exhibit
P/PT-type habitability.
It is also found that in seven cases for the CHZ, as well as two cases
for the GHZ, the RHZs expire prior to the truncation of habitability
due to the planetary orbital stability requirement.  In those cases,
only P-type habitability exists; no PT-type habitability is found
as the orbital stability constraint bears no relevance.

Figure 8 and 9 show various combinations of equal-mass and nonequal-mass 
binary systems; they all show numerous similarities, though the spatial
scales are noticeably different as they are defined through the stellar
luminosities.  If equal-mass binary systems are considered, taking
$M_1 = M_2 = 1.0~M_\odot$ and $M_1 = M_2 = 0.5~M_\odot$ as examples, 
the extent of both P and PT-type habitability increase with increasing
stellar mass or luminosity.  The distances for P and PT habitability are
found to almost coincide indicating that the orbital stability constraint
affects the inner and outer limit of P-type habitability in about the
same manner.  Furthermore, the inner and outer limits of both S and ST
habitability are shifted to larger distances from each stellar component
for stars of higher luminosity, as expected.  Moreover, for stars
of higher luminosity, there is a larger spatial domain where S-type
habitability is truncated due to the additional constraint of planetary
orbital stability.  For equal-mass systems of $1.0~M_\odot$, ST and
S-type habitability is identified at distances of 6.85 and 13.46~AU,
whereas for $0.5~M_\odot$, ST and S-type habitability is identified
at 1.38 and 3.00~AU.

Figure 9 depicts two selected cases of nonequal-mass binary systems.
In both cases the stellar primary is chosen as 1.0~$M_\odot$, whereas
the stellar secondary is chosen as 0.75~$M_\odot$ and 0.5~$M_\odot$,
respectively; the corresponding stellar luminosities of the secondaries
are 0.357 and 0.045~$L_\odot$, respectively (see Table 5).  A reduced
luminosity of the secondary binary component adversely affects the
extent of the RHZ, as expected.  Interestingly, a reduction of mass
for the stellar secondary has a nontrivial impact on the orbital
stability domains, which considerably depend on the mass ratio $\mu$
(see Eqs. 46 and 47).  If $\mu$ is reduced, the permissible stability
domain for P-type orbits is increased, whereas the permissible
stability domain for S-type orbits is decreased.  Thus, the assessment
of S, P, ST, and PT habitability for nonequal-mass binaries requires
a detailed computational analysis.  In the example of Fig.~9, in
regard to $M_2 = 0.75~M_\odot$, S, P, ST, and PT habitability occurs
at distances of 0.82, 1.16, 6.18, and 12.19~AU, respectively, and for
$M_2 = 0.5~M_\odot$, S, P, ST, and PT habitability occurs at distances
of 0.94, 1.01, 5.50, and 10.86~AU, respectively.  The respective
differences are notable, but not drastic; more pronounced differences
occur for systems of more luminous stars (see Table~10).  Note that both
Figs. 8 and 9 refer to habitability assessments pertaining to the GHZ
(see Table~2).  They are given to exemplarily showcase the structure,
extent and location of the S-type and P-type RHZs as well as the relevance
of the orbital stability limits for both S-type and P-type habitability,
thus allowing us to uniquely identify the spatial domains of S, P, ST,
and PT-type habitability for each system.

\clearpage

Computationally, the occurrence of P, PT, ST, and S-type habitability can be
identified as follows.  For sufficiently small binary separations
$2a$, it is found that both the inner and outer limit of the P-type RHZ are
located beyond of the P-type orbital stability limit (see Eq.~47), which
constitutes a lower limit of planetary orbital stability (see Sect.~4).
If the distance of binary separation is increased, the inner and outer limits
of the P-type RHZ decrease, whereas the P-type orbital stability limit increases.
Thus, starting at a certain value of $2a$, only a fraction of the width of the
P-type RHZ will be available for providing habitability; in this case, PT-type
habitability is attained.  If the binary separation is further increased,
the entire width of the P-type RHZ will be unavailable for providing
habitability because habitability would be incompatible with the orbital
stability constraint.  Eventually, the P-type RHZ expires; see also
information provided in Table~9.

For mid-sized values of the binary separation distance, the S-type RHZ is
encountered to exist, but it is unable to provide habitability because of the
S-type orbital stability limit (see Eq.~46), which constitutes an upper
limit of planetary orbital stability (see Sect.~4).  The S-type RHZs continue
to exist further out; note that the inner and outer limits of the S-type RHZs
essential continue to run parallel as function of the binary separation distance
for most systems.  If the binary separation distance $2a$ is further increased,
again as part of our numerical experiment,
the S-type orbital stability will increase and will cross the inner limit of
the S-type RHZ; in this case, ST-type habitability is encountered as some, but
not all of the width of the S-type RHZ is available for providing habitability.
Eventually, for sufficiently large binary separations, the S-type orbital
stability limit also crosses the outer limit of the S-type RHZ.  In this case,
the full width of the S-type RHZ is available for facilitating habitability,
consistent with the definition of S-type habitability.

Another application is displayed in Fig.~10.  It shows for a given spectral type
of equal-star binaries the stellar separation distances $2a$ for which CHZs, GHZs,
and EHZs are able to exist.  The CHZs, GHZs, and EHZs can be either S or ST-type,
on one hand, or P or PT-type, on the other hand, to qualify for depiction.  The
results are given as function of stellar spectral type, for stars between
spectral type F0 to M0.  The figure shows that P/PT-type habitable regions are able
to exist for a relatively large range of separation distances in case of relatively
luminous stars (i.e., spectral type F), but only for a relatively small range of
separation distances for lesser luminous stars (i.e., spectral types K and M).
Regarding S/ST-type habitable regions the situation is reversed.  Figure 10 also
indicates a notable domain of binary separations where no habitable regions are
found owing to the lack of RHZs, the lack of planetary orbital stability, or both.
Moreover, no domain of binary separation distances is identified where S/ST-type
and P/PT-type habitable regions overlap.


\section{Summary and Conclusions}

In this study we present a new method about a comprehensive assessment
of S-type and P-type habitability in stellar binary systems.  P-type
orbits occur when the planet orbits both binary components, whereas in case
of S-type orbits the planet orbits only one of the binary components with
the second component considered a perturbator.  An important characteristic
of the new method is that it combines the orbital stability constraint for
a system planet with the necessity that a habitable region given by
the stellar radiative energy fluxes (``radiative habitable zone") must
exist.  The requirement to combine these two properties has also been
recognized in previous studies \citep[e.g.,][]{tak08,kop10,egg12,kan13}.

Another element of the present study is to introduce a habitability
classification regarding stellar binary systems, consisting of habitability
types S, P, ST, and PT.  This type of classification also considers whether
or not S-type and P-type radiative habitable zones are reduced in size due
to the additional constraint of planetary orbital stability.  In summary,
five different cases were identified, which are:  S-type and P-type
habitability provided by the full extent of the RHZ; habitability, where
the RHZ is truncated by the additional constraint of planetary orbital
stability (labelled as ST and PT-type, respectively); and cases of
no habitability at all.  This classification scheme can be applied
to both equal-mass and non-equal mass binary systems, as well as to
systems with binaries in elliptical orbits, which will be the focus
of the forthcoming Paper~II of this series.  As part of the current
study a significant array of results are given for a notable range
of main-sequence stars, which are of both observational and theoretical
interest.

A key aspect of the proposed method is the introduction of a combined algebraic
formalism for the assessment of both S-type and P-type habitability; in particular,
mathematical criteria are presented allowing to determine for which systems
S-type and P-type RHZs are realized.  In this regard, a priori choices about
the presence of S-type and P-type RHZs are neither necessary nor possible
as the existence of S-type as well as P-type RHZs is proliferated through
well-defined mathematical conditions pertaining to the underlying fourth-order
algebraic equation.  The coefficients of the polynomial are given by the binary
separation distance ($2a$), the solar system-based parameter for the limit of
habitability ($s_\ell$), and the modified values for the luminosities
($L_{1\ell}'$, $L_{2\ell}'$) of the stellar binary components, referred to as recast
stellar luminosities.  Regarding the binary system habitable zone, we consider
conservative, general and extended zones of habitability, noting that their
inner and outer limits are informed by previous solar system investigations
\citep[e.g.,][]{kas93,und03,sel07}.

\clearpage

In our segment of applications, we examined the existence of habitable
S-type orbits for selected examples.  We found that regarding the RHZs,
owing to the typically relatively large separation of the stellar
components, the effect of the stellar secondary on the extents of
the RHZs is usually very minor.  The secondary's main influence on
circumstellar habitability consists in imposing restrictions regarding
planetary orbital stability implemented as an upper stability limit
around each stellar component, which often truncates or nullifies
S-type planetary habitability.  In the framework of our study, we
specifically considered the radiative EHZ, which is most outwardly
extented (i.e., up to 2.4~AU in case of the Sun).  It was found that
this kind of zone is most affected by the limitation of planetary
orbital stability as it is located closest to the secondary stellar
component.

Furthermore, we also examined the existence of habitable P-type orbits.
In this case, relatively complicated scenarios emerge.  In general,
it was found that the best prospects for circumbinary habitability
emerge for (1) systems with stellar components of relatively high
luminosities (no surprise here!), (2) systems where the stellar
luminosities are relatively similar (for main-sequence stars, as
implied by their stellar masses), and (3) systems of relatively
small binary separations.  If conditions (2) or (3) are not met,
it may occur that the outer RHL is located inside of the inner RHL,
thus nullifying the RHZ irrespectively of planetary orbital stability
considerations.  On the other hand, an unequal intrabinary distribution
of masses entails a lower limit of planetary orbital stability
(i.e., positioned closer to the binary system) thus implying an
enhanced opportunity for circumbinary habitability.  However, this
aspect is of lesser significance for most systems compared to the
restrictions for the RHZs due to the imbalance given by the stellar
luminosities.

Various applications in this study concern stars of masses between
0.75 and 1.5~$M_\odot$.  This approach is motivated to unequivocally
demonstrate the effects of stellar binarity on the extent and
structure of circumstellar habitability, which is most pronounced
for massive, i.e., highly luminous stars.  Nonetheless, most
stars in binaries are expected to be low-mass stars, i.e., stars
of spectral types K and M, owing to the skewness of the Galactic
initial mass function \citep[e.g.,][]{kro01,kro02,cha03}.
For example, we compared pairs of systems given by
(1.0~$M_\odot$, 1.0~$M_\odot$) and (1.5~$M_\odot$, 0.5~$M_\odot$).
Obviously, the overall luminosity is by far greatest in the
(1.5~$M_\odot$, 0.5~$M_\odot$) system following the
mass--luminosity relationship, i.e., $L_\ast \propto M_\ast^4$
\citep[e.g.,][]{rei87}.  However, this system is found to be
the highly unfavorable for the facilitation of circumbinary
habitability.  Particularly, it is found that the P-type GHZ in the
(1.0~$M_\odot$, 1.0~$M_\odot$) system extends to 0.91~AU, whereas
it extends only to 0.65~AU in the (1.5~$M_\odot$, 0.5~$M_\odot$)
system.  Furthermore, smaller spatial extents are identified
for P-type CHZs, as this type of HZ is in highest jeopardy
owing to the relative small $(s_4-s_2)$ bracket compared to
the $(s_5-s_1)$ bracket for GHZs (see Table~2).  In fact, a
considerable number of systems do not offer CHZs at all,
which again is a consequence of the radiative imbalance in
those systems.  Also, the nullification of CHZs in binary systems
is most likely to occur in systems of relatively large separation
distance.  In contrast, the best opportunities for facilitating
circumbinary habitability is given in the context of EHZs, as
expected.

Future work will deal with a significant augmentation of our
method to other systems, including systems with binary components
in elliptical orbits (see Paper~II).  This will allow us to compare
applications of our method, including results for
individual systems, to other findings in the literature.  We also
expect our method to be applicable to general binary systems with
main-sequence stars as well as to systems containing evolved stars;
this latter effort is motivated by observational evidence and
supporting theoretical efforts indicating that planets are also
able to exist around stars that have left the main-sequence
\citep[e.g.,][]{sato03,ram09,ebe10,doy11,sato13}.  Particularly, it is
highly desirable to augment our method to systems of higher order,
as motivated by the steady progress in theory as well as ongoing
and future observational discoveries of exosolar planetary systems.

\acknowledgments
This work has been supported in part by the SETI institute.  The author
acknowledges comments by B. Quarles and Z. E. Musielak as well as
assistance with computer graphics by S. Sato, S. Satyal, and M. Sosebee.
The paper also benefited from detailed comments by an anonymous referee. 
This study made use of the software applications Fortran$^\registered$,
MATHEMATICA$^\registered$, and MATLAB$^\registered$.
The author anticipates the development of a black box code, called
{\tt BinHab}, to be hosted at The University of Texas at Arlington,
which will allow the assessment of habitability in binary systems
based on the developed method.


\clearpage


\begin{figure*} 
\centering
\begin{tabular}{c}
\includegraphics[scale=1.2]{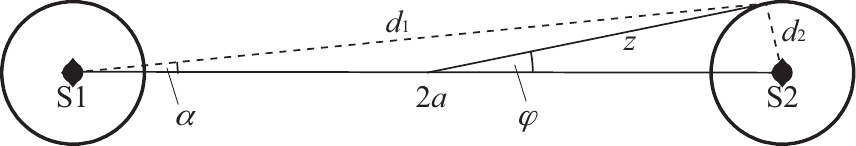} \\  \\ \\ \\ \\
\noalign{\bigskip}
\noalign{\bigskip}
\includegraphics[scale=1.6]{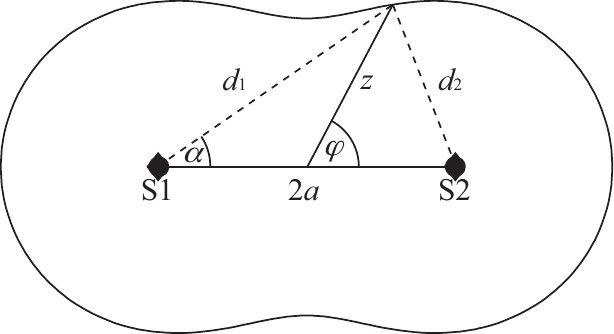} \\ \\ \\ \\ \\
\end{tabular}
\caption{Set-up for the mathematical treatment of S-type (top) and
P-type (bottom) habitable zones of binary systems as given by the
stellar radiative fluxes.  Note that the stars S1 and S2 have been
depicted as identical for convenience.
}
\end{figure*}

%

\clearpage


\begin{figure*} 
\centering
\begin{tabular}{c}
\includegraphics[scale=0.5]{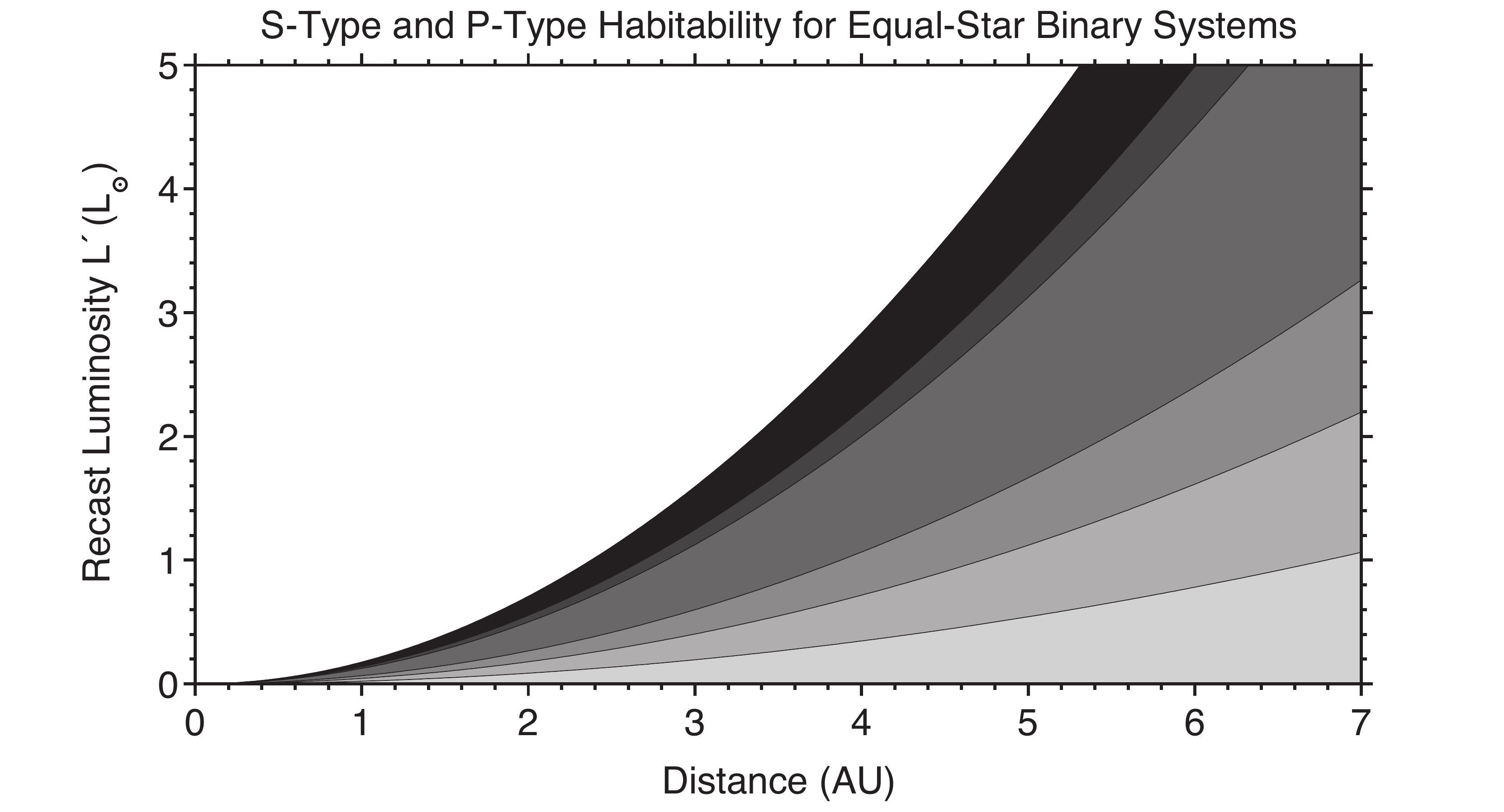}  \\
\end{tabular}
\caption{Habitability limits for equal-mass binaries regarding S-type and P-type
planetary orbits pertaining to the RHZ.  The various limits, separated by grayish areas,
refer to different values of $s_\ell$ given as 0.84, 0.95, 1.00, 1.37, 1.67, and 2.40~AU
(from left to right; corresponding to $\ell$ = 1 to 6); see Sect.~2 and 3.2 for details.
}
\end{figure*}

\clearpage


\begin{figure*} 
\centering
\begin{tabular}{cc}
\includegraphics[scale=0.30]{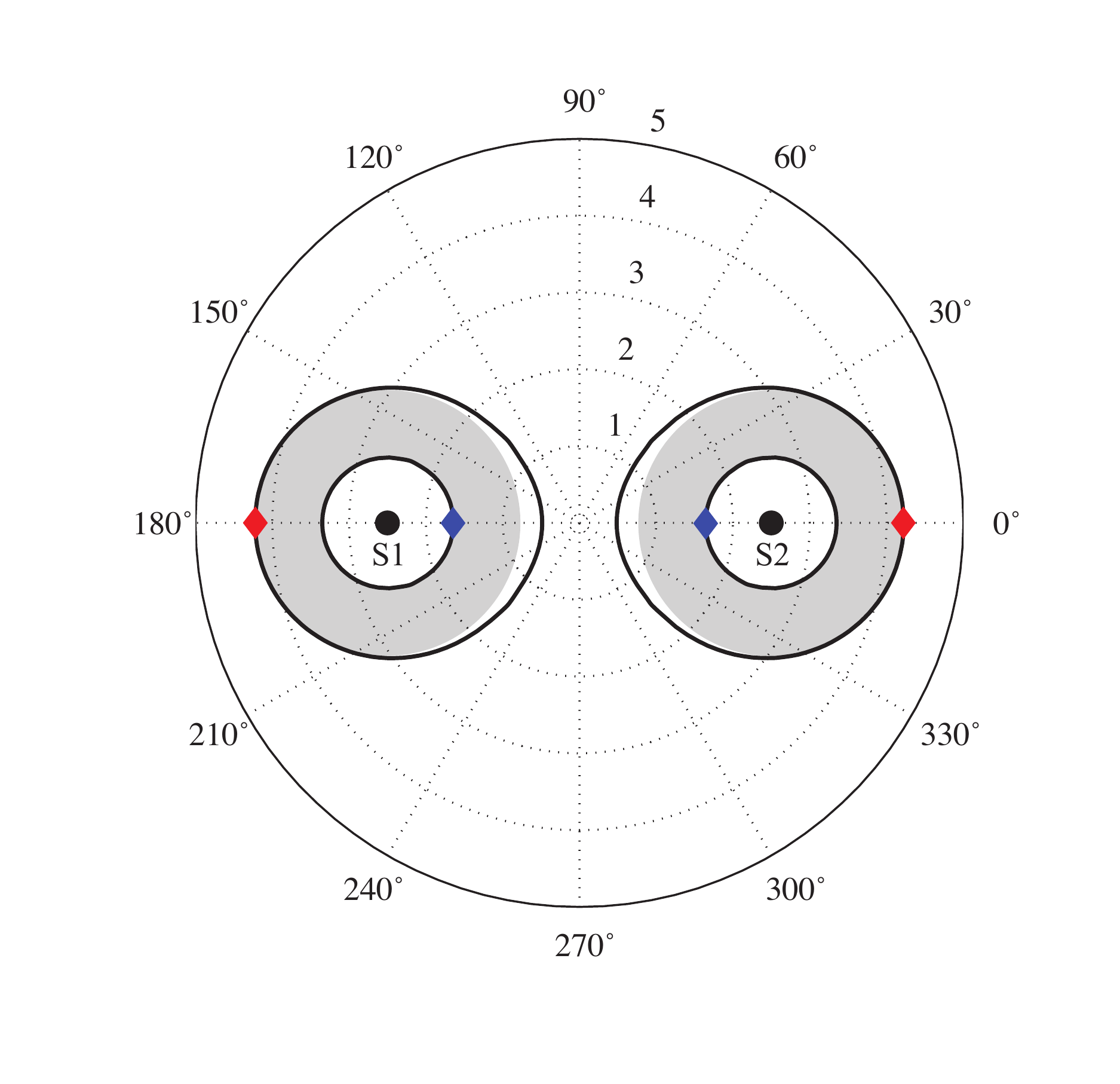} &
\includegraphics[scale=0.30]{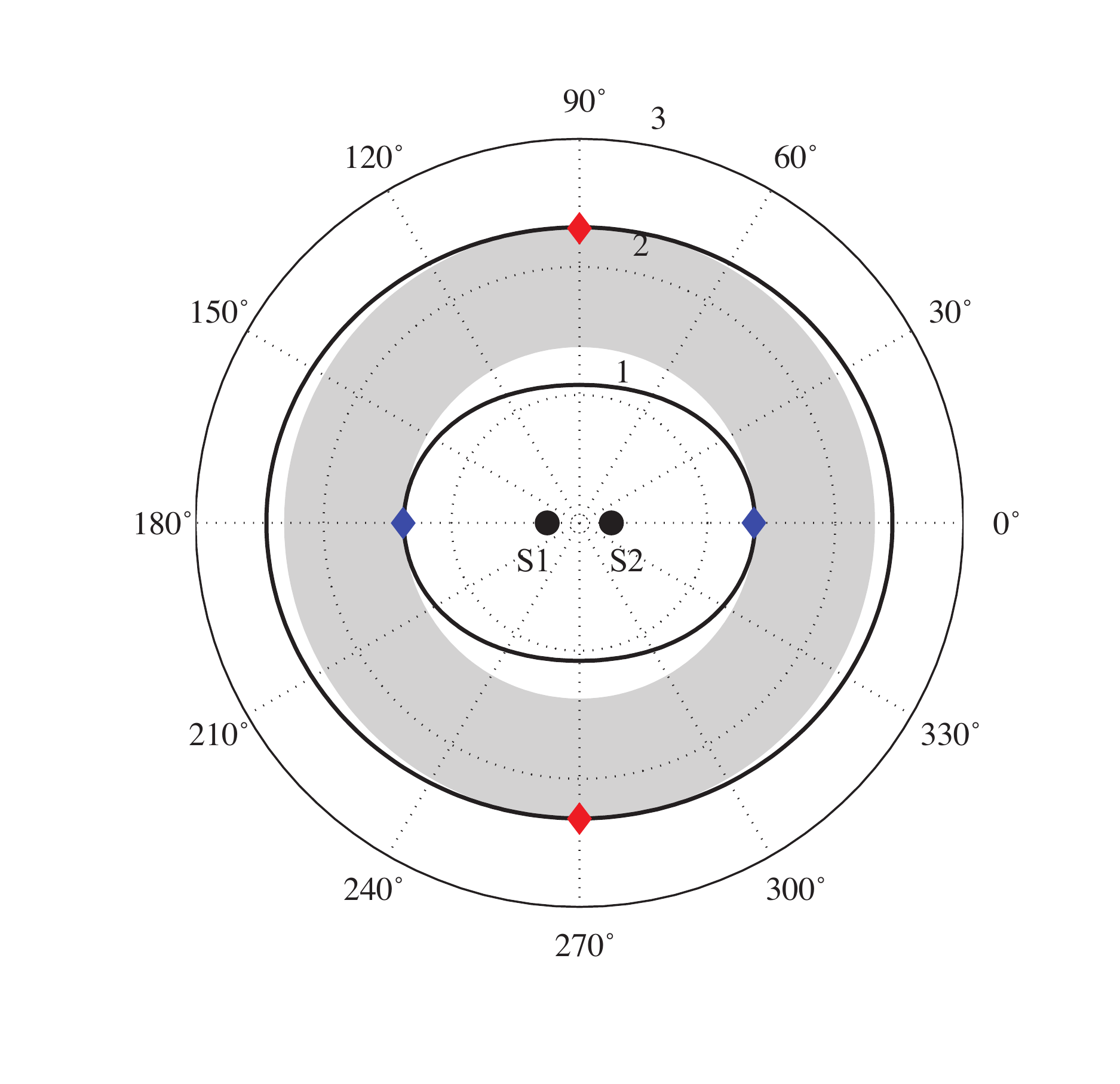} \\
\includegraphics[scale=0.30]{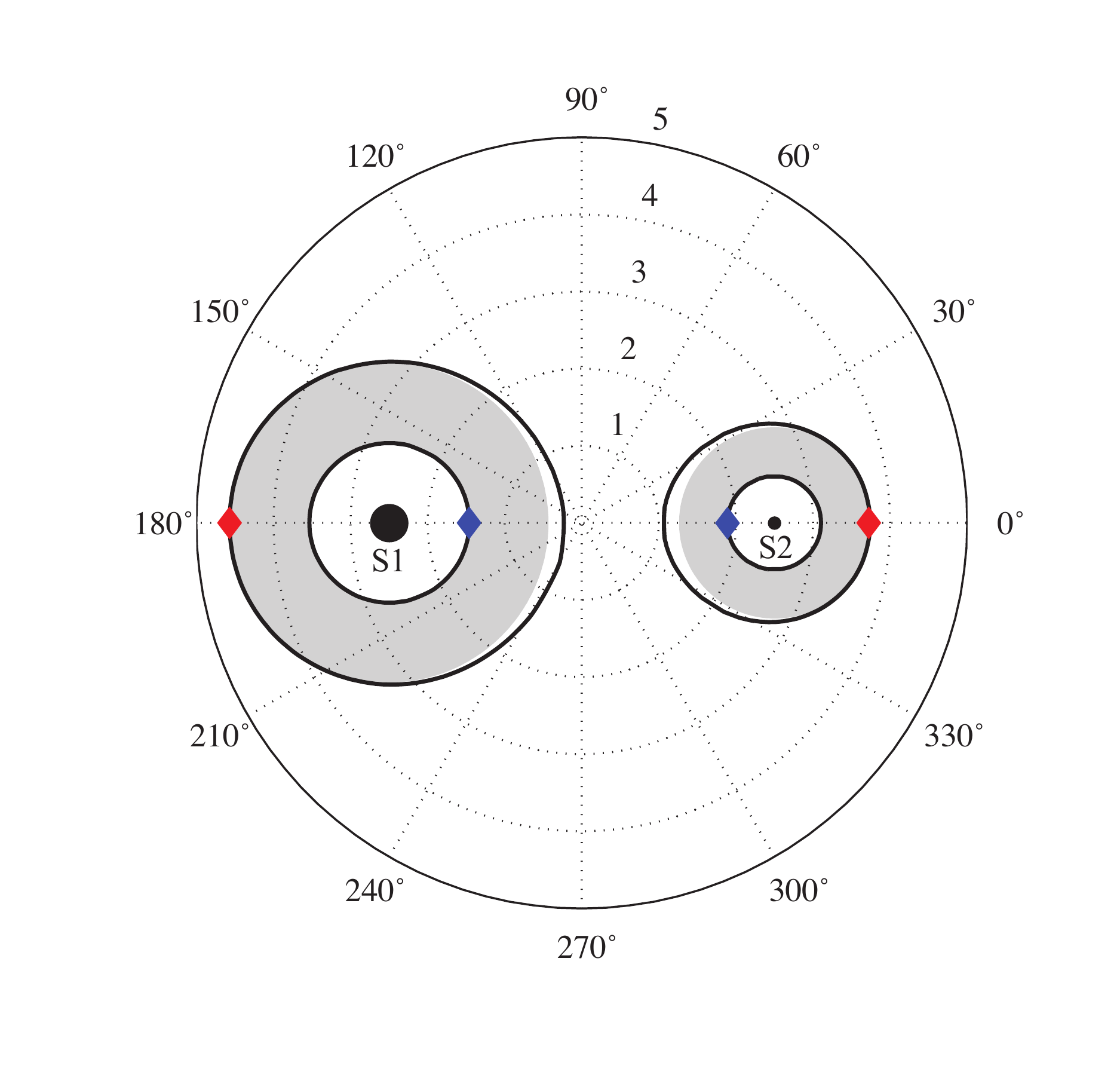} &
\includegraphics[scale=0.30]{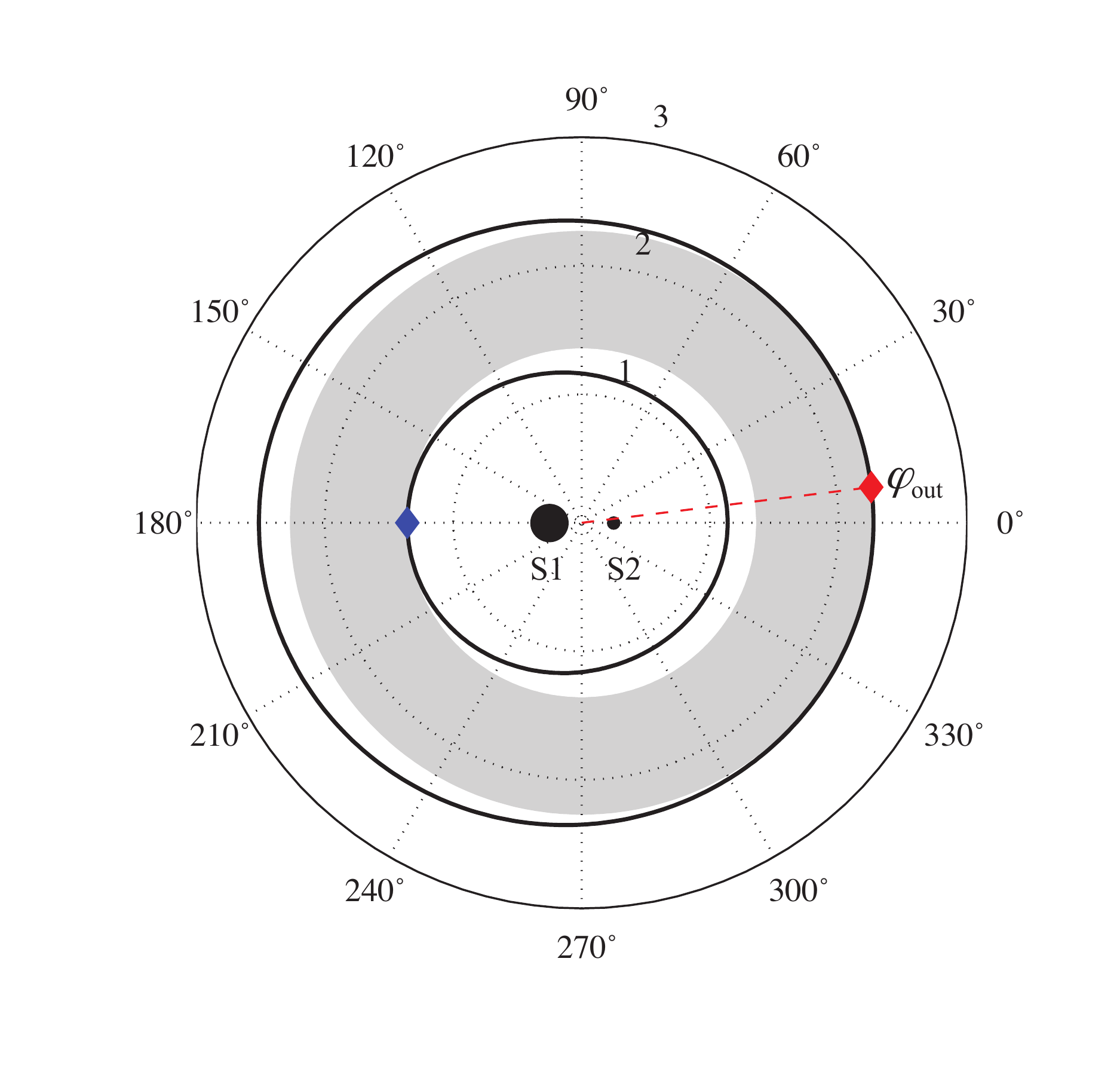} \\
\end{tabular}
\caption{Examples of S-type and P-type RHZs for different systems obtained
through solutions of the governing fourth-order polynomial equations.
The solutions are given in polar coordinates with the radial coordinate
depicted in units of AU.
The thick solid lines indicate the RHLs, corresponding to the inner and
outer limit of habitability based on $s_\ell = 0.84$ and 1.67~AU, respectively.
Note that in the S-type case the RHLs are bended toward the center, whereas
in the P-type case they are of elliptical shape.  The top row displays
systems with $L_1 = L_2 = L_\odot$, whereas the bottom top row displays
systems with $L_1 = 1.5~L_\odot$ and $L_2 = 0.5~L_\odot$.
The left column assumes separation distances $2a$ of 5.0~AU
rendering P-type RHZs, whereas the right column assumes separation
distances of 0.5~AU resultant in S-type RHZs.  The gray areas indicate the
appropriate circular regions (annuli), referred to as RHZs, for each case.
The touching points between the RHZs and the inner and outer RHLs (utilized for
the definition of the RHZs) are depicted as blue and red diamonds, respectively.
They are positioned at angles of 0, $\pi/2$, $\pi$, and $3\pi/2$, except for
the outer RHL in case of P-type RHZs for non-equal star binary systems
(depicted as $\varphi_{\rm out}$).
}
\end{figure*}


\clearpage


\begin{figure*} 
\centering
\begin{tabular}{c}
\includegraphics[scale=0.4]{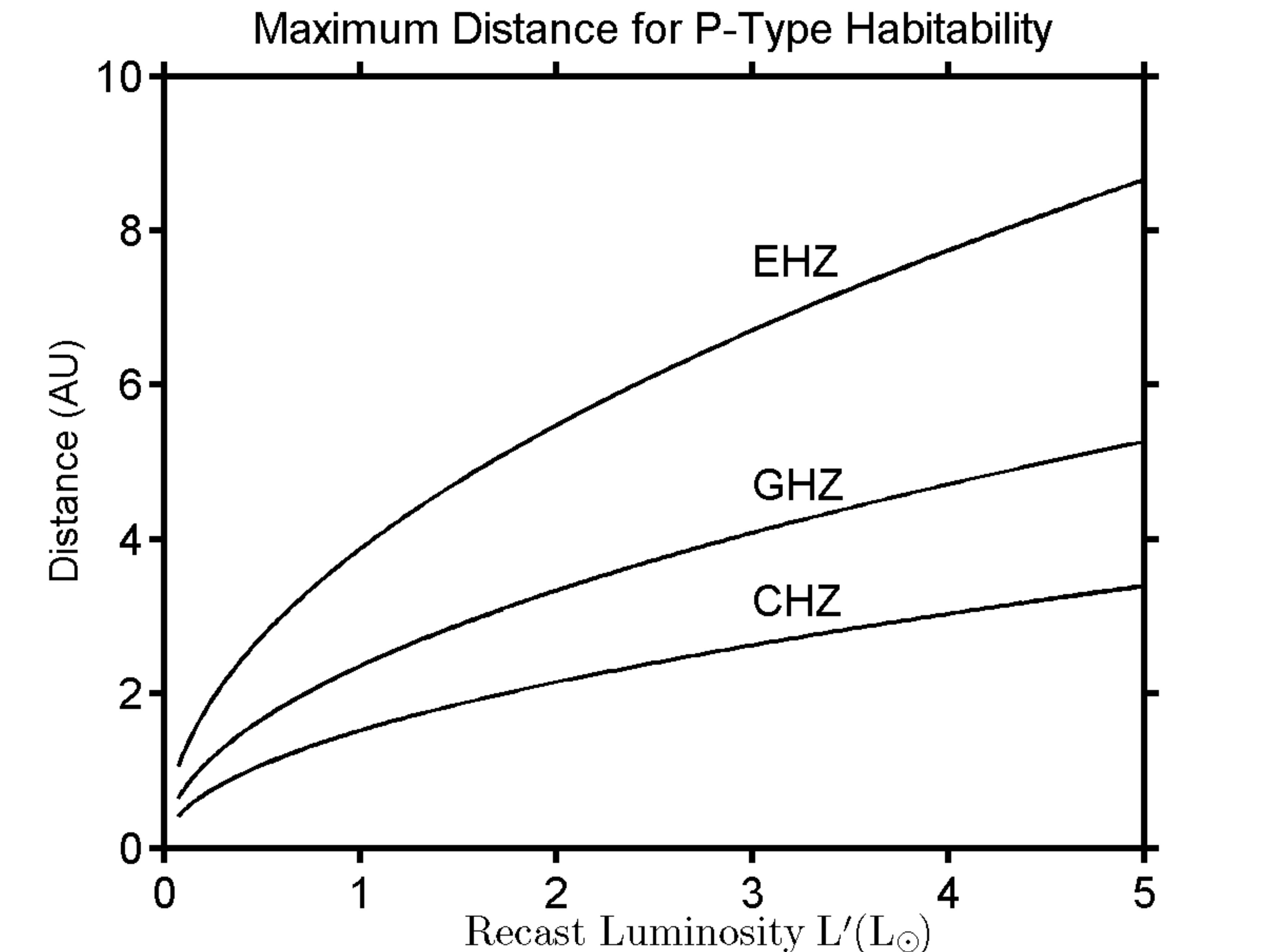}
\end{tabular}
\caption{Maximum of permissible separation distance $2a$ in equal-mass
binary systems still permitting P-type RHZs (see Eq.~45).  Results are
given for the CHZ, GHZ, and EHZ.
}
\end{figure*}

\clearpage


\begin{figure*}
\centering
\begin{tabular}{cc}

\includegraphics[scale=0.30]{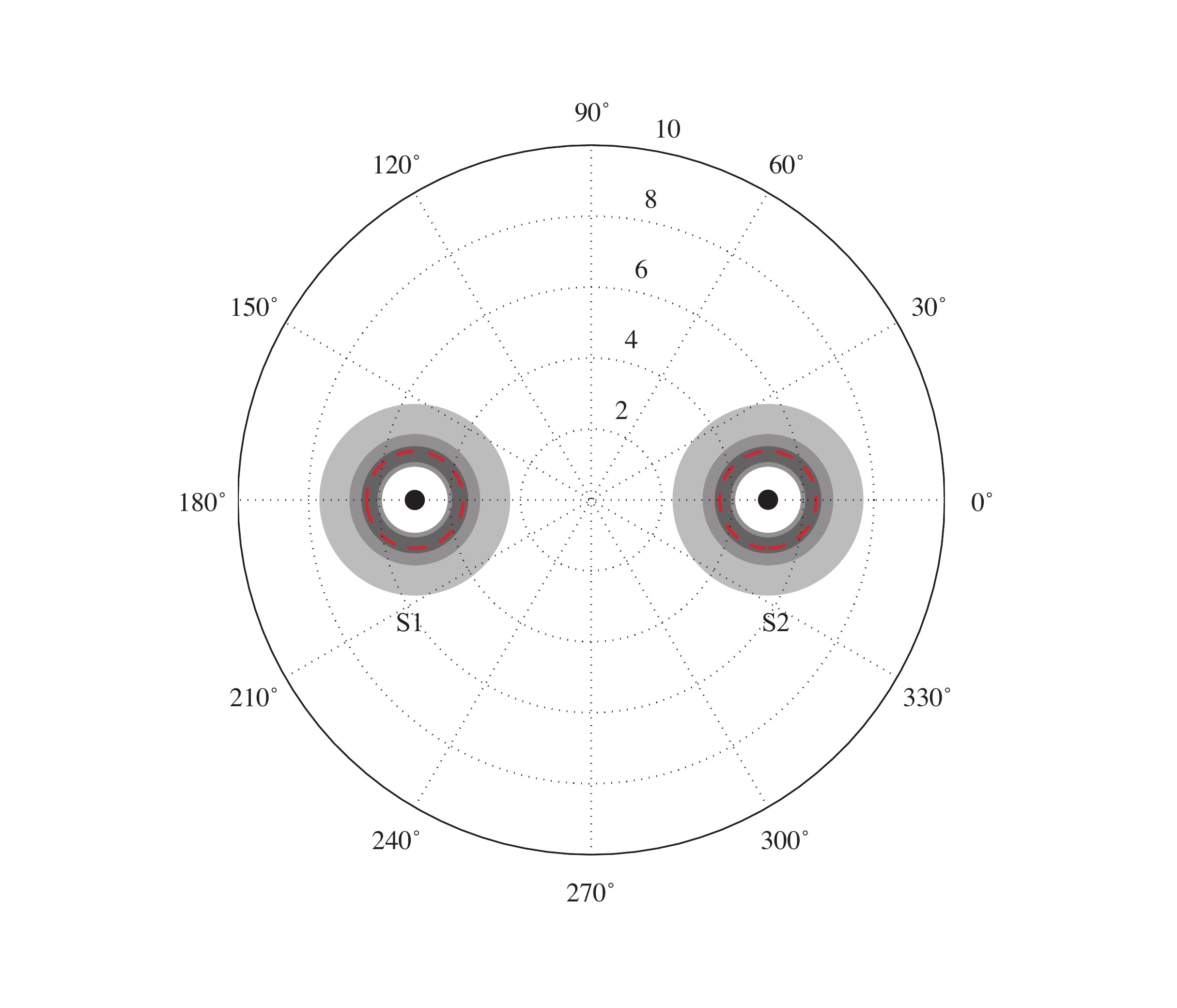} &
\includegraphics[scale=0.30]{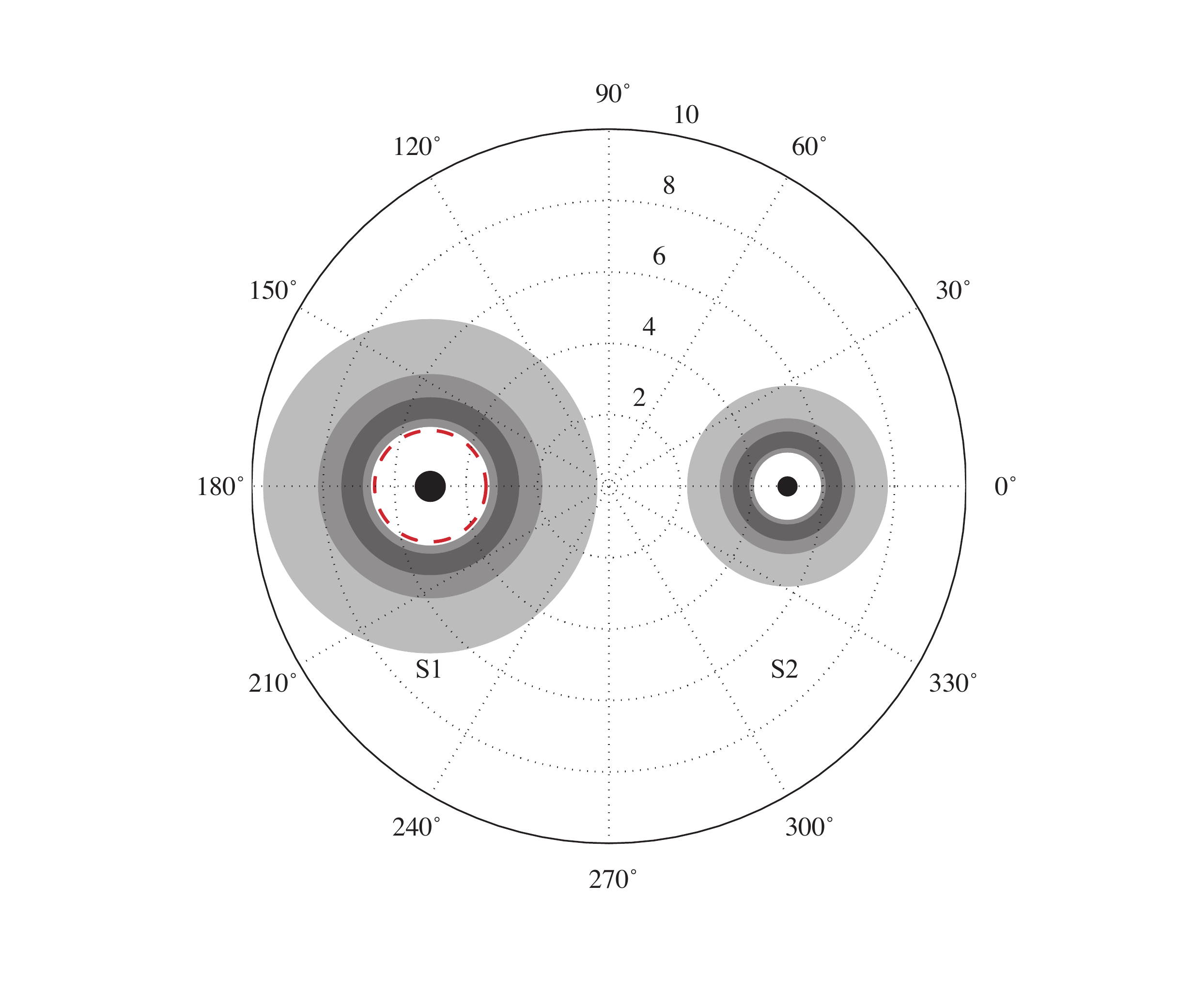} \\
\includegraphics[scale=0.30]{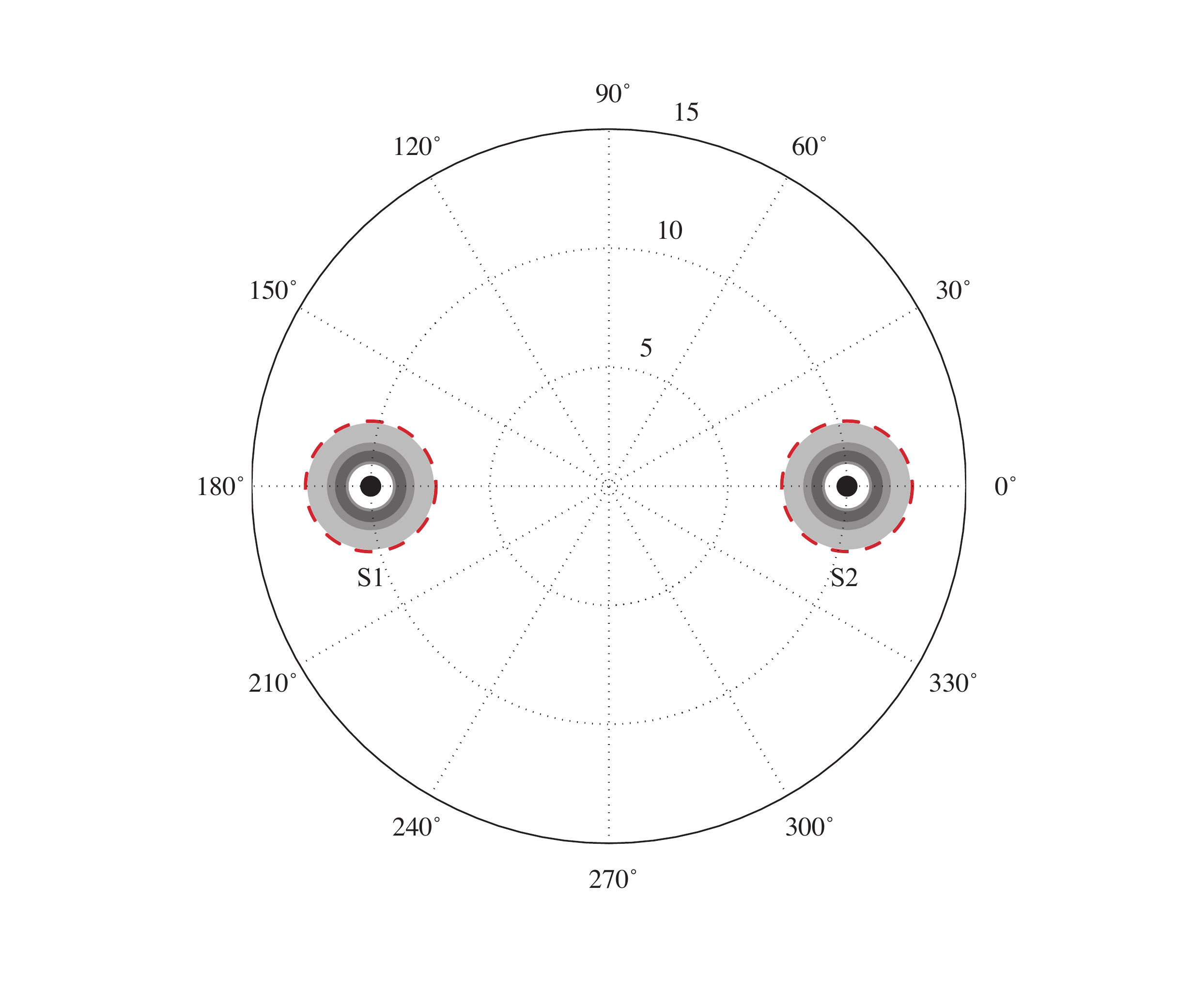} &
\includegraphics[scale=0.30]{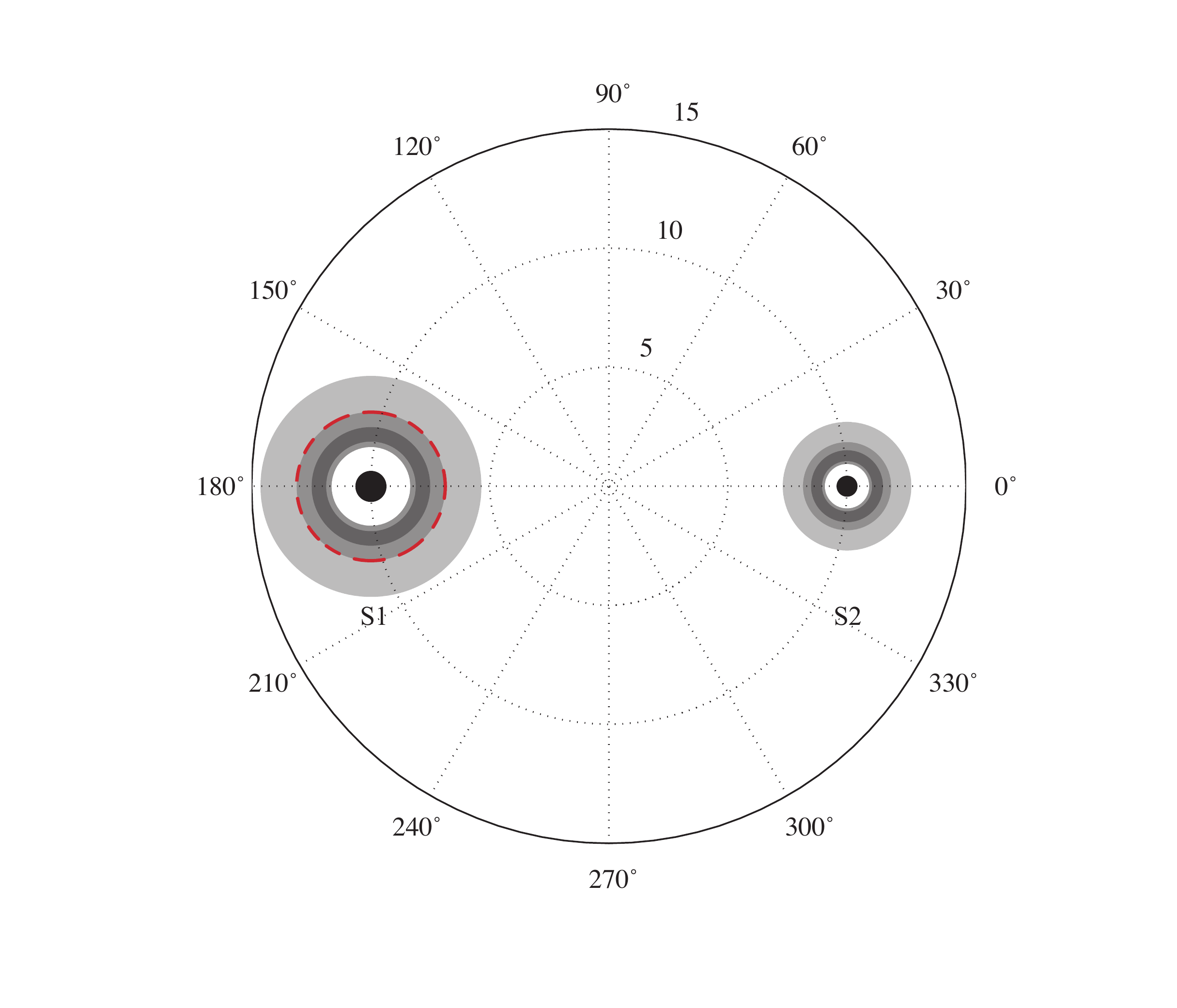} \\

\end{tabular}
\caption{Domains of S-type habitability for main-sequence star binary systems.
The solutions are given in polar coordinates with the radial coordinate depicted in units of AU.
The left column depicts systems with masses of $M_1 = M_2 = 1.0~M_\odot$, whereas
the right column depicts systems with masses of $M_1 = 1.5~M_\odot$ and $M_2 = 1.0~M_\odot$.
The separation distances $2a$ are given as 10~AU (top) and 20~AU (bottom).
Results for the conservative, general and extended RHZs are given as dark gray, medium gray,
and light gray areas, respectively.  The orbital stability limits are indicated by
red dashed lines.
}
\end{figure*}


\clearpage


\begin{figure*} 
\centering
\begin{tabular}{cc}

\includegraphics[scale=0.30]{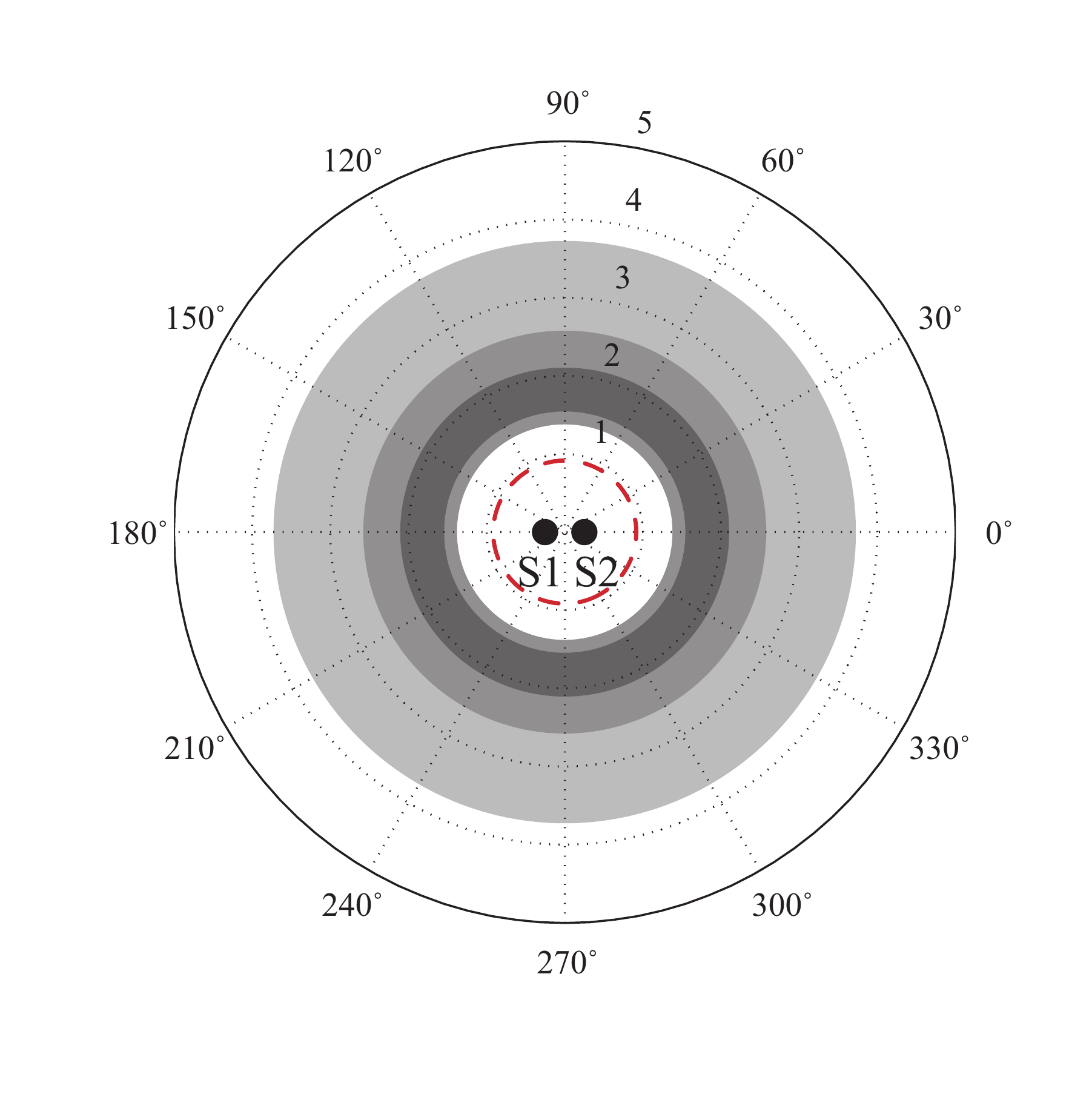} &
\includegraphics[scale=0.30]{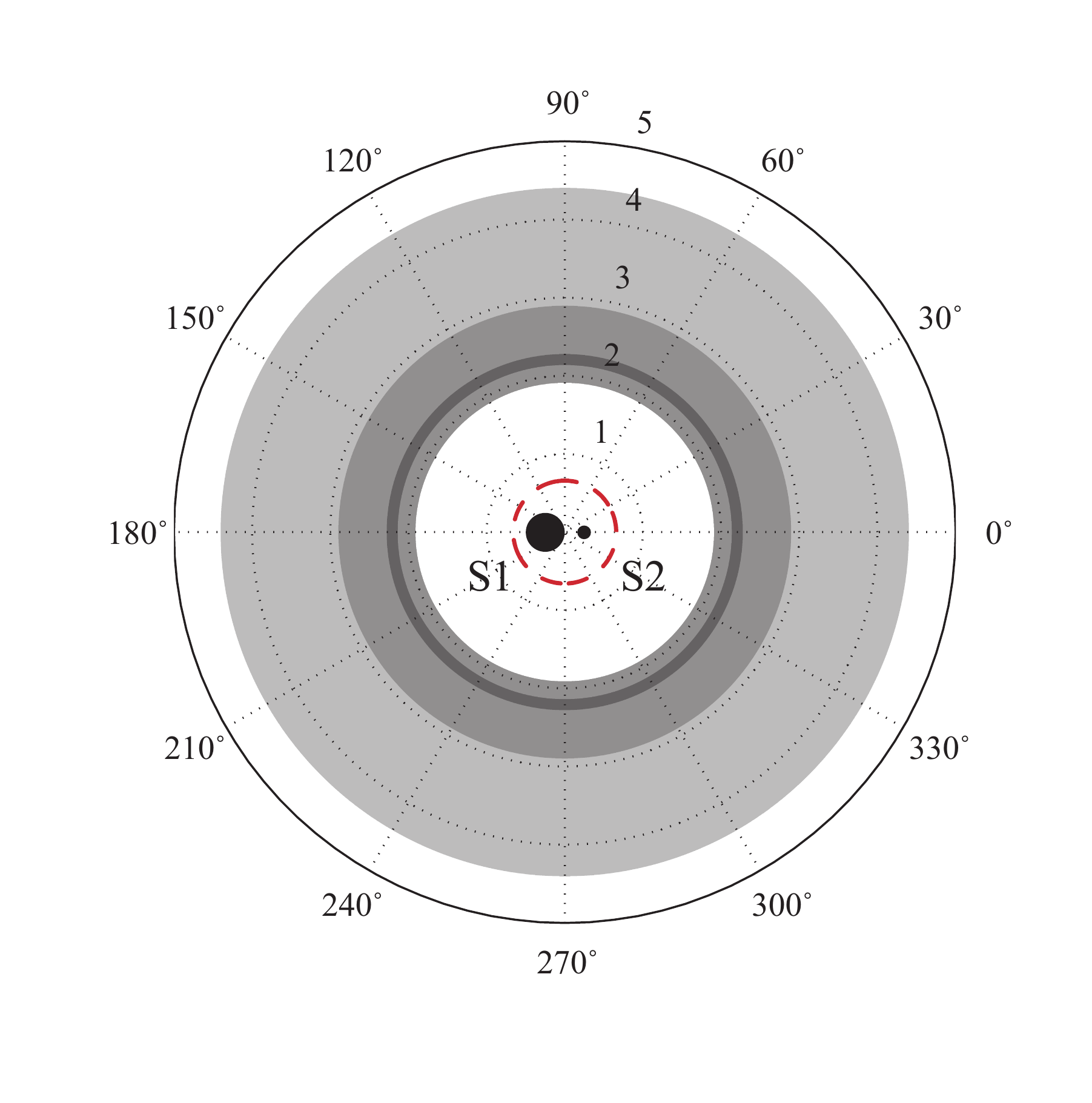} \\
\includegraphics[scale=0.30]{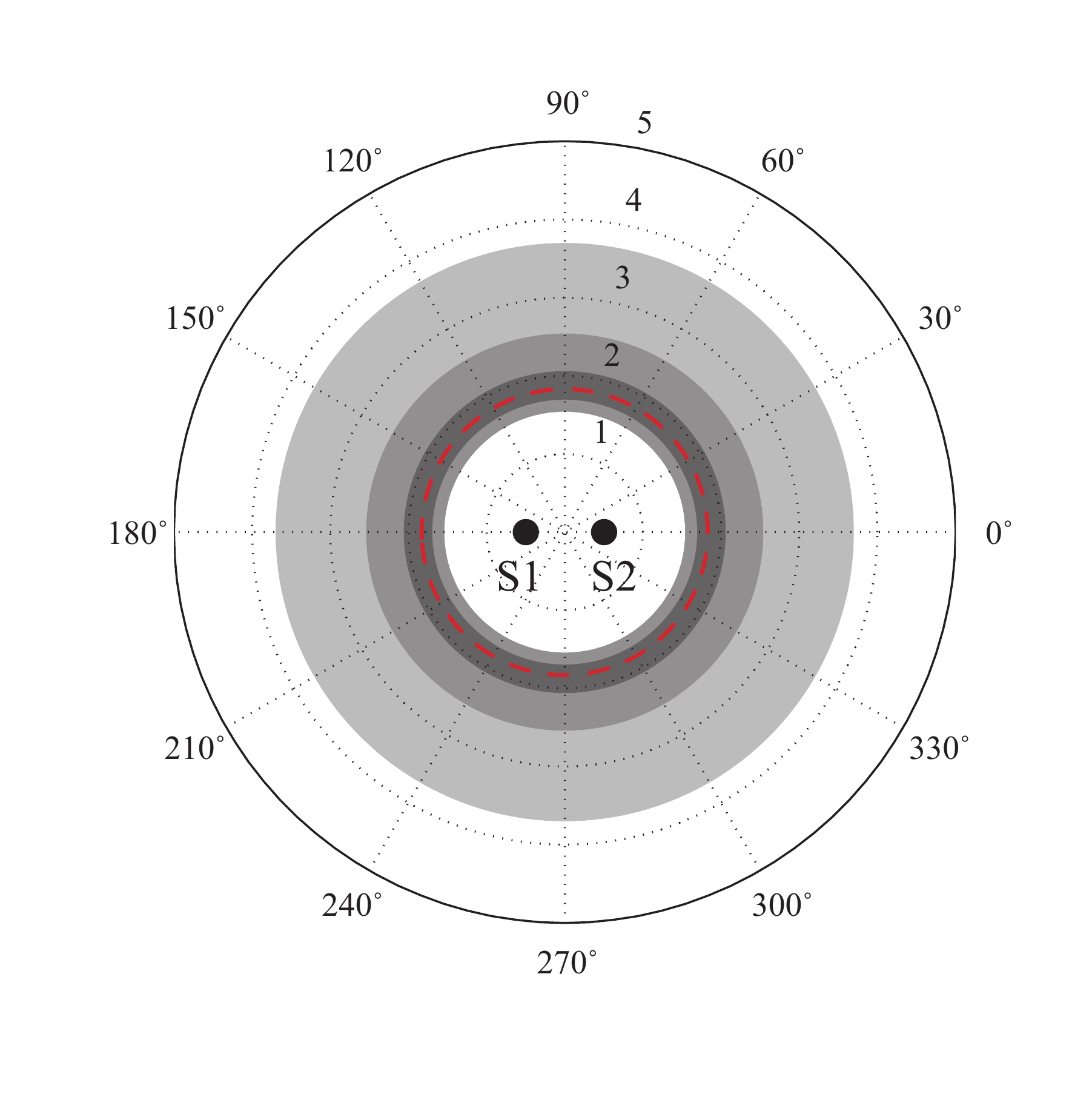} &
\includegraphics[scale=0.30]{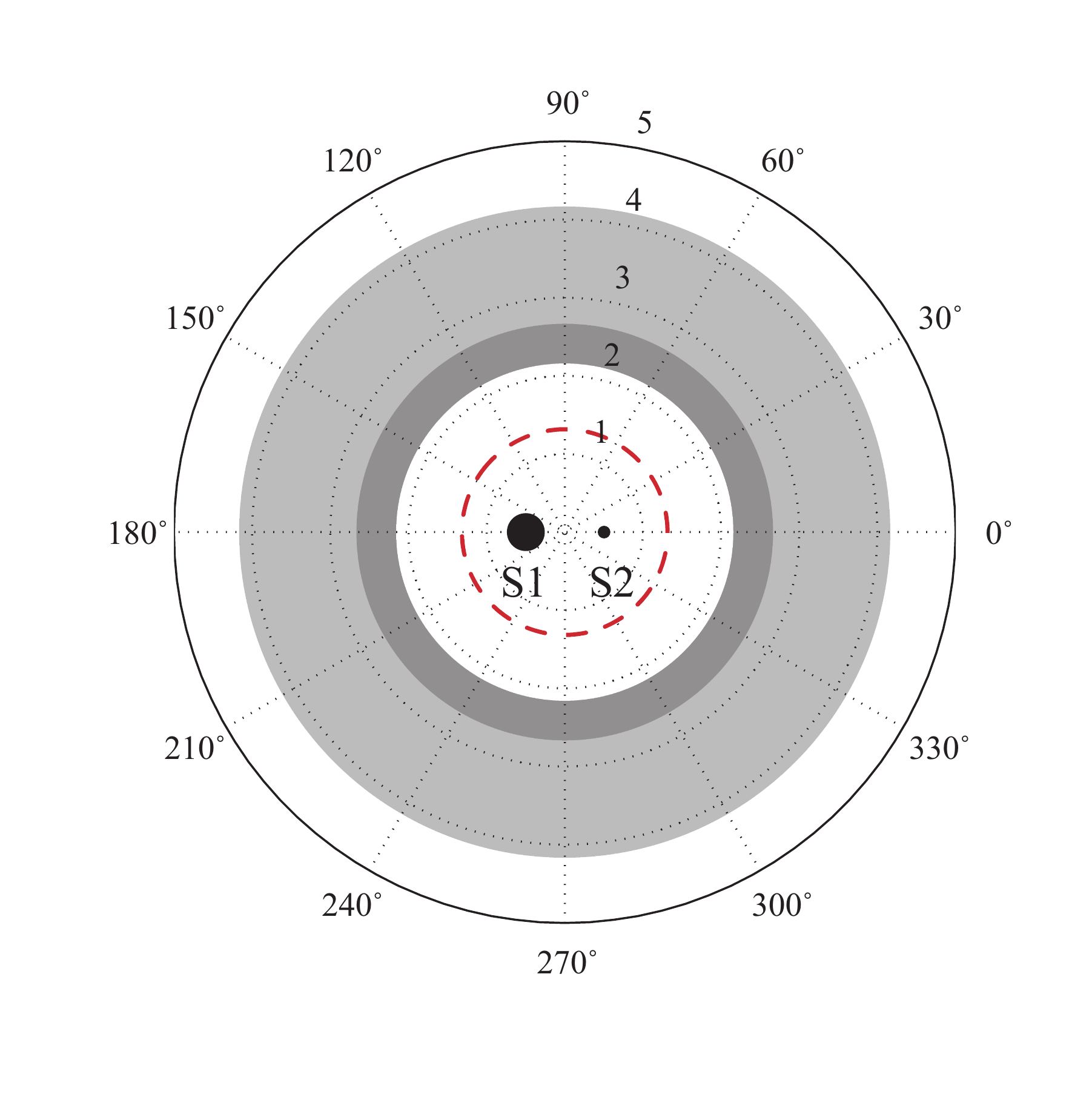} \\
\includegraphics[scale=0.30]{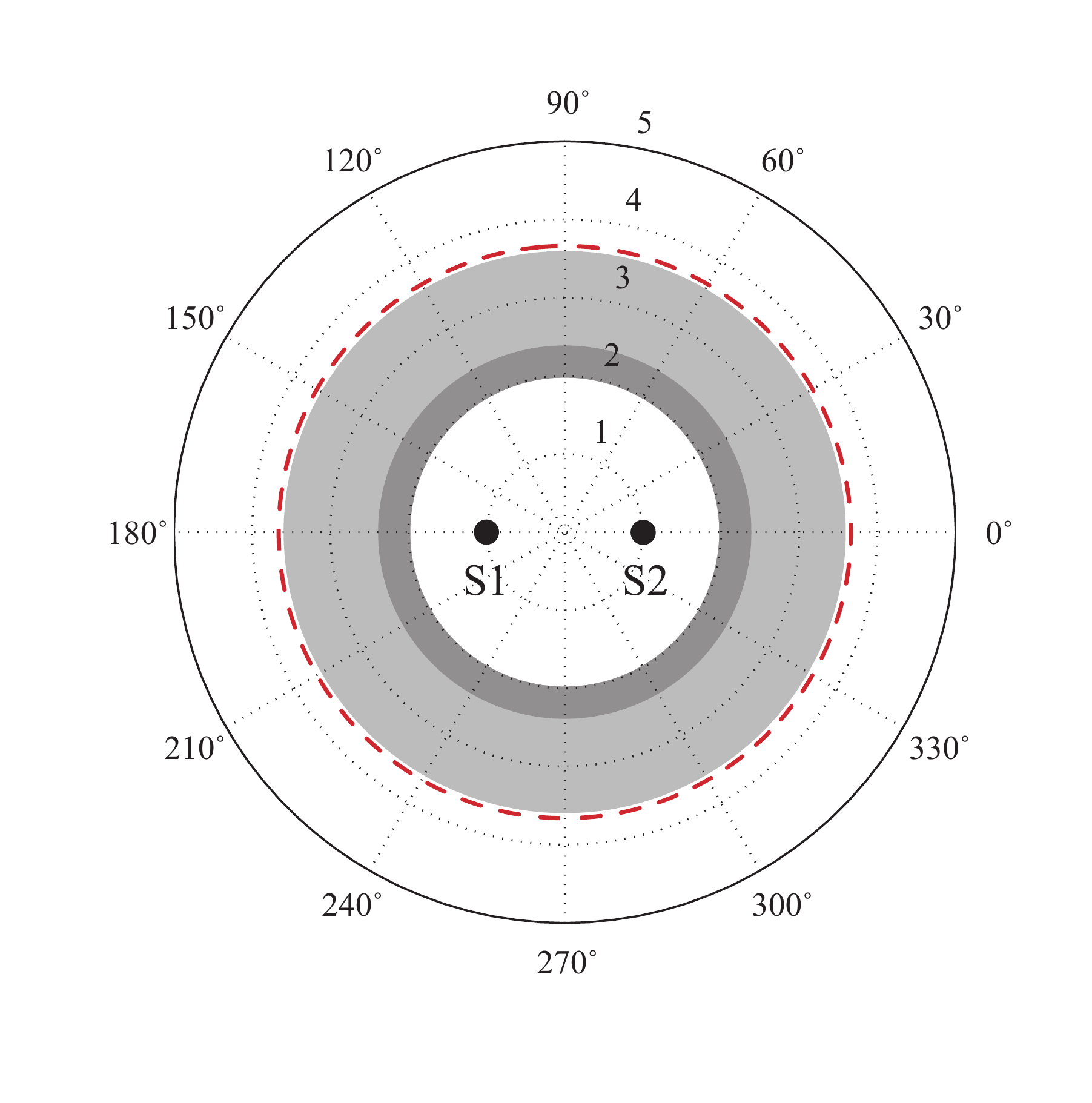} &
\includegraphics[scale=0.30]{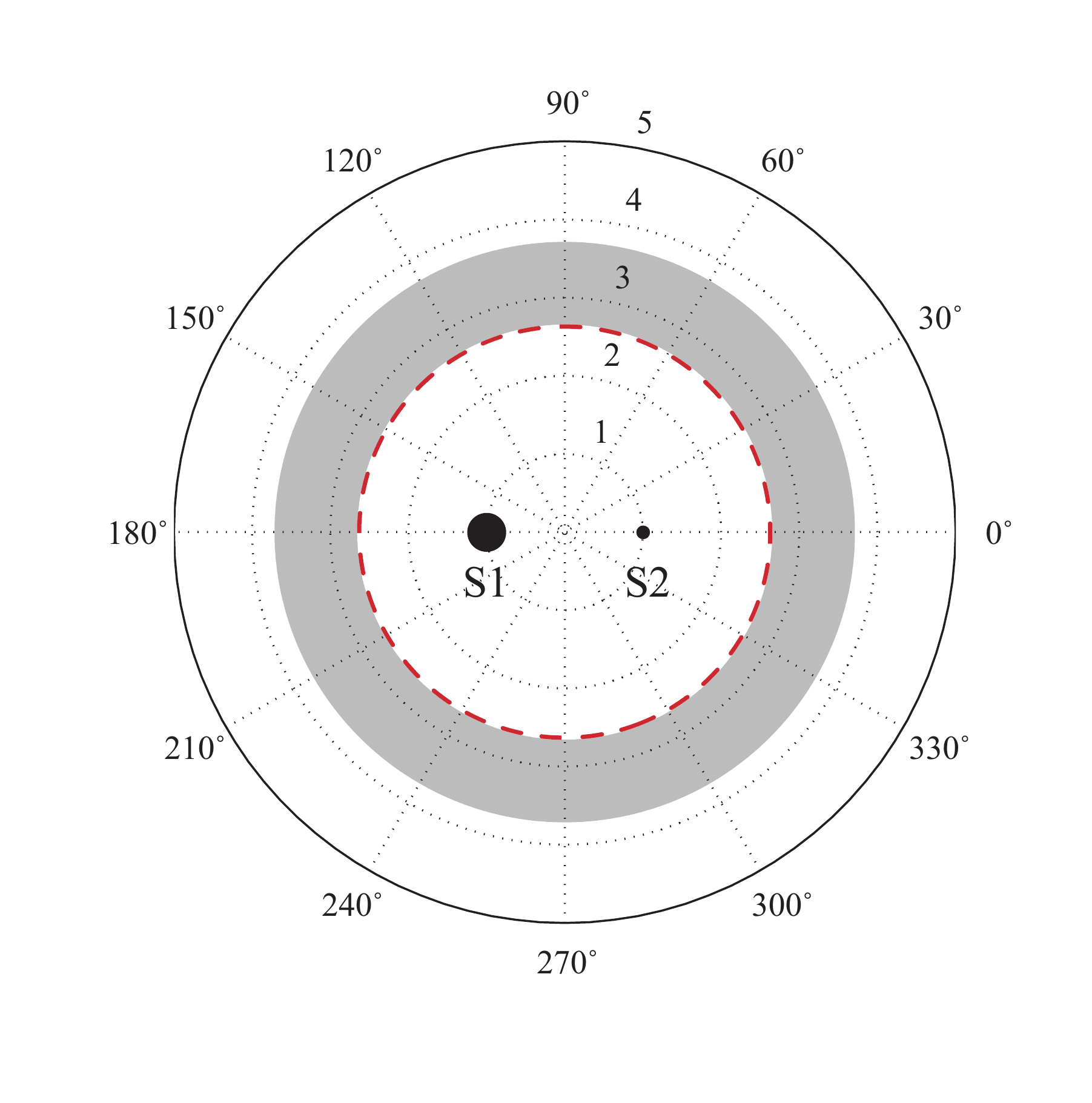} \\

\end{tabular}
\caption{
Domains of P-type habitability for main-sequence star binary systems.
The solutions are given in polar coordinates with the radial coordinate depicted in units of AU.
The left column depicts systems with masses of $M_1 = M_2 = 1.0~M_\odot$, whereas
the right column depicts systems with masses of $M_1 = 1.5~M_\odot$ and $M_2 = 0.5~M_\odot$.
The separation distances $2a$ are given as 0.5~AU (top), 1.0~AU (middle), and 2.0~AU (bottom).
Results for the conservative, general and extended RHZs are given as dark gray, medium gray,
and light gray areas, respectively.  The orbital stability limits are indicated by
red dashed lines.
}
\end{figure*}


\clearpage


\begin{figure*} 
\centering
\begin{tabular}{cc}

\includegraphics[scale=0.40]{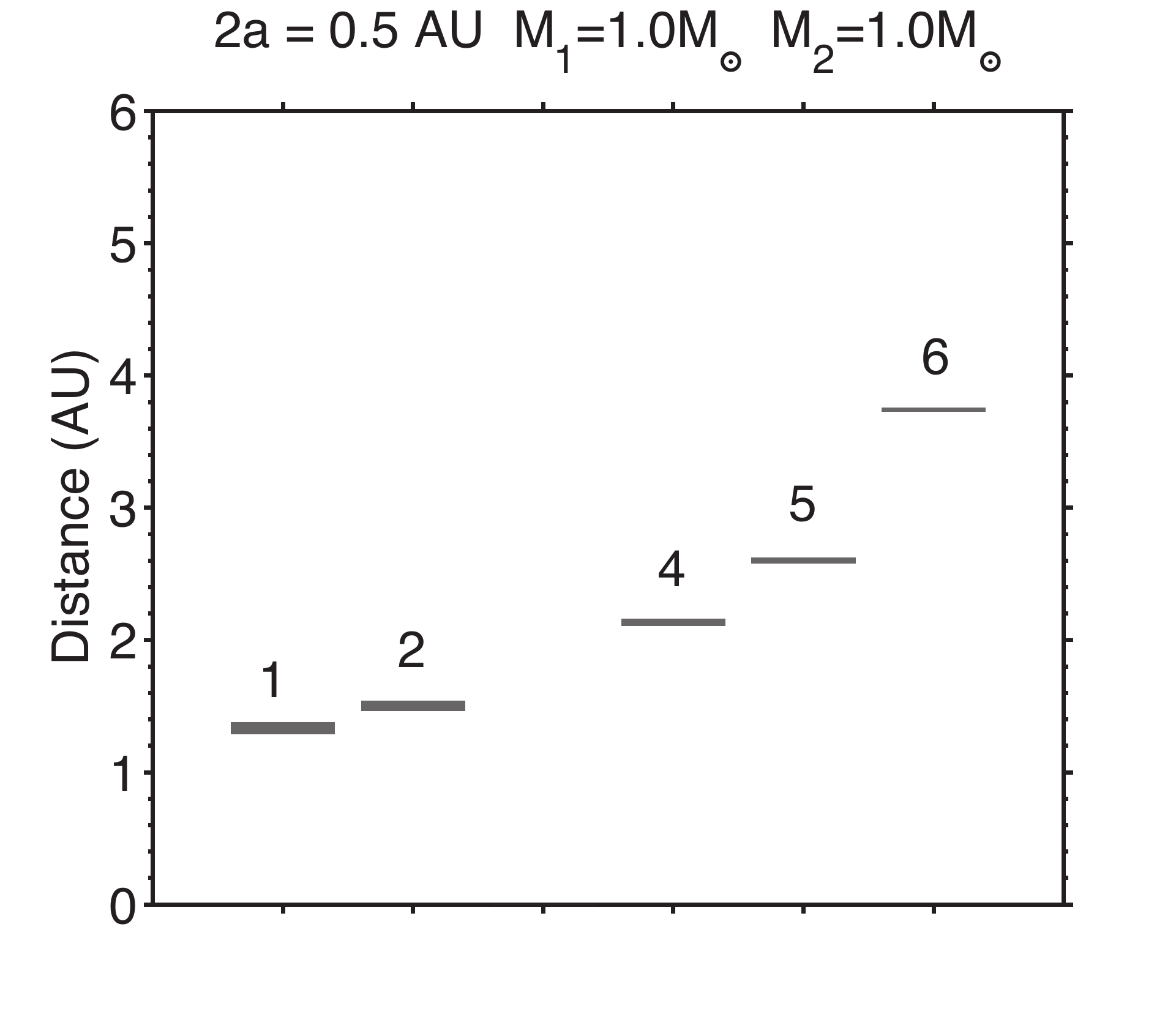} &
\includegraphics[scale=0.40]{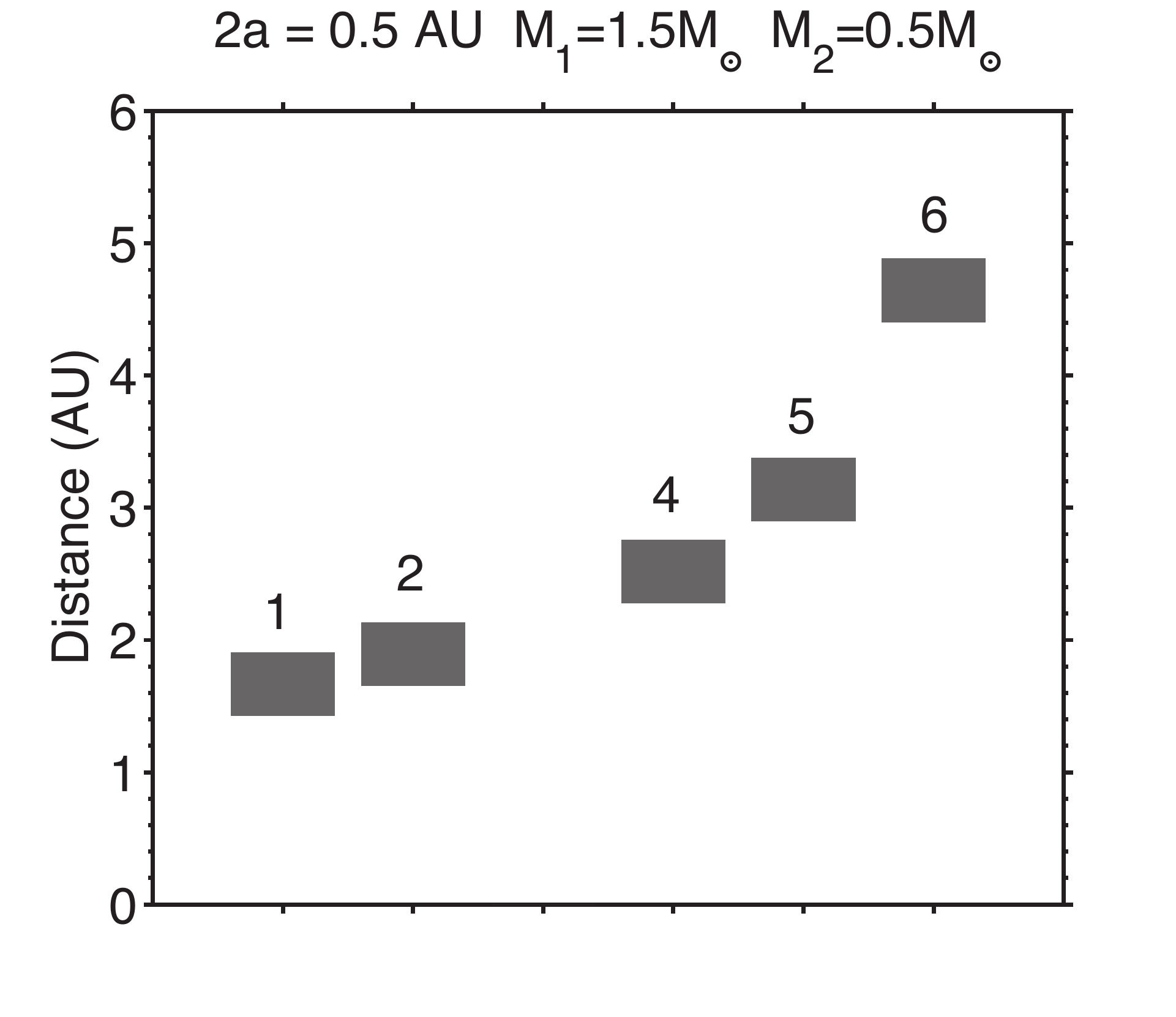} \\
\includegraphics[scale=0.40]{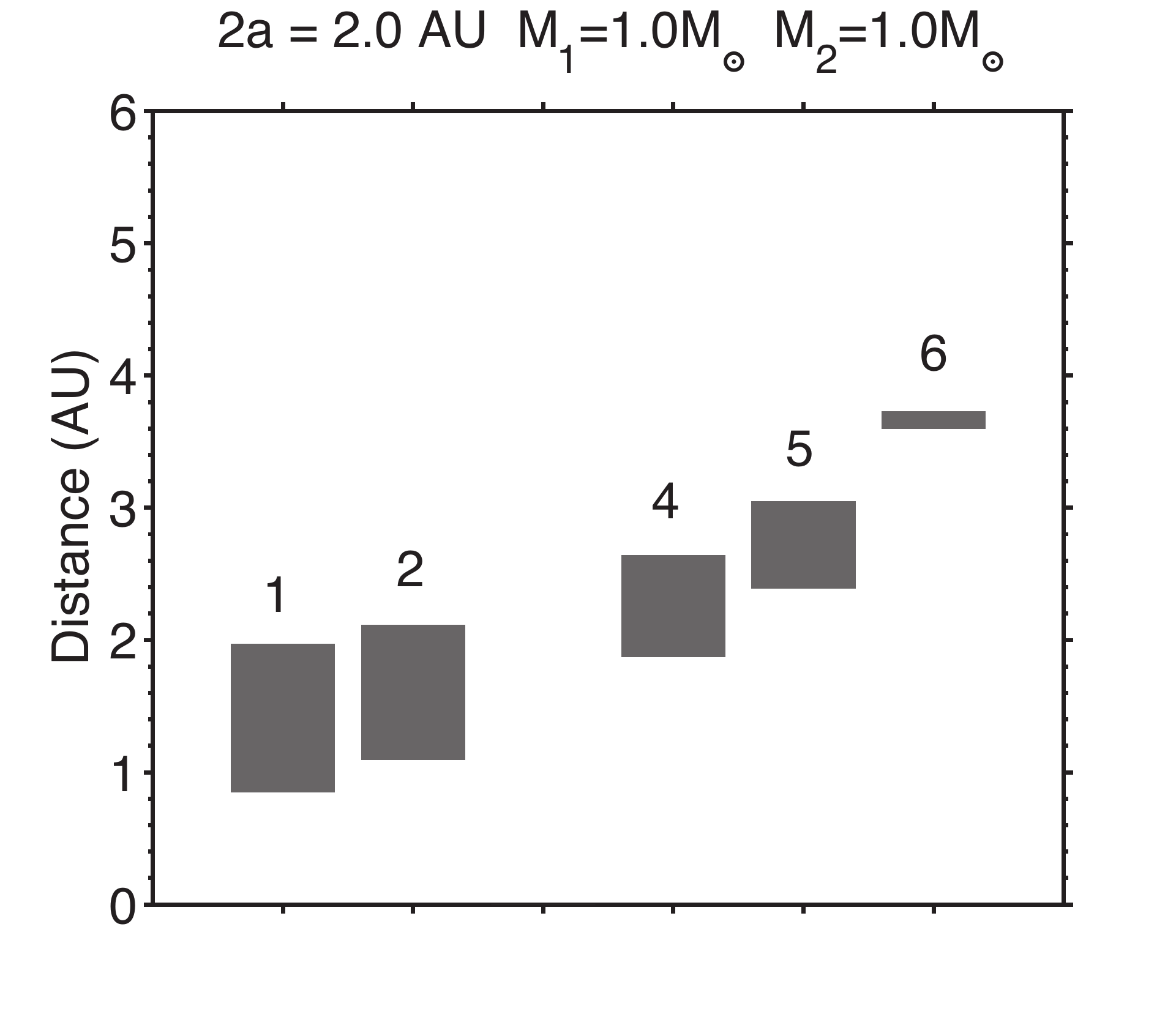} &
\includegraphics[scale=0.40]{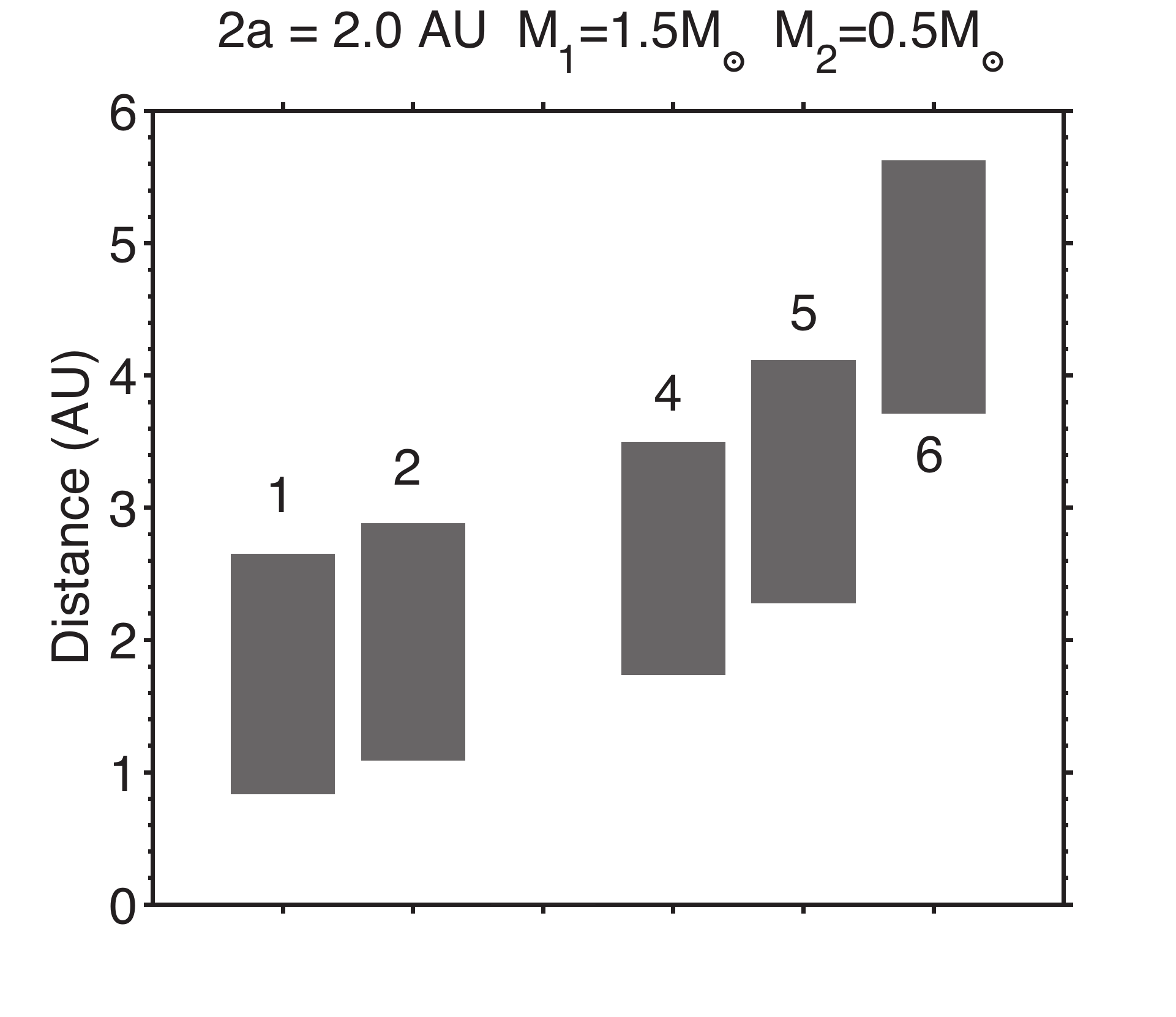} \\

\end{tabular}
\caption{Ranges of the inner and outer limits of RHZs
also referred to as ${\rm RHZ}_{\rm in}$ and ${\rm RHZ}_{\rm out}$,
respectively, pertaining to $s_\ell = 0.84$, 0.95, 1.37, 1.67,
and 2.40~AU (from left to right, with labels 1, 2, 4, 5, and 6).
The figure panels depict binary systems with separation distances
$2a$ of 0.5~AU (top) and 2.0~AU (bottom) and with masses of
$M_1 = M_2 = 1.0~M_\odot$ (left) and $M_1 = 1.5~M_\odot$ and
$M_2 = 0.5~M_\odot$ (right); see also Tables 6 to 8 for
additional information.
}
\end{figure*}


\clearpage


\begin{figure*} 
\centering
\begin{tabular}{c}
\includegraphics[scale=0.45]{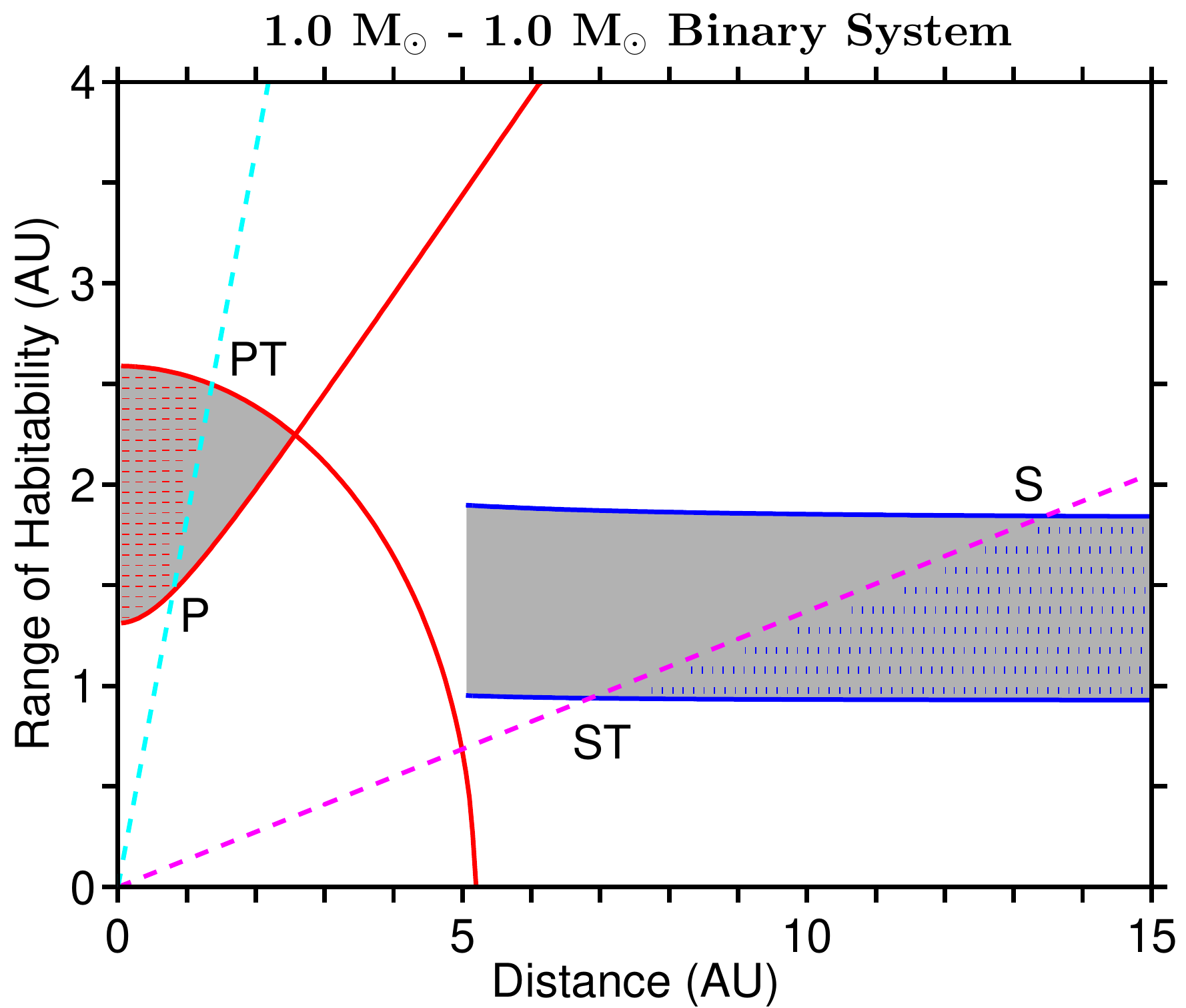} \\
\includegraphics[scale=0.45]{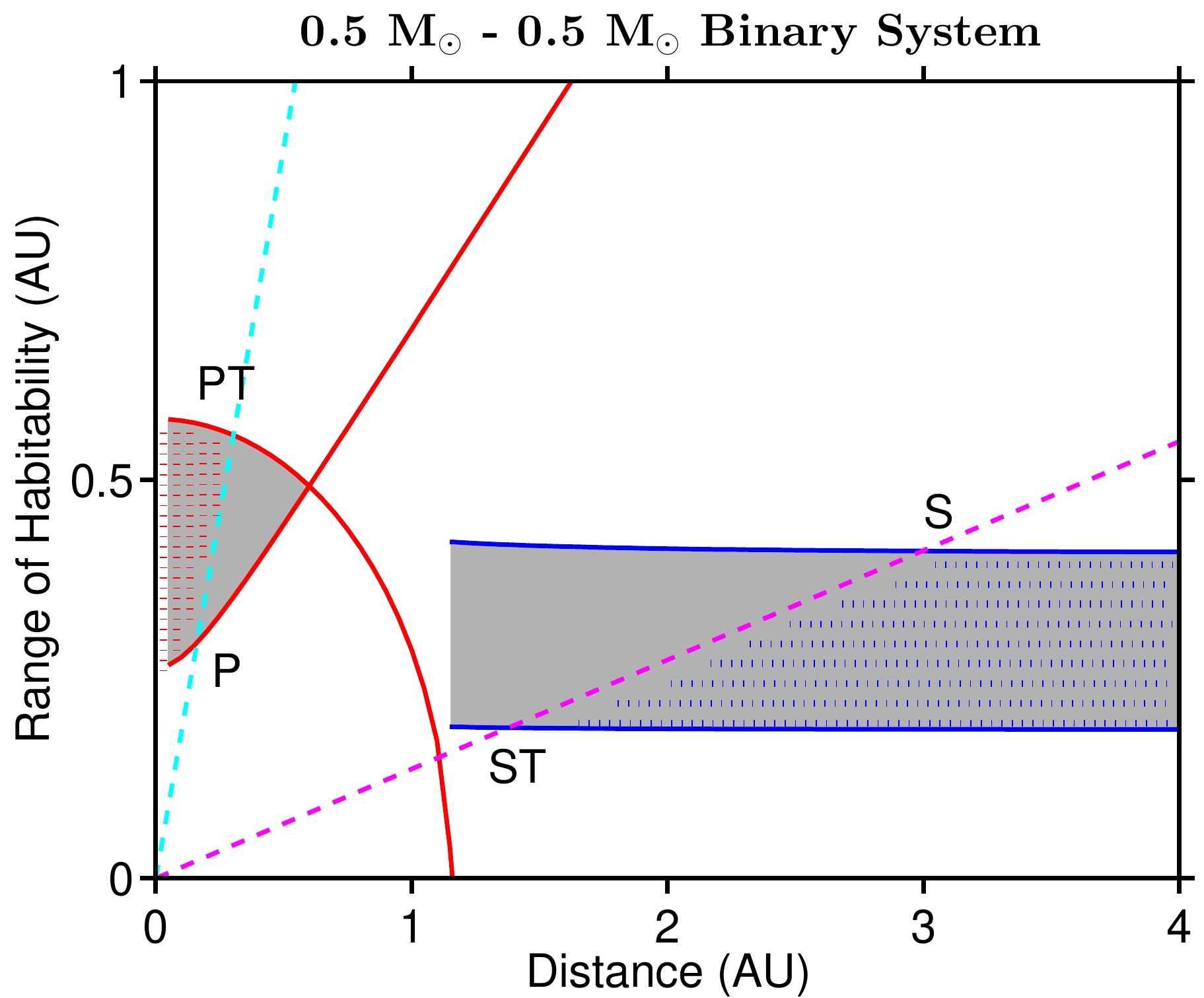} \\

\end{tabular}
\caption{Range of habitability in equal-star binary systems with the
stellar components given as 1.0 and $0.5~M_\odot$, respectively.
Results are obtained as function of the binary separation distance
$2a$ pertaining to the GHZ.  The two red lines indicate the
limits of the P-type RHZ (i.e., RHZ$_{\rm in}$ and RHZ$_{\rm out}$;
see Eqs. (44a) and (44b)), whereas the two blue lines indicate the
limits of the S-type RHZ (see Eqs. (41a) and (41b)).  The available
S-type and P-type RHZs are depicted as grayish areas.  The cyan
dashed line indicates the P-type orbital stability limit, whereas
the violet dashed line indicates the S-type orbital stability limit;
see Eqs. (47) and (46), respectively.  Note that the P-type orbital
stability limit constitutes a lower limit, whereas the S-type orbital
stability limit constitutes an upper limit; thus, the available
ranges of habitability within the RHZs are indicated as red-hatched
and blue-hatched areas, respectively.  Hence, P-type habitability is
attained in the range beneath the P intersection point, PT-type habitability
between the intersection points P and PT, ST-type habitability between
the intersection points ST and S, and S-type habitability beyond the
S intersection point.  No habitability is found between the intersection
points PT and ST.
}
\end{figure*}

\clearpage


\begin{figure*} 
\centering
\begin{tabular}{c}

\includegraphics[scale=0.55]{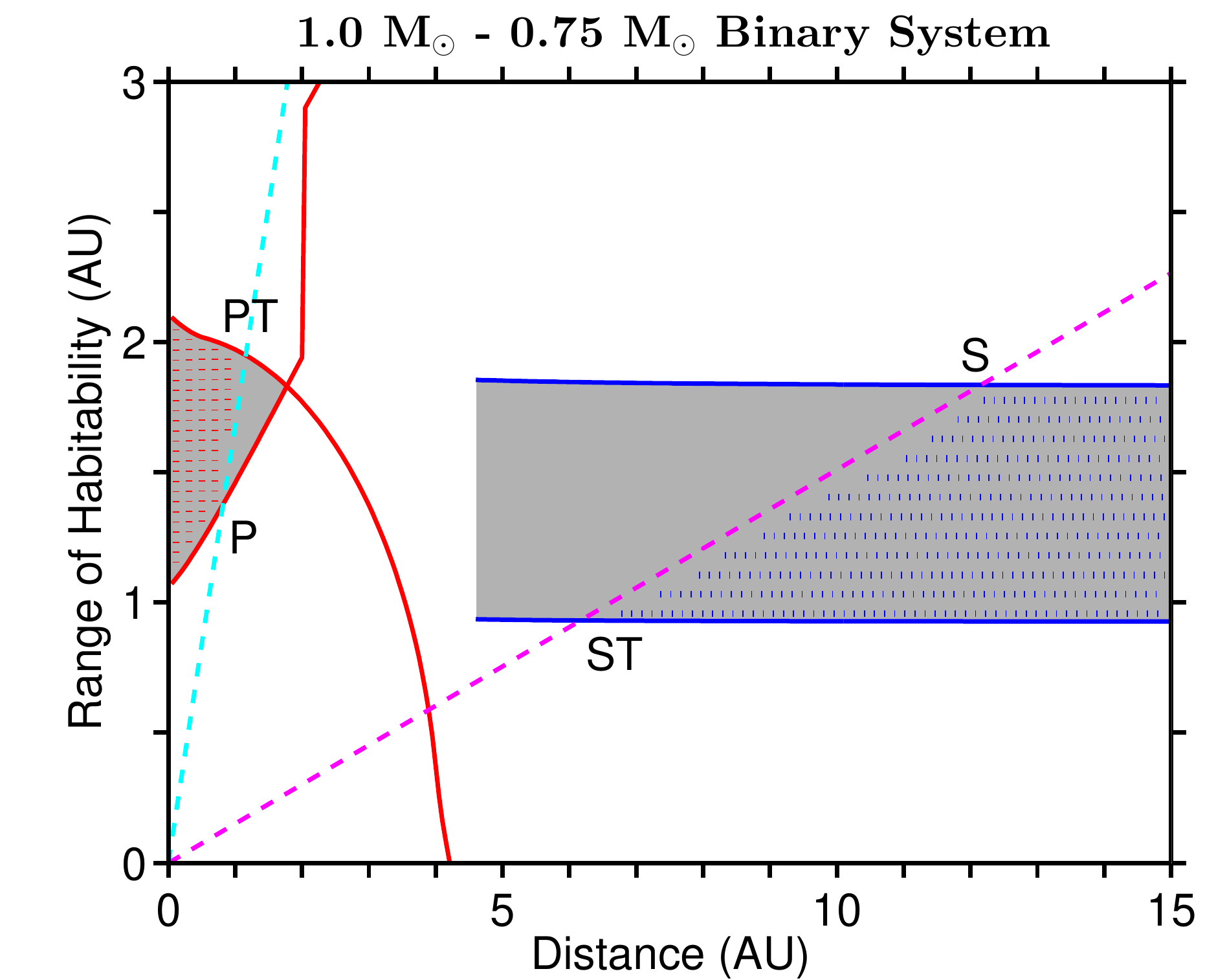} \\
\includegraphics[scale=0.55]{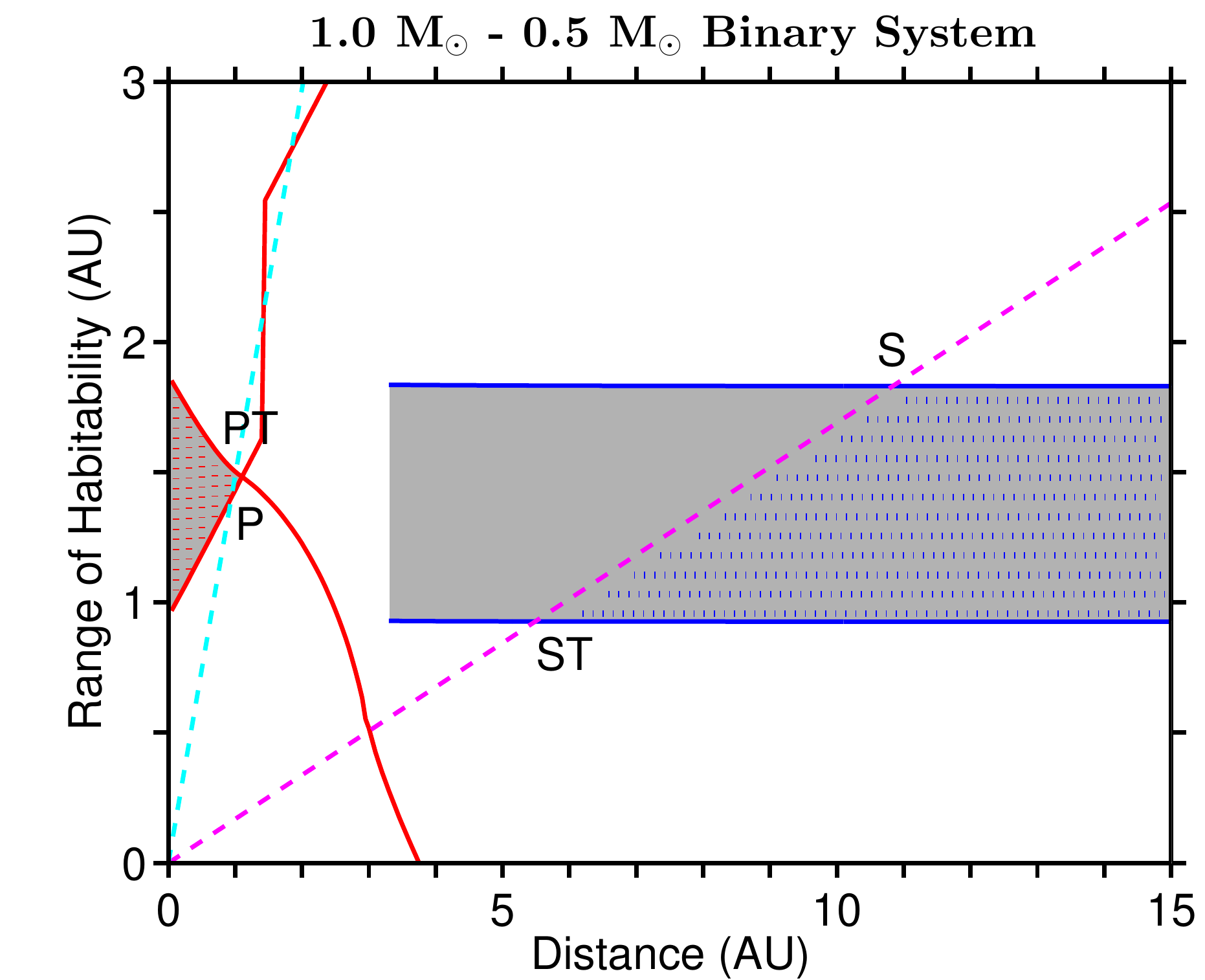} \\

\end{tabular}
\caption{Same as Fig.~8, but now for nonequal-star binary systems.
The primary is chosen as $1.0~M_\odot$, whereas the secondary is
chosen as 0.75 and $0.5~M_\odot$, respectively.  Note that the
algebraic solution for P-type ${\rm RHZ}_{\rm in}$ outside the scope
of relevance may be ill defined.
}
\end{figure*}

\clearpage


\begin{figure*} 
\centering
\begin{tabular}{c}

\includegraphics[scale=0.65]{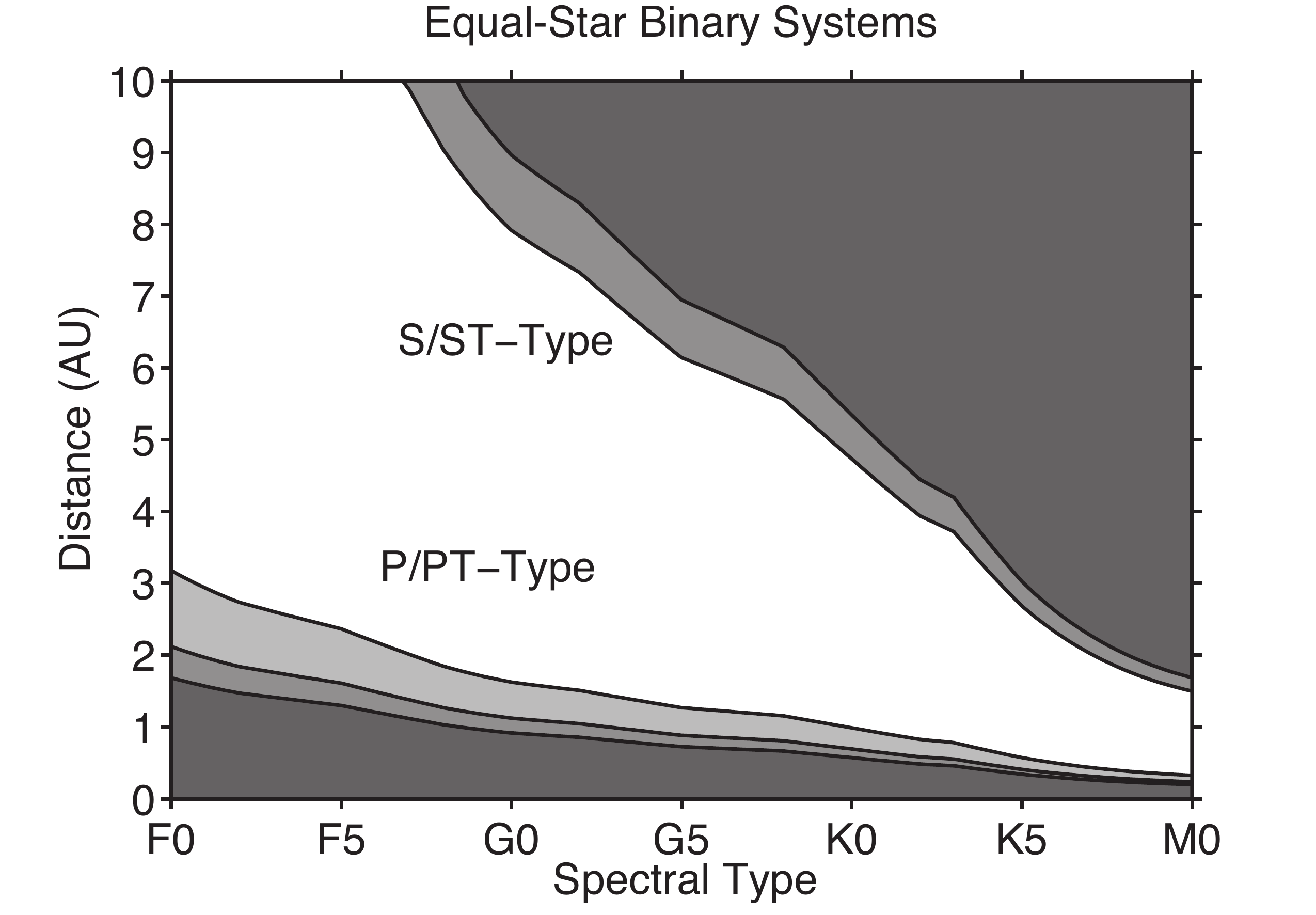} \\
\end{tabular}
\caption{Depiction of S/ST-type and P/PT-type habitability based on
the joint constraint of planetary orbital stability and the availability
of a habitable region provided by the stellar radiative energy fluxes.
The domains of the CHZs, GHZs and EHZs are depicted as dark gray, medium gray,
and light gray areas, respectively.  The results refer to the particular case
of equal-mass binaries, i.e., main-sequence stars of identical spectral types.
Note that no separate S/ST-type contour is attained for the EHZ, owing to the
fact that the inner boundary of the EHZ (i.e., $s_1$; see Table~2) agrees with
the inner boundary of the GHZ, thus rendering the same mathematical criterion
for the inner limit of habitability.  For the F0~V, F2~V, and F5~V stars, the
S/ST-type contours extend beyond the figure frame.  For the GHZ and EHZ, they
are given as 15.9, 13.6, and 11.7~AU, and for the CHZ, they are given as
18.1, 15.5, and 13.3~AU, respectively.
}
\end{figure*}

\clearpage


\begin{deluxetable}{lccccc}
\tablecaption{Stellar Parameters}
\tablewidth{0pt}
\tablehead{
Sp. Type & $T_{\rm eff}$ & $R_\ast$    & $L_\ast$    & $S_{{\rm rel},\ast}$ & $M_\ast$ \\
       ... &  (K)        & ($R_\odot$) & ($L_\odot$) & ...                  & ($M_\odot$)
}
\startdata
    F0  &  7178  &  1.62  &  6.255  &  1.145  &  1.60 \\
    F2  &  6909  &  1.48  &  4.481  &  1.113  &  1.52 \\
    F5  &  6528  &  1.40  &  3.196  &  1.072  &  1.40 \\
    F8  &  6160  &  1.20  &  1.862  &  1.037  &  1.19 \\
    G0  &  5943  &  1.12  &  1.405  &  1.019  &  1.05 \\
    G2  &  5811  &  1.08  &  1.194  &  1.009  &  0.99 \\
    G5  &  5657  &  0.95  &  0.830  &  0.997  &  0.91 \\
    G8  &  5486  &  0.91  &  0.673  &  0.985  &  0.84 \\
    K0  &  5282  &  0.83  &  0.481  &  0.971  &  0.79 \\
    K2  &  5055  &  0.75  &  0.330  &  0.957  &  0.74 \\
    K5  &  4487  &  0.64  &  0.149  &  0.926  &  0.67 \\
    K8  &  4006  &  0.53  &  0.066  &  0.905  &  0.58 \\
    M0  &  3850  &  0.48  &  0.045  &  0.900  &  0.51 \\
\enddata
\tablecomments{
$S_{{\rm rel},\ast}$ is calculated for $s_\ell = s_3$.
}
\end{deluxetable}

\clearpage


\begin{deluxetable}{lccc}
\tablecaption{Habitability Limits for the Sun}
\tablewidth{0pt}
\tablehead{
$\ell$ & $s_\ell$ & HZ Limit & Description  \\
...    & (AU)     & ...      & ...
}
\startdata
 1 &  0.84   &  GHZ / EHZ & Runaway greenhouse effect               \\
 2 &  0.95   &  CHZ       & Start of water loss                     \\
 3 &  1.00   &  ...       & Earth-equivalent position               \\
 4 &  1.37   &  CHZ       & First CO$_2$ condensation               \\
 5 &  1.67   &  GHZ       & Maximum greenhouse effect, no clouds    \\
 6 &  2.40   &  EHZ       & Maximum greenhouse effect, 100\% clouds \\
\enddata
\tablecomments{See text for references.}
\end{deluxetable}

\clearpage


\begin{deluxetable}{lcccccc}
\tablecaption{Habitable Zones of Single Main-Sequence Stars}
\tablewidth{0pt}
\tablehead{
\colhead{Sp. Type} & \multicolumn{6}{c}{Habitable Zone} \\
\noalign{\smallskip}
\hline
\noalign{\smallskip}
... & HZ$(s_1)$ & HZ$(s_2)$ & HZ$(s_3)$ & HZ$(s_4)$ & HZ$(s_5)$ & HZ$(s_6)$ \\
... & (AU)      & (AU)      & (AU)      & (AU)      & (AU)      & (AU)
}
\startdata
  F0  &  1.98  &  2.25  &  2.33  &  2.91  &  3.66  &  5.49  \\
  F2  &  1.70  &  1.93  &  2.00  &  2.54  &  3.18  &  4.72  \\
  F5  &  1.46  &  1.65  &  1.72  &  2.24  &  2.78  &  4.08  \\
  F8  &  1.13  &  1.28  &  1.34  &  1.78  &  2.19  &  3.19  \\
  G0  &  0.99  &  1.12  &  1.17  &  1.58  &  1.94  &  2.80  \\
  G2  &  0.91  &  1.03  &  1.09  &  1.48  &  1.81  &  2.61  \\
  G5  &  0.77  &  0.87  &  0.91  &  1.25  &  1.53  &  2.19  \\
  G8  &  0.69  &  0.78  &  0.83  &  1.15  &  1.39  &  1.99  \\
  K0  &  0.59  &  0.67  &  0.71  &  0.99  &  1.20  &  1.71  \\
  K2  &  0.49  &  0.55  &  0.59  &  0.84  &  1.01  &  1.43  \\
  K5  &  0.34  &  0.38  &  0.40  &  0.59  &  0.71  &  0.99  \\
  K8  &  0.22  &  0.25  &  0.27  &  0.41  &  0.49  &  0.67  \\
  M0  &  0.19  &  0.21  &  0.23  &  0.35  &  0.41  &  0.56  \\
\enddata
\end{deluxetable}

\clearpage


\begin{deluxetable}{lccccccc}
\tablecaption{Recast Stellar Luminosity}
\tablewidth{0pt}
\tablehead{
Sp. Type & $T_{\rm eff}$ & $L'_{i1}$   & $L'_{i2}$   & $L'_{i3}$   & $L'_{i4}$   & $L'_{i5}$   & $L'_{i6}$  \\
     ... &  (K)          & ($L_\odot$) & ($L_\odot$) & ($L_\odot$) & ($L_\odot$) & ($L_\odot$) & ($L_\odot$)
}
\startdata
    F0  &  7178  &  5.545 &  5.625 &  5.464 &  4.509 &  4.801 &  5.223  \\
    F2  &  6909  &  4.075 &  4.121 &  4.027 &  3.445 &  3.621 &  3.873  \\
    F5  &  6528  &  3.005 &  3.027 &  2.981 &  2.681 &  2.770 &  2.897  \\
    F8  &  6160  &  1.802 &  1.809 &  1.794 &  1.692 &  1.722 &  1.764  \\
    G0  &  5943  &  1.382 &  1.384 &  1.379 &  1.337 &  1.349 &  1.366  \\
    G2  &  5811  &  1.185 &  1.186 &  1.184 &  1.168 &  1.172 &  1.179  \\
    G5  &  5657  &  0.832 &  0.832 &  0.832 &  0.837 &  0.836 &  0.834  \\
    G8  &  5486  &  0.683 &  0.682 &  0.684 &  0.703 &  0.697 &  0.690  \\
    K0  &  5282  &  0.494 &  0.493 &  0.496 &  0.523 &  0.515 &  0.505  \\
    K2  &  5055  &  0.343 &  0.341 &  0.345 &  0.374 &  0.366 &  0.354  \\
    K5  &  4487  &  0.150 &  0.149 &  0.151 &  0.177 &  0.170 &  0.161  \\
    K8  &  4006  &  0.071 &  0.071 &  0.072 &  0.089 &  0.085 &  0.079  \\
    M0  &  3850  &  0.050 &  0.049 &  0.051 &  0.064 &  0.060 &  0.055  \\
\enddata
\tablecomments{
$L'_{i\ell}$ denotes general binaries with $i=1,2$ for star S1 and S2, respectively,
with $\ell = 1$ to 6. 
}
\end{deluxetable}

\clearpage


\begin{deluxetable}{lccccc}
\tablecaption{Target List, Sections 5 and 6}
\tablewidth{0pt}
\tablehead{
$M_\ast$    & Spectral Type & $T_{\rm eff}$ & $L_\ast$    & $L'_{i3}$   & HZ$(s_3)$ \\
($M_\odot$) & ...           & (K)           & ($L_\odot$) & ($L_\odot$) & (AU) 
}
\startdata
 1.50  &  F2~V &  6842  &  4.233  &  3.830  &  1.96  \\
 1.25  &  F7~V &  6256  &  2.173  &  2.078  &  1.44  \\
 1.00  &  G2~V &  5833  &  1.228  &  1.216  &  1.10  \\
 0.75  &  K2~V &  5100  &  0.357  &  0.372  &  0.61  \\
 0.50  &  M0~V &  3857  &  0.045  &  0.050  &  0.22  \\
\enddata
\tablecomments{
$L'_{i3}$ denotes general binaries with $i=1,2$ for star S1 and S2, respectively. 
}
\end{deluxetable}

\clearpage


\begin{deluxetable}{lcccc}
\tablecaption{P-Type Stellar Habitability; $2a = 0.5$~AU}
\tablewidth{0pt}
\tablehead{
\multicolumn{2}{c}{HZ} & \multicolumn{3}{c}{Binary System Data} \\
\noalign{\smallskip}
\hline
\noalign{\smallskip}
$\ell$ & $s_\ell$ & ${\rm RHZ}_{\rm in}$ & ${\rm RHZ}_{\rm out}$ & $a_{\rm cr}$ \\
...    & (AU)     & (AU)                 & (AU)                  & (AU)  
}
\startdata
\multicolumn{5}{c}{Model:~~$M_1 = 1.0~M_{\odot}$, $M_2 = 1.0~M_{\odot}$} \\
\noalign{\smallskip}
\hline
\noalign{\smallskip}
 1  &  0.84  &  1.29  &  1.38  &  0.92  \\
 2  &  0.95  &  1.46  &  1.54  &  0.92  \\
 3  &  1.00  &  1.54  &  1.62  &  0.92  \\
 4  &  1.37  &  2.10  &  2.16  &  0.92  \\
 5  &  1.67  &  2.58  &  2.62  &  0.92  \\
 6  &  2.40  &  3.72  &  3.76  &  0.92  \\
\noalign{\smallskip}
\hline
\noalign{\smallskip}
\multicolumn{5}{c}{Model:~~$M_1 = 1.25~M_{\odot}$, $M_2 = 0.75~M_{\odot}$} \\
\noalign{\smallskip}
\hline
\noalign{\smallskip}
 1  &  0.84  &  1.19  &  1.52  &  0.79  \\
 2  &  0.95  &  1.36  &  1.69  &  0.79  \\
 3  &  1.00  &  1.45  &  1.79  &  0.79  \\
 4  &  1.37  &  1.96  &  2.28  &  0.79  \\
 5  &  1.67  &  2.43  &  2.76  &  0.79  \\
 6  &  2.40  &  3.58  &  3.91  &  0.79  \\
\noalign{\smallskip}
\hline
\noalign{\smallskip}
\multicolumn{5}{c}{Model:~~$M_1 = 1.5~M_{\odot}$, $M_2 = 0.5~M_{\odot}$} \\
\noalign{\smallskip}
\hline
\noalign{\smallskip}
 1  &  0.84  &  1.42  &  1.91  &  0.66  \\
 2  &  0.95  &  1.65  &  2.14  &  0.66  \\
 3  &  1.00  &  1.73  &  2.22  &  0.66  \\
 4  &  1.37  &  2.28  &  2.76  &  0.66  \\
 5  &  1.67  &  2.90  &  3.38  &  0.66  \\
 6  &  2.40  &  4.40  &  4.89  &  0.66  \\
\enddata
\end{deluxetable}

\clearpage


\begin{deluxetable}{lcccc}
\tablecaption{P-Type Stellar Habitability; $2a = 1.0$~AU}
\tablewidth{0pt}
\tablehead{
\multicolumn{2}{c}{HZ} & \multicolumn{3}{c}{Binary System Data} \\
\noalign{\smallskip}
\hline
\noalign{\smallskip}
$\ell$ & $s_\ell$ & ${\rm RHZ}_{\rm in}$ & ${\rm RHZ}_{\rm out}$ & $a_{\rm cr}$ \\
...    & (AU)     & (AU)                 & (AU)                  & (AU)  
}
\startdata
\multicolumn{5}{c}{Model:~~$M_1 = 1.0~M_{\odot}$, $M_2 = 1.0~M_{\odot}$} \\
\noalign{\smallskip}
\hline
\noalign{\smallskip}
 1  &  0.84  &  1.21  &  1.54  &  1.83  \\
 2  &  0.95  &  1.40  &  1.69  &  1.83  \\
 3  &  1.00  &  1.48  &  1.76  &  1.83  \\
 4  &  1.37  &  2.06  &  2.28  &  1.83  \\
 5  &  1.67  &  2.54  &  2.72  &  1.83  \\
 6  &  2.40  &  3.70  &  3.83  &  1.83  \\
\noalign{\smallskip}
\hline
\noalign{\smallskip}
\multicolumn{5}{c}{Model:~~$M_1 = 1.25~M_{\odot}$, $M_2 = 0.75~M_{\odot}$} \\
\noalign{\smallskip}
\hline
\noalign{\smallskip}
 1  &  0.84  &  1.11  &  1.75  &  1.57  \\
 2  &  0.95  &  1.29  &  1.92  &  1.57  \\
 3  &  1.00  &  1.38  &  2.02  &  1.57  \\
 4  &  1.37  &  1.90  &  2.49  &  1.57  \\
 5  &  1.67  &  2.35  &  2.96  &  1.57  \\
 6  &  2.40  &  3.46  &  4.12  &  1.57  \\
\noalign{\smallskip}
\hline
\noalign{\smallskip}
\multicolumn{5}{c}{Model:~~$M_1 = 1.5~M_{\odot}$, $M_2 = 0.5~M_{\odot}$} \\
\noalign{\smallskip}
\hline
\noalign{\smallskip}
 1  &  0.84  &  1.21  &  2.16  &  1.32  \\
 2  &  0.95  &  1.43  &  2.38  &  1.32  \\
 3  &  1.00  &  1.51  &  2.47  &  1.32  \\
 4  &  1.37  &  2.05  &  3.00  &  1.32  \\
 5  &  1.67  &  2.66  &  3.62  &  1.32  \\
 6  &  2.40  &  4.17  &  5.13  &  1.32  \\
\enddata
\end{deluxetable}

\clearpage


\begin{deluxetable}{lcccc}
\tablecaption{P-Type Stellar Habitability; $2a = 2.0$~AU}
\tablewidth{0pt}
\tablehead{
\multicolumn{2}{c}{HZ} & \multicolumn{3}{c}{Binary System Data} \\
\noalign{\smallskip}
\hline
\noalign{\smallskip}
$\ell$ & $s_\ell$ & ${\rm RHZ}_{\rm in}$ & ${\rm RHZ}_{\rm out}$ & $a_{\rm cr}$ \\
...    & (AU)     & (AU)                 & (AU)                  & (AU)  
}
\startdata
\multicolumn{5}{c}{Model:~~$M_1 = 1.0~M_{\odot}$, $M_2 = 1.0~M_{\odot}$} \\
\noalign{\smallskip}
\hline
\noalign{\smallskip}
 1  &  0.84  &  0.85  &  1.98  &  3.66  \\
 2  &  0.95  &  1.10  &  2.11  &  3.66  \\
 3  &  1.00  &  1.20  &  2.18  &  3.66  \\
 4  &  1.37  &  1.87  &  2.64  &  3.66  \\
 5  &  1.67  &  2.39  &  3.05  &  3.66  \\
 6  &  2.40  &  3.60  &  4.09  &  3.66  \\
\noalign{\smallskip}
\hline
\noalign{\smallskip}
\multicolumn{5}{c}{Model:~~$M_1 = 1.25~M_{\odot}$, $M_2 = 0.75~M_{\odot}$} \\
\noalign{\smallskip}
\hline
\noalign{\smallskip}
 1  &  0.84  &  0.70  &  2.23  &  3.15  \\
 2  &  0.95  &  0.95  &  2.40  &  3.15  \\
 3  &  1.00  &  1.07  &  2.50  &  3.15  \\
 4  &  1.37  &  1.69  &  2.95  &  3.15  \\
 5  &  1.67  &  2.18  &  3.42  &  3.15  \\
 6  &  2.40  &  3.32  &  4.55  &  3.15  \\
\noalign{\smallskip}
\hline
\noalign{\smallskip}
\multicolumn{5}{c}{Model:~~$M_1 = 1.5~M_{\odot}$, $M_2 = 0.5~M_{\odot}$} \\
\noalign{\smallskip}
\hline
\noalign{\smallskip}
 1  &  0.84  &  0.83  &  2.66  &  2.63  \\
 2  &  0.95  &  1.09  &  2.88  &  2.63  \\
 3  &  1.00  &  1.18  &  2.96  &  2.63  \\
 4  &  1.37  &  1.74  &  3.50  &  2.63  \\
 5  &  1.67  &  2.28  &  4.12  &  2.63  \\
 6  &  2.40  &  3.71  &  5.63  &  2.63  \\
\enddata
\end{deluxetable}

\clearpage


\begin{deluxetable}{lccccc}
\tablecaption{Maximum Binary Separation Distance $2a$ Permitting P-type RHZs}
\tablewidth{0pt}
\tablehead{
Primary & \multicolumn{5}{c} {Secondary} \\
...               & $M_2 = 1.5~M_\odot$  & $1.25~M_\odot$ & $1.0~M_\odot$  & $0.75~M_\odot$ & $0.5~M_\odot$ \\
\noalign{\smallskip}
\hline
\noalign{\smallskip}
$M_1$        & \multicolumn{5}{c} {Maximum Binary Separation Distance} \\
$(M_\odot)$  & (AU) & (AU) & (AU) & (AU) & (AU)  
}
\startdata
\multicolumn{6}{c}{ CHZ given as $(s_2,s_4)$ } \\
\noalign{\smallskip}
\hline
\noalign{\smallskip}
  1.5   &  2.47 &  1.90  &  1.50  &  0.87  &  0.65  \\
  1.25  &  ...  &  2.01  &  1.61  &  0.96  &  0.58  \\
  1.0   &  ...  &  ...   &  1.64  &  1.01  &  0.53  \\
  0.75  &  ...  &  ...   &  ...   &  1.00  &  0.47  \\
  0.5   &  ...  &  ...   &  ...   &  ...   &  0.43  \\
\noalign{\smallskip}
\hline
\noalign{\smallskip}
\multicolumn{6}{c}{ GHZ given as $(s_1,s_5)$ } \\
\noalign{\smallskip}
\hline
\noalign{\smallskip}
  1.5   &  4.25 &  3.48  &  2.95  &  2.14  &  1.55  \\
  1.25  &  ...  &  3.26  &  2.74  &  1.93  &  1.27  \\
  1.0   &  ...  &  ...   &  2.57  &  1.77  &  1.10  \\
  0.75  &  ...  &  ...   &  ...   &  1.50  &  0.84  \\
  0.5   &  ...  &  ...   &  ...   &  ...   &  0.60  \\
\noalign{\smallskip}
\hline
\noalign{\smallskip}
\multicolumn{6}{c}{ EHZ given as $(s_1,s_6)$ } \\
\noalign{\smallskip}
\hline
\noalign{\smallskip}
  1.5   &  7.38 &  7.15  &  5.96  &   4.43  &  3.29  \\
  1.25  &  ...  &  5.50  &  5.48  &   3.80  &  2.64  \\
  1.0   &  ...  &  ...   &  4.25  &   3.41  &  2.19  \\
  0.75  &  ...  &  ...   &  ...   &   2.40  &  1.55  \\
  0.5   &  ...  &  ...   &  ...   &   ...   &  0.92  \\
\enddata
\end{deluxetable}

\clearpage


\thispagestyle{empty} 
\begin{landscape}
\begin{deluxetable}{lcccccccccccccccccccc} 
\tabletypesize{\scriptsize}
\tablecaption{Habitability Classification for Stellar Binary Systems}
\tablewidth{0pt}
\tablehead{
Primary & \multicolumn{20}{c} {Secondary} \\
...     &           \multicolumn{4}{c}{$M_2 = 1.5~M_\odot$} & \multicolumn{4}{c}{$M_2 = 1.25~M_\odot$} &
                    \multicolumn{4}{c}{$M_2 = 1.0~M_\odot$} & \multicolumn{4}{c}{$M_2 = 0.75~M_\odot$} &
                    \multicolumn{4}{c}{$M_2 = 0.5~M_\odot$} \\
\noalign{\smallskip}
\hline
\noalign{\smallskip}
 $M_1$       & \multicolumn{20}{c} {Habitability Classification} \\
 ...         &  P   &  PT  &  ST  &  S   &  P   &  PT  &  ST  &  S   &  P   &  PT  &  ST  &  S   &  P   &  PT  &  ST  &  S   &  P   &  PT  &  ST  &  S   \\
$(M_\odot)$  & (AU) & (AU) & (AU) & (AU) & (AU) & (AU) & (AU) & (AU) & (AU) & (AU) & (AU) & (AU) & (AU) & (AU) & (AU) & (AU) & (AU) & (AU) & (AU) & (AU)
}
\startdata
\multicolumn{21}{c}{ CHZ given as $(s_2,s_4)$ } \\
\noalign{\smallskip}
\hline
\noalign{\smallskip}
  1.5   &  1.63 &  1.86 & 13.89 & 18.31 &  1.64 &  1.72 & 13.01 & 17.18 &  1.50 & ...$^\dagger$ & 12.11 & 16.02 &  0.87 & ...$^\dagger$ & 11.16 & 14.79 &  0.65 & ...$^\dagger$ & 10.19 & 13.50  \\
  1.25  &  ...  &  ...  &  ...  &  ...  &  1.19 &  1.42 & 10.17 & 14.01 &  1.23 &  1.35         &  9.41 & 12.98 &  0.96 & ...$^\dagger$ &  8.59 & 11.88 &  0.58 & ...$^\dagger$ &  7.75 & 10.73  \\
  1.0   &  ...  &  ...  &  ...  &  ...  &  ...  &  ...  &  ...  &  ...  &  0.91 &  1.12         &  7.75 & 11.01 &  0.93 &  0.95         &  7.00 &  9.98 &  0.53 & ...$^\dagger$ &  6.22 &  8.89  \\
  0.75  &  ...  &  ...  &  ...  &  ...  &  ...  &  ...  &  ...  &  ...  &  ...  &  ...          &  ...  &  ...  &  0.50 &  0.65         &  4.26 &  6.37 &  0.47 & ...$^\dagger$ &  3.70 &  5.57  \\
  0.5   &  ...  &  ...  &  ...  &  ...  &  ...  &  ...  &  ...  &  ...  &  ...  &  ...          &  ...  &  ...  &  ...  &  ...          &  ...  &  ...  &  0.18 &  0.26         &  1.55 &  2.53  \\
\noalign{\smallskip}
\hline
\noalign{\smallskip}
\multicolumn{21}{c}{ GHZ given as $(s_1,s_5)$ } \\
\noalign{\smallskip}
\hline
\noalign{\smallskip}
  1.5   &  1.43 &  2.32 & 12.22 & 22.85 &  1.44 &  2.14 & 11.44 & 21.44 &  1.53 &  2.06 & 10.65 & 19.99 &  1.70 &  1.87 &  9.82 & 18.45 &  1.55 & ...$^\dagger$ &  8.96 & 16.85  \\
  1.25  &  ...  &  ...  &  ...  &  ...  &  1.05 &  1.75 &  8.97 & 17.27 &  1.08 &  1.65 &  8.30 & 15.99 &  1.16 &  1.45 &  7.58 & 14.64 &  1.27 & ...$^\dagger$ &  6.83 & 13.22  \\
  1.0   &  ...  &  ...  &  ...  &  ...  &  ...  &  ...  &  ...  &  ...  &  0.80 &  1.36 &  6.85 & 13.46 &  0.82 &  1.16 &  6.18 & 12.19 &  0.94 &  1.01         &  5.50 & 10.86  \\
  0.75  &  ...  &  ...  &  ...  &  ...  &  ...  &  ...  &  ...  &  ...  &  ...  &  ...  &  ...  &  ...  &  0.44 &  0.78 &  3.77 &  7.69 &  0.46 &  0.60         &  3.28 &  6.72  \\
  0.5   &  ...  &  ...  &  ...  &  ...  &  ...  &  ...  &  ...  &  ...  &  ...  &  ...  &  ...  &  ...  &  ...  &  ...  &  ...  &  ...  &  0.16 &  0.30         &  1.38 &  3.00  \\
\noalign{\smallskip}
\hline
\noalign{\smallskip}
\multicolumn{21}{c}{ EHZ given as $(s_1,s_6)$ } \\
\noalign{\smallskip}
\hline
\noalign{\smallskip}
  1.5   &  1.43 &  3.44 & 12.22 & 33.89 &  1.44 &  3.14 & 11.44 & 31.80 &  1.53 &  3.02 & 10.65 & 29.64 &  1.70 &  2.75 &  9.82 & 27.36 &  2.03 &  2.63 &  8.96 & 24.99  \\
  1.25  &  ...  &  ...  &  ...  &  ...  &  1.05 &  2.55 &  8.97 & 25.18 &  1.08 &  2.40 &  8.30 & 23.32 &  1.16 &  2.10 &  7.58 & 21.34 &  1.37 &  1.92 &  6.83 & 19.28  \\
  1.0   &  ...  &  ...  &  ...  &  ...  &  ...  &  ...  &  ...  &  ...  &  0.80 &  1.97 &  6.85 & 19.41 &  0.82 &  1.66 &  6.18 & 17.58 &  0.94 &  1.45 &  5.50 & 15.66  \\
  0.75  &  ...  &  ...  &  ...  &  ...  &  ...  &  ...  &  ...  &  ...  &  ...  &  ...  &  ...  &  ...  &  0.44 &  1.10 &  3.77 & 10.90 &  0.46 &  0.85 &  3.28 &  9.51  \\
  0.5   &  ...  &  ...  &  ...  &  ...  &  ...  &  ...  &  ...  &  ...  &  ...  &  ...  &  ...  &  ...  &  ...  &  ...  &  ...  &  ...  &  0.16 &  0.42 &  1.38 &  4.14  \\
\enddata
\tablecomments{
P-type habitability is given at distances below P, PT-type habitability at distances between P and PT, ST-habitability
at distances between ST and S, and S-type habitability at distances beyond S.  No habitability is identified in the
distance range between PT and ST.  See Figs. 8 and 9 for additional details for selected cases.
In some cases the limit for P-type habitability is given by the expiration of the RHZ (see Table 9), i.e., where the
outer and inner limits of the RHLs intercept, an occurrence unrelated to the planetary orbital stability requirement.
In these cases no PT-type habitability is found as indicated by a dagger ($\dag$).  The results are given in terms of
stellar separation distances $2a$.  
}
\end{deluxetable}
\end{landscape}


\clearpage

\begin{thebibliography}{}

\bibitem[Abramowitz \& Stegun(1972)]{abr72}
Abramowitz, M., \& Stegun, I. A. 1972, Handbook of Mathematical Functions
with Formulas, Graphs, and Mathematical Tables, 9th printing (New York: Dover)

\bibitem[Asghari et al.(2004)]{asg04}
Asghari, N., Broeg, C., Carone, L., et al., 2004, \aap, 426, 353

\bibitem[Beyer(1987)]{bey87}
Beyer, W. H. 1987, Handbook of Mathematical Sciences, 6th edn.
(Boca Raton: CRC Press)

\bibitem[Bonavita \& Desidera(2007)]{bon07}
Bonavita, M., \& Desidera, S. 2007, \aap, 468, 721

\bibitem[Bronshtein \& Semendyayev(1997)]{bro97}
Bronshtein, I. N., \& Semendyayev, K. A. 1997, Handbook of Mathematics, 3rd edn.
(New York: Springer)

\bibitem[Castelli \& Kurucz(2004)]{cas04}
Castelli, F., \& Kurucz, R. L. 2004, in IAU Symp. 210, Modelling of
Stellar Atmospheres ed. N. E. Piskunov, W. W. Weiss, \& D. F. Gray
(San Francisco: ASP), CD-ROM Poster 20; arXiv:astro-ph/0405087

\bibitem[Chabrier(2003)]{cha03}
Chabrier, G. 2003, \pasp, 115, 763

\bibitem[Cuntz \& Yeager (2009)]{cun09}
Cuntz, M., \& Yeager, K. E. 2009, \apjl, 697, L86

\bibitem[Cuntz et al.(2007)]{cun07}
Cuntz, M., Eberle, J., \& Musielak, Z. E. 2007, \apjl, 669, L105

\bibitem[Cuntz et al.(2012)]{cun12}
Cuntz, M., von Bloh, W., Schr\"oder, K.-P., Bounama, C., \& Franck, S. 2012,
Int. J. Astrobiol., 11, 15

\bibitem[Doyle et al.(2011)]{doy11}
Doyle, L. R., Carter, J. A., Fabrycky, D. C., et al. 2011, Science, 333, 1602

\bibitem[Duquennoy \& Mayor(1991)]{duq91}
Duquennoy, A., \& Mayor, M. 1991, \aap, 248, 485

\bibitem[Dvorak(1982)]{dvo82}
Dvorak, R. 1982, \"Osterreichische Akademie der Wissenschaften,
Mathematisch-Naturwissenschaftliche Klasse, Sitzungsberichte Abt. 2,
191 (10), 423

\bibitem[Dvorak(1984)]{dvo84}
Dvorak, R. 1984, Celest. Mech. Dyn. Astron., 34, 369

\bibitem[Dvorak(1986)]{dvo86}
Dvorak, R. 1986, \aap, 167, 379

\bibitem[Dvorak et al.(2010)]{dvo10} 
Dvorak, R., Pilat-Lohinger, E., Bois, E., et al. 2010, Astrobiology, 10, 33

\bibitem[Eberle \& Cuntz(2010)]{ebe10}
Eberle, J., \& Cuntz, M. 2010, \apjl, 721, L168

\bibitem[Eberle et al.(2008)]{ebe08}
Eberle, J., Cuntz, M., \& Musielak, Z. E. 2008, \aap, 489, 1329

\bibitem[Eggenberger et al.(2004)]{egg04}
Eggenberger, A., Udry, S., \& Mayor, M. 2004, \aap, 417, 353

\bibitem[Eggenberger et al.(2007)]{egg07}
Eggenberger, A., Udry, S., Chauvin, G., Beuzit, J.-L., Lagrange, A.-M.,
S\'egransan, D., \& Mayor, M. 2007, \aap, 474, 273

\bibitem[Eggl et al.(2012)]{egg12}
Eggl, S., Pilat-Lohinger, E., Georgakarakos, N., Gyergyovits, M., \&
Funk, B. 2012, \apj, 752, A74

\bibitem[Fatuzzo et al.(2006)]{fat06}
Fatuzzo, M., Adams, F. C., Gauvin, R., \& Proszkow, E. M. 2006,
\pasp, 118, 1510

\bibitem[Forget(2013)]{for13}
Forget, F. 2013, Int. J. Astrobiol., 12, 177

\bibitem[Forget \& Pierrehumbert(1997)]{for97}
Forget, F., \& Pierrehumbert, R. T. 1997, Science, 278, 1273

\bibitem[Gray(2005)]{gra05}
Gray, D. F. 2005, The Observation and Analysis of Stellar Photospheres,
2nd edn. (Cambridge: Cambridge University Press)

\bibitem[Haghighipour et al.(2010)]{hag10}
Haghighipour, N., Dvorak, R., \& Pilat-Lohinger, E. 2010,
in Planets in Binary Systems, ed. N. Haghighipour,
Astrophysics and Space Science Library, Vol. 366
(New York: Springer Science + Business Media), p. 285

\bibitem[Halevy et al.(2009)]{hal09}
Halevy, I., Pierrehumbert, R. T., \& Schrag, D. P. 2009, J. Geophys. Res.,
114, D18112

\bibitem[Holman \& Wiegert(1999)]{hol99}
Holman, M. J., \& Wiegert, P. A. 1999, \aj, 117, 621

\bibitem[Horner \& Jones(2010)]{hor10}
Horner, J., \& Jones, B. W. 2010, Int. J. Astrobiol., 9, 273

\bibitem[Jackson(1999)]{jac99}
Jackson, J. D. 1999, Classical Electrodynamics, 3rd edn.,
John Wiley \& Sons, Inc., Hoboken, NJ

\bibitem[Jones et al.(2001)]{jon01}
Jones, B. W., Sleep, P. N., \& Chambers, J. E. 2001, \aap, 366, 254

\bibitem[Jones \& Sleep(2010)]{jon10}
Jones, B. W., \& Sleep, P. N. 2010, \mnras, 407, 1259

\bibitem[Kaltenegger et al.(2010)]{kal10}
Kaltenegger, L., Eiroa, C., Ribas, I., et al. 2010, Astrobiology, 10, 103

\bibitem[Kane \& Hinkel(2013)]{kan13}
Kane, S. R., \& Hinkel, N. R. 2013, \apj, 762, A7

\bibitem[Kasting \& Catling(2003)]{kas03}
Kasting, J. F., \& Catling, D. 2003, \araa, 41, 429

\bibitem[Kasting et al.(1993)]{kas93}
Kasting, J. F., Whitmire, D. P., \& Reynolds, R. T. 1993,
Icarus, 101, 108

\bibitem[Kopparapu \& Barnes(2010)]{kop10}
Kopparapu, R. K., \& Barnes, R. 2010, \apj, 716, 1336

\bibitem[Kroupa(2001)]{kro01}
Kroupa, P. 2001, \mnras, 322, 231

\bibitem[Kroupa(2002)]{kro02}
Kroupa, P. 2002, Science, 295, 82

\bibitem[Kurucz(2005)]{kur05}
Kurucz, R. L. 2005, Mem. S. A. It., 8, 14

\bibitem[Lada(2006)]{lad06}
Lada, C. J. 2006, \apjl, 640, L63

\bibitem[Lammer et al.(2009)]{lam09}
Lammer, H., Bredeh\"oft, J. H., Coustenis, A., et al. 2009,
\aapr, 17, 181

\bibitem[Lammer et al.(2010)]{lam10}
Lammer, H., Selsis, F., Chassefi{\`e}re, E., et al. 2010, Astrobiology, 10, 45

\bibitem[Lucarini et al.(2013)]{luc13}
Lucarini, V., Pascale, S., Boschi, R., Kirk, E., \& Iro, N. 2013, AN, 334, 576

\bibitem[Menou \& Tabachnik(2003)]{men03}
Menou, K., \& Tabachnik, S. 2003, \apj, 583, 473

\bibitem[Mischna et al.(2000)]{mis00}
Mischna, M. A., Kasting, J. F., Pavlov, A., \& Freedman, R. 2000,
Icarus, 145, 546

\bibitem[Musielak et al.(2005)]{mus05}
Musielak, Z. E., Cuntz, M., Marshall, E. A., \& Stuit, T. D. 2005,
\aap, 434, 355

\bibitem[Noble et al.(2002)]{nob02}
Noble, M., Musielak, Z. E., \& Cuntz, M. 2002, \apj, 572, 1024

\bibitem[Orosz et al.(2012)]{oro12}
Orosz, J. A., Welsh, W. F., Carter, J. A., et al.  2012, Science,
337, 1511

\bibitem[Patience et al.(2002)]{pat02}
Patience, J., White, R. J., Ghez, A. M., et al. 2002, \apj, 581, 654

\bibitem[Perryman(2011)]{per11}
Perryman, M. 2011, The Exoplanet Handbook (Cambridge: Cambridge University Press)

\bibitem[Quarles et al.(2011)]{qua11}
Quarles, B., Eberle, J., Musielak, Z. E., \& Cuntz, M. 2011, \aap, 533, A2

\bibitem[Quarles et al.(2012)]{qua12}
Quarles, B., Musielak, Z. E., \& Cuntz, M. 2012, \apj, 750, A14

\bibitem[Quintana \& Lissauer(2010)]{qui10}
Quintana, E. V., \& Lissauer, J. J. 2010, in Planets in Binary Systems,
ed. N. Haghighipour, Astrophysics and Space Science Library, Vol. 366
(New York: Springer Science + Business Media), p. 265

\bibitem[Raghavan et al.(2006)]{rag06}
Raghavan, D., Henry, T. J., Mason, B. D., et al. 2006, \apj, 646, 523

\bibitem[Raghavan et al.(2010)]{rag10}
Raghavan, D., McAlister, H. A., Henry, T. J., et al. 2010, \apjs, 190, 1

\bibitem[Ramm et al.(2009)]{ram09}
Ramm, D. J., Pourbaix, D., Hearnshaw, J. B., \& Komonjinda, S., 2009,
\mnras, 394, 1695

\bibitem[Reid(1987)]{rei87}
Reid, N. 1987, \mnras, 225, 873

\bibitem[Roell et al.(2012)]{roe12}
Roell, T., Neuh\"auser, R., Seifahrt, A., \& Mugrauer, M. 2012, \aap,
542, A92

\bibitem[S\'andor et al.(2007)]{san07}
S\'andor, Zs., S{\H u}li, \'A., \'Erdi, B., Pilat-Lohinger, E., Dvorak, R.
2007, \mnras, 375, 1495

\bibitem[Scalo et al.(2007)]{sca07}
Scalo, J., Kaltenegger, L., Segura, A. G., et al. 2007, Astrobiology, 7, 85

\bibitem[Sato et al.(2003)]{sato03}
Sato, B., Ando, H., Kambe, E., et al., 2003, \apj, 597, L157

\bibitem[Sato et al.(2013)]{sato13}
Sato, B., Omiya, M., Harakawa, H., et al., 2013, \pasj, 65, 85

\bibitem[Satyal et al.(2013)]{saty13}
Satyal, S., Quarles, B., \& Hinse, T. C. 2013, \mnras, 433, 2215

\bibitem[Selsis et al.(2007)]{sel07}
Selsis, F., Kasting, J. F., Levrard, B., Paillet, J., Ribas, I., \&
Delfosse, X. 2007, \aap, 476, 1373

\bibitem[Selsis et al.(2008)]{sel08}
Selsis, F., Kaltenegger, L., Paillet, J., et al. 2008, Physica Scripta, 130, 014032

\bibitem[Setiawan et al.(2003)]{set03}
Setiawan, J., Hatzes, A. P., von der L\"uhe, et al. 2003, \aap, 398, L19

\bibitem[Slawson et al.(2011)]{sla11}
Slawson, R. W., Pr{\v s}a, A., Welsh, W. F., et al. 2011, \aj, 142, 160

\bibitem[Stix(2004)]{sti04}
Stix, M. 2004, The Sun: An Introduction, 2nd edn. (A\&A Library, Berlin: Springer)

\bibitem[Szenkovits \& Mak\'o(2008)]{sze08}
Szenkovits, F., \& Mak\'o, Z. 2008, Celest. Mech. Dyn. Astron., 101, 273

\bibitem[Takeda et al.(2008)]{tak08}
Takeda, G., Kita, R., \& Rasio, F. A. 2008, \apj, 683, 1063

\bibitem[Tarter et al.(2007)]{tar07}
Tarter, J. C., Backus, P. R., Mancinelli, R. L., et al. 2007, Astrobiology, 7, 30

\bibitem[Trilling et al.(2007)]{tri07}
Trilling, D. E., Stansberry, J. A., Stapelfeldt, K. R., et al. 2007, \apj, 658, 1289

\bibitem[Underwood et al.(2003)]{und03}
Underwood, D. R., Jones, B. W., \& Sleep, P. N. 2003, Int. J. Astrobiol., 2, 289

\bibitem[Williams \& Pollard(2002)]{wil02}
Williams, D. M., \& Pollard, D. 2002, Int. J. Astrobiol., 1, 61

\bibitem[Zahnle et al.(2007)]{zah07}
Zahnle, K., Arndt, N., Cockell, C., Halliday, A., Nisbet, E., Selsis, F., \& Sleep, N. H.
2007, \ssr, 129, 35

\end{thebibliography}
\end{document}